\documentclass[preprint,prd,aps,showpacs,showkeys,nofootinbib]{revtex4}
\usepackage{graphicx}
\usepackage{dcolumn}
\usepackage{bm}
\begin{document}

%\preprint{hep-ph/0412147}

\title{Renormalization and two loop electroweak corrections to lepton
anomalous dipole moments in the standard model and beyond (I):
heavy fermion contributions}

\author{Tai-Fu Feng\footnote{email:fengtf@dlut.edu.cn}, Xiu-Yi Yang}

\affiliation{Department of Physics, Dalian University of
Technology, Dalian, 116024, China}

\date{\today}

\begin{abstract}

Applying effective Lagrangian method and on-shell scheme,
we analyze the electroweak corrections to anomalous dipole
moments of lepton from some special two loop diagrams in which a closed
heavy fermion loop is attached to the virtual gauge bosons or Higgs fields.
As the masses of virtual fermions in
inner loop are much heavier than the electroweak scale, we verify
the final results satisfying the decoupling theorem explicitly
if the interactions among Higgs and heavy fermions do not contain
the nondecoupling couplings.
At the decoupling limit, we also present the leading corrections
to lepton anomalous dipole moments from those two loop diagrams
in some popular extensions of the standard model, such as the
fourth generation, supersymmetry, universal extra dimension,
and the littlest Higgs with T-parity.

\end{abstract}


\maketitle

\section{Introduction\label{sec1}}
\indent\indent
At both aspects of experiment and theory, the magnetic dipole moments (MDMs)
of leptons draw great attention of physicists because of their obvious
importance. The anomalous dipole moments of lepton not only
provide a potential window to detect new physics beyond the standard model (SM),
but also can be used for testing loop effect in electroweak theories.
The current experimental world average of the muon MDM is\cite{exp}
\begin{eqnarray}
&&a_{_\mu}^{exp}=11\;659\;208\;\pm\;6\;\times 10^{-10}\;.
\label{data}
\end{eqnarray}

Contributions to the MDM of muon are generally
divided into three sectors: QED loops, hadronic contributions as
well as electroweak corrections.  The largest uncertainty of the SM prediction
originates from the evaluation of hadronic vacuum polarization and
light-by-light corrections. Depending on which evaluation
of hadronic vacuum polarization is chosen, the differences between
the SM predictions and experimental result lie in the range
$1.3\sigma\sim3.8\sigma$\cite{Rafael1,Jegerlehner}.

For the electroweak corrections, the standard one loop contribution
amounts to $19.5\times10^{-10}$, and the one loop corrections from new physics
sector are generally suppressed by $\Lambda_{_{\rm EW}}^2/\Lambda_{_{\rm NP}}^2$.
Here $\Lambda_{_{\rm EW}}$ denotes the electroweak energy scale, and
$\Lambda_{_{\rm NP}}$ denotes the energy scale of new physics.
Comparing with the analysis at one loop level, the two loop analysis is
more complicated and less advanced. Utilizing the heavy mass expansion
approximation (HME) together with the projection operator method,
the authors of Ref.\cite{czarnecki} have evaluated the two loop standard
electroweak corrections to muon MDM. Within the framework of CP
conservation, Ref.\cite{heinemeyer} presents the
supersymmetric corrections from some special two-loop diagrams
where a closed chargino (neutralino) or scalar fermion loop is
inserted into those two-Higgs-doublet one-loop diagrams. Ref.\cite{geng}
discusses the contributions to muon MDM from the effective
vertices $H^\pm W^\mp\gamma, h_0(H_0)\gamma\gamma$ which are induced
by the scalar quarks of the third generation. Furthermore, the
contributions from two loop Bar-Zee-type diagrams to the electric
dipole moments (EDMs) of light fermions are discussed extensively in
literature \cite{Pilaftsis}.

In this paper, we calculate the corrections to the anomalous dipole
moments of lepton from some special diagrams in which a closed heavy
fermion loop is attached to the virtual electroweak gauge or Higgs fields.
The effective Lagrangian method can yield the one loop electroweak corrections to
lepton MDMs and EDMs exactly in the SM and beyond, and
has been adopted to calculate the two loop supersymmetric corrections
for the branching ratio of $b\rightarrow s\gamma$ \cite{Feng1}, neutron EDM \cite{Feng2}
and lepton MDMs and EDMs \cite{Feng3,Feng4}. In concrete
calculation, we assume that all external leptons and photon
are off-shell, then expand the amplitude of corresponding triangle
diagrams according to the external momenta of leptons and photon.
Using loop momentum translating invariance, we formulate the sum of
amplitude from those triangle diagrams corresponding to same self energy in
the form which explicitly satisfies the Ward identity required by
the QED gauge symmetry, then get all dimension 6 operators
together with their coefficients. After the equations of
motion are applied to external leptons, higher dimensional operators, such
as dimension 8 operators, also contribute to the lepton MDMs  and
EDMs in principle. However, the contributions of dimension 8
operators contain the additional suppression factor
$m_l^2/\Lambda_{_{\rm EW}}^2$ comparing with that of dimension
6 operators, where $m_l$ is the mass of lepton. Setting
$\Lambda_{_{\rm EW}}\sim100{\rm GeV}$,
one obtains easily that this suppression factor is
about $10^{-6}$ for the muon lepton. Under current experimental precision, it implies
that the contributions of all higher dimension operators ($D\ge8$)
can be neglected safely.

We adopt the naive dimensional regularization with the
anticommuting $\gamma_{_5}$ scheme, where there is no distinction
between the first 4 dimensions and the remaining $D-4$ dimensions.
Since the bare effective Lagrangian contains the ultraviolet
divergence which is induced by divergent subdiagrams, we give the
renormalized results in the on-mass-shell scheme \cite{onshell}.
Additional, we adopt the nonlinear $R_\xi$ gauge with $\xi=1$ for
simplification \cite{nonlinear-R-xi}. This special gauge-fixing term
guarantees explicit electromagnetic gauge invariance throughout the calculation,
not just at the end because the choice of gauge-fixing term
eliminates the $\gamma W^\pm G^\mp$ vertex in the Lagrangian.

This paper is composed of the sections as follows.
In section \ref{sec2}, we introduce the effective Lagrangian
method and our notations. We will demonstrate how to obtain
the identities among two loop integrals from the loop momentum
translating invariance through an example, then obtain the
corrections from the relevant diagrams to the lepton MDMs and EDMs.
Section \ref{sec3} is devoted to the analysis and
discussion in some concrete electroweak models.
In section \ref{sec4}, we give our conclusion. Some
tedious formulae are collected in the appendices.

\section{The corrections from the relating diagrams\label{sec2}}
\indent\indent
The lepton MDMs and EDMs can actually be written as the operators
\begin{eqnarray}
&&{\cal L}_{_{MDM}}={e\over4m_{_l}}\;a_{_l}\;\bar{l}\sigma^{\mu\nu}
l\;F_{_{\mu\nu}}
\;,\nonumber\\
&&{\cal L}_{_{EDM}}=-{i\over2}\;d_{_l}\;\bar{l}\sigma^{\mu\nu}\gamma_5
l\;F_{_{\mu\nu}}\;.
\label{adm}
\end{eqnarray}
Here, $\sigma_{\mu\nu}=i[\gamma_\mu,\gamma_\nu]/2$, $l$ denotes the lepton
fermion which is on-shell, $F_{_{\mu\nu}}$ is the electromagnetic field strength,
$m_{_l}$ is the lepton mass and $e$ represents the electric charge, respectively.

%%%%%%%%%%%%%%%%%%%%%%%%%BEGIN MODIFICATION%%%%%%%%%%%%%%%%%%%%%%%%%%%%%
%In Ref.\cite{heinemeyer}, the MDMs of leptons are defined through the
%current algebra method. Including the radiative corrections, one can
%generally express the effective coupling among the photon and on-shell
%leptons in momentum space as
%\begin{eqnarray}
%&&\Gamma_{_\rho}=ie\;\bar{l}(p^\prime)\;I_{_\rho}l(p)
%\;,\nonumber\\
%&&I_{_\rho}=F_{_V}(k^2)\gamma_\rho+{[/\!\!\!k,\gamma_\rho]\over4m_{_l}}F_{_M}(k^2)
%-{i[/\!\!\!k,\gamma_\rho]\gamma_5\over2e}F_{_P}(k^2)+\cdots\;.
%\label{current-algebra}
%\end{eqnarray}
%Where $k\rightarrow0$ denotes the incoming momentum of photon,
%and $p,\;p^\prime=p+k$ denote the momenta of incoming and outgoing
%leptons separately. The lepton MDMs and EDMs are formulated as:
%\begin{eqnarray}
%&&a_{_l}=F_{_M}(0)\;,
%\nonumber\\
%&&d_{_l}=F_{_P}(0)\;.
%\label{mdm-current}
%\end{eqnarray}
%Applying the equations of motion for leptons $\overline{u}_l(p^\prime)
%(/\!\!\!p^\prime-m_{_l})=0,\;(/\!\!\!p-m_{_l})u_l(p)=0$, we can verify
%the equivalence of two definitions directly.
%%%%%%%%%%%%%%%%%%%%%%%%%%%END MODIFICATION%%%%%%%%%%%%%%%%%%%%%%%%%%%%%

%%%%%%%%%%%%%%%%%%%%%%%%%BEGIN MODIFICATION%%%%%%%%%%%%%%%%%%%%%%%%%%%%%
It is convenient to get the corrections from loop diagrams
to lepton MDMs and EDMs in terms of the effective Lagrangian method,
if the loop diagrams contain the virtual fields which are much heavier
than the external lepton, i.e. $m_{_{V}}\gg m_{_l}$ with $m_{_{V}}$
denoting the mass scale of virtual fields.
Since $/\!\!\!p^\prime=/\!\!\!p=m_{_l}\ll m_{_{V}}$ for on-shell leptons and
$/\!\!\!k\rightarrow0\ll m_{_{V}}$ for photon, we can expand the amplitude of
corresponding triangle diagrams according to the external momenta of
leptons and photon. The two loop diagrams also contain some
virtual light freedoms generally, such as virtual neutrinos, charged leptons or
photon, and it is unsuitable to expand the propagators
of light freedoms in powers of external momenta obviously.
In order to obtain the corrections to lepton MDM and EDM properly,
we should firstly match the amplitude of two loop diagrams from full theory to
that of corresponding diagrams from effective theory which is composed by the QED
Lagrangian and some higher dimension operators of light freedoms, then extract the Wilson
coefficients of those high dimension operators which are only depend on
the masses of virtual heavy freedoms as well as the possible
matching scales. Finally, we strictly analyze the amplitude of
corresponding diagrams from effective theory to obtain the contributions
from the virtual light freedoms to lepton MDMs and EDMs.
%%%%%%%%%%%%%%%%%%%%%%%%%%%END MODIFICATION%%%%%%%%%%%%%%%%%%%%%%%%%%%%%
As discussed in the section \ref{sec1}, it is enough to
retain only those dimension 6 operators in later calculations:
\begin{eqnarray}
&&{\cal O}_{_1}^\mp={1\over(4\pi)^2}\;\bar{l}\;(i/\!\!\!\!{\cal D})^3
\omega_\mp\;l\;,\nonumber\\
&&{\cal O}_{_2}^\mp={eQ_{_f}\over(4\pi)^2}\;\overline{(i{\cal D}_{_\mu}l)}
\gamma^\mu F\cdot\sigma\omega_\mp l\;,\nonumber\\
&&{\cal O}_{_3}^\mp={eQ_{_f}\over(4\pi)^2}\;\bar{l}F\cdot\sigma\gamma^\mu
\omega_\mp(i{\cal D}_{_\mu}l)\;,\nonumber\\
&&{\cal O}_{_4}^\mp={eQ_{_f}\over(4\pi)^2}\;\bar{l}(\partial^\mu F_{_{\mu\nu}})
\gamma^\nu\omega_\mp l\;,\nonumber\\
&&{\cal O}_{_5}^\mp={m_{_l}\over(4\pi)^2}\;\bar{l}\;(i/\!\!\!\!{\cal D})^2
\omega_\mp\;l\;,\nonumber\\
&&{\cal O}_{_6}^\mp={eQ_{_f}m_{_l}\over(4\pi)^2}\;\bar{l}\;F\cdot\sigma
\omega_\mp\;l\;,\nonumber\\
\label{ops}
\end{eqnarray}
with ${\cal D}_{_\mu}=\partial_{_\mu}+ieA_{_\mu}$ and $\omega_\mp=(1\mp\gamma_5)/2$.

%%%%%%%%%%%%%%%%%%%%%%%%%BEGIN MODIFICATION%%%%%%%%%%%%%%%%%%%%%%%%%%%%%
Certainly, all dimension 6 operators in Eq.(\ref{ops}) induce the effective
couplings among photons and leptons. The effective vertices with one
external photon are written as
\begin{eqnarray}
&&{\cal O}_{_1}^\mp={ieQ_{_f}\over(4\pi)^2}\Big\{\Big((p+k)^2+p^2\Big)\gamma_\rho
+(/\!\!\!p+/\!\!\!k)\gamma_\rho/\!\!\!p\Big\}\omega_\mp\;,\nonumber\\
&&{\cal O}_{_2}^\mp={ieQ_{_f}\over(4\pi)^2}(/\!\!\!p+/\!\!\!k)
[/\!\!\!k,\gamma_\rho]\omega_\mp\;,\nonumber\\
&&{\cal O}_{_3}^\mp={ieQ_{_f}\over(4\pi)^2}[/\!\!\!k,\gamma_\rho]/\!\!\!p
\omega_\mp\;,\nonumber\\
&&{\cal O}_{_4}^\mp={ieQ_{_f}\over(4\pi)^2}\Big(k^2\gamma_\rho
-/\!\!\!kk_\rho\Big)\omega_\mp\;,\nonumber\\
&&{\cal O}_{_5}^\mp={ieQ_{_f}\over(4\pi)^2}\Big\{(/\!\!\!p+/\!\!\!k)
\gamma_\rho+\gamma_\rho/\!\!\!p\Big\}m_{_l}\omega_\mp\;,\nonumber\\
&&{\cal O}_{_6}^\mp={ieQ_{_f}\over(4\pi)^2}[/\!\!\!k,\gamma_\rho]
m_{_l}\omega_\mp\;.
\label{Feynman-rule-opers}
\end{eqnarray}
%%%%%%%%%%%%%%%%%%%%%%%%%%%END MODIFICATION%%%%%%%%%%%%%%%%%%%%%%%%%%%%%

%----------------------------BEGIN NEW ADDING----------------------------------------
If the full theory is invariant under the combined transformation of charge
conjugation, parity and time reversal (CPT), the induced effective
theory preserves the symmetry after the heavy freedoms are integrated out.
The fact implies the Wilson coefficients of the operators ${\cal O}_{_{2,3,6}}^\mp$
satisfying the relations
\begin{eqnarray}
&&C_2^\mp=C_3^{\mp*},\;C_6^+=C_6^{-*}\;,
\label{cpt1}
\end{eqnarray}
where $C_i^\mp\;(i=1,2,\cdots,6)$ represent the Wilson coefficients
of the corresponding operators ${\cal O}_{_i}^\mp$ in the effective Lagrangian. After applying
the equations of motion to external leptons, we find that the concerned terms
in the effective Lagrangian are transformed into
\begin{eqnarray}
&&C_2^\mp{\cal O}_{_2}^\mp+C_2^{\mp*}{\cal O}_{_3}^\mp+C_6^+{\cal O}_{_6}^+
+C_6^{+*}{\cal O}_{_6}^-
\nonumber\\
&&\hspace{-0.6cm}\Rightarrow
(C_2^++C_2^{-*}+C_6^+){\cal O}_{_6}^++(C_2^{+*}+C_2^-+C_6^{+*}){\cal O}_{_6}^-
\nonumber\\
&&\hspace{-0.6cm}=
{eQ_{_f}m_{_l}\over(4\pi)^2}\Big\{\Re(C_2^++C_2^{-*}+C_6^+)\;\bar{l}\;\sigma^{\mu\nu}\;l
+i\Im(C_2^++C_2^{-*}+C_6^+)\;\bar{l}\;\sigma^{\mu\nu}\gamma_5\;l\Big\}F_{\mu\nu}\;.
\label{cpt2}
\end{eqnarray}
Here, $\Re(\cdots)$ denotes the operation to take the real part of a complex number,
and $\Im(\cdots)$ denotes the operation to take the imaginary part of a complex number.
Applying Eq.(\ref{adm}) and Eq.(\ref{cpt2}), we finally get
\begin{eqnarray}
&&a_l={4Q_{_f}m_{_l}^2\over(4\pi)^2}\Re(C_2^++C_2^{-*}+C_6^+)\;,
\nonumber\\
&&d_l=-{2eQ_{_f}m_{_l}\over(4\pi)^2}\Im(C_2^++C_2^{-*}+C_6^+)\;.
\label{cpt3}
\end{eqnarray}
In other words, the MDM of lepton is proportional to real part of
the effective coupling $C_2^++C_2^{-*}+C_6^+$, as well as the EDM of lepton
is proportional to imaginary part of the effective coupling $C_2^++C_2^{-*}+C_6^+$.
%----------------------------END NEW ADDING------------------------------------------

%%%%%%%%%%%%%%%%%%%%%%%%%BEGIN MODIFICATION%%%%%%%%%%%%%%%%%%%%%%%%%%%%%
After expanding the amplitude of corresponding triangle diagrams,
we extract the Wilson coefficients of operators in Eq.(\ref{Feynman-rule-opers})
which are formulated in the linear combinations of one and two loop
vacuum integrals in momentum space, then obtain the MDMs and EDMs of leptons.
Taking those diagrams in which a closed heavy fermion loop is inserted
into the propagator of charged gauge boson as an example, we show in detail
how to obtain the MDMs and EDMs of leptons through the effective Lagrangian
method.
%%%%%%%%%%%%%%%%%%%%%%%%%%%END MODIFICATION%%%%%%%%%%%%%%%%%%%%%%%%%%%%%

\subsection{The corrections from the diagrams where a closed heavy fermion
loop is inserted into the self energy of $W^\pm$ gauge boson}
\begin{figure}[t]
\setlength{\unitlength}{1mm}
\begin{center}
\begin{picture}(0,120)(0,0)
\put(-62,-10){\includegraphics{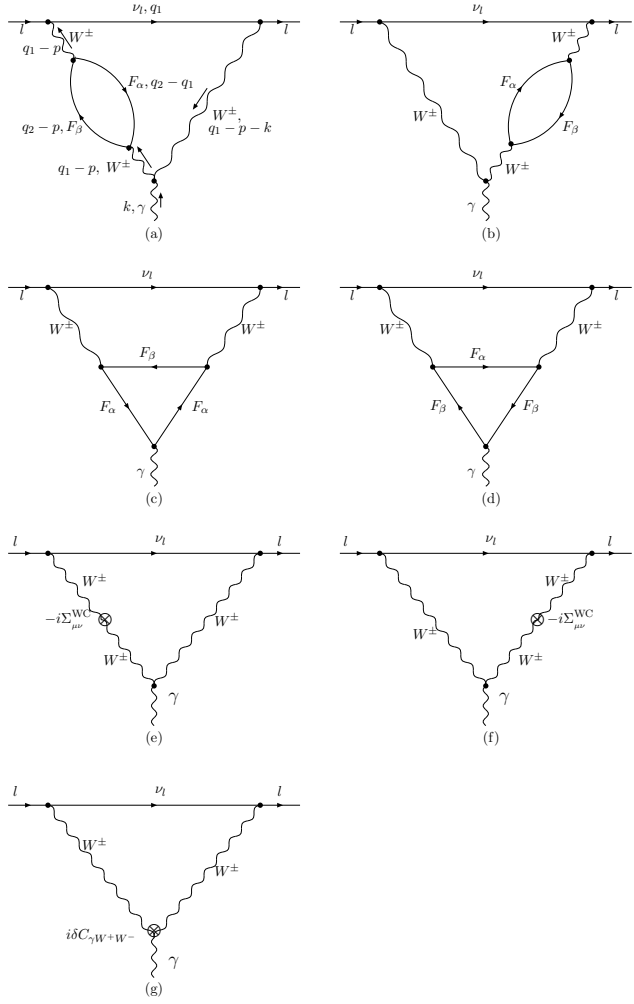}}
\end{picture}
\caption[]{The relating two-loop diagrams in which a closed heavy
fermion loop is attached to virtual $W^\pm$ bosons, where
the diagrams (e,f,g) contribute
the counter terms to cancel the ultraviolet
divergence arisen by divergent subdiagrams in (a,b,c,d) respectively.}
\label{fig1}
\end{center}
\end{figure}

\indent\indent
In order to get the amplitude of the diagrams in Fig.\ref{fig1}, one can
write the renormalizable interaction among the charged electroweak gauge
boson $W^\pm$ and the heavy fermions $F_{\alpha,\beta}$ in a more universal
form as
\begin{eqnarray}
&&{\cal L}_{_{WFF}}={e\over s_{_{\rm w}}}W^{-,\mu}\bar{F}_\alpha\gamma_\mu
(\zeta^L_{_{\alpha\beta}}\omega_-+\zeta^R_{_{\alpha\beta}}\omega_+)F_\beta
+h.c.\;,
\label{WFF}
\end{eqnarray}
where the concrete expressions of $\zeta^{L,R}_{_{\alpha\beta}}$ depend on the
models employed in our calculation. The conservation of electric charge
requires $Q_\beta-Q_\alpha=1$, where $Q_{\alpha,\beta}$ denote the electric
charges of the heavy fermions $F_{\alpha,\beta}$ respectively.

%%%%%%%%%%%%%%%%%%%%%%%%%%%%%%%%%%%%%%%%%%%%%%%%%%%%%%%%%%%%%%%%%%%
%%%%%%%%%%%%%%%%%%%%%%%%%%%%%%%%%%%%%%%%%%%%%%%%%%%%%%%%%%%%%%%%%%%

%%%%%%%%%%%%%%%%%%%%%%%%%BEGIN MODIFICATION%%%%%%%%%%%%%%%%%%%%%%%%%%%%%
Applying Eq.(\ref{WFF}), we write firstly the amplitude of those two loop
diagrams in Fig.\ref{fig1}. For example, the amplitude of Fig.\ref{fig1}(a)
can be formulated as
\begin{eqnarray}
%%%%%%%%%%%%%%%%%%%%%%%%%%%%%%%%%%%%%%%%%%%%%%%%%%%%%%%%%%%%%%%%%%%%%%%%%%%%%%%%%%%
&&i{\cal A}_{_{{\rm ww},\rho}}^{1(a)}(p,k)=-\overline{\psi}_{_f}\int{d^D q_1\over(2\pi)^D}
{d^D q_2\over(2\pi)^D}\Big(-i{e\Lambda_{_{\rm RE}}^\varepsilon\over\sqrt{2}
s_{_{\rm w}}}\Big)\gamma^\mu\omega_-{i/\!\!\!q_1\over
q_1^2}\Big(-i{e\Lambda_{_{\rm RE}}^\varepsilon\over
\sqrt{2}s_{_{\rm w}}}\Big)\gamma^\nu\omega_-\psi_{_f}
\nonumber\\
&&\hspace{2.8cm}\times
{-i\over(q_1-p-k)^2-m_{_{\rm w}}^2}
\Big\{ie\Big[-g_{\mu\sigma}(2p+k-2q_1)_\rho
+2(g_{\rho\mu}k_\sigma-g_{\rho\sigma}k_\mu)\Big]\Big\}
\nonumber\\
&&\hspace{2.8cm}\times
{-i\over(q_1-p)^2-m_{_{\rm w}}^2}{-i\over(q_1-p)^2-m_{_{\rm w}}^2}
{\bf Tr}\Bigg[\Big(i{e\Lambda_{_{\rm RE}}^\varepsilon\over s_{_{\rm w}}}\Big)
\gamma^\sigma\Big\{\zeta^{L*}_{_{\alpha\beta}}\omega_-
+\zeta^{R*}_{_{\alpha\beta}}\omega_+\Big\}
\nonumber\\
&&\hspace{2.8cm}\times
{i(/\!\!\!q_2-/\!\!\!q_1+m_{_{F_\alpha}})
\over(q_2-q_1)^2-m_{_{F_\alpha}}^2}\Big(i{e\Lambda_{_{\rm RE}}^\varepsilon\over
s_{_{\rm w}}}\Big)\gamma_\nu\Big\{\zeta^L_{_{\alpha\beta}}\omega_-
+\zeta^R_{_{\alpha\beta}}\omega_+\Big\}
{i(/\!\!\!q_2-/\!\!\!\!p+m_{_{F_\beta}})\over(q_2-p)^2-m_{_{F_\beta}}^2}\Bigg]\;.
%%%%%%%%%%%%%%%%%%%%%%%%%%%%%%%%%%%%%%%%%%%%%%%%%%%%%%%%%%%%%%%%%%%%%%%%%%%%%%%%%%%
\label{eq-wa1}
\end{eqnarray}
Here $\Lambda_{_{\rm RE}}$ denotes the renormalization scale that can take
any value in the range from the electroweak scale $\Lambda_{_{\rm EW}}$ to the
new physics scale $\Lambda_{_{\rm NP}}$ naturally, and
we adopt the shortcut notations: $c_{_{\rm w}}=\cos\theta_{_{\rm w}},\;s_{_{\rm w}}
=\sin\theta_{_{\rm w}},\;$ with $\theta_{_{\rm w}}$ denoting the Weinberg angle.
Additionally, $p,\;k$ are the incoming momenta of lepton and photon fields,
$\rho$ is the Lorentz index of photon. Certainly,
the amplitude does not depend on how to mark the momenta of virtual fields for
the invariance of loop momentum translation.
It can be checked easily that the sum of amplitude for diagrams in Fig.\ref{fig1}
satisfies the Ward identity required by the QED gauge invariance
\begin{eqnarray}
&&k^\rho{\cal A}_{_{{\rm ww},\rho}}(p,k)=e[\Sigma_{_{\rm ww}}(p+k)
-\Sigma_{_{\rm ww}}(p)]\;,
\label{WTI-ww}
\end{eqnarray}
where ${\cal A}_{_{{\rm ww},\rho}}$
denotes the sum of amplitudes for the diagrams (a), (b), (c) and (d) in Fig.\ref{fig1},
as well as $\Sigma_{_{\rm ww}}$ denotes the amplitude of corresponding self energy
diagram, respectively.

According the external momenta of leptons and photon, we expand
the amplitude in Eq.(\ref{eq-wa1}) as
\begin{eqnarray}
%%%%%%%%%%%%%%%%%%%%%%%%%%%%%%%%%%%%%%%%%%%%%%%%%%%%%%%%%%%%%%%%%%%%%%%%%%%%%%%%%%%
&&i{\cal A}_{_{{\rm ww},\rho}}^{1(a)}(p,k)
=-i{e^5\over2s_{_{\rm w}}^4}\cdot\Lambda_{_{\rm RE}}^{4\epsilon}
\int{d^D q_1\over(2\pi)^D}{d^Dq_2\over(2\pi)^D}{1\over q_1^2
(q_1^2-m_{_{\rm w}}^2)^3((q_2-q_1)^2-m_{_{F_\alpha}}^2)(q_2^2-m_{_{F_\beta}}^2)}
\nonumber\\
&&\hspace{2.8cm}\times
\Bigg\{1+{2q_1\cdot(3p+k)\over q_1^2-m_{_{\rm w}}^2}+{2q_1\cdot p\over q_2^2-m_{_{F_\beta}}^2}
-{2p^2+(p+k)^2\over q_1^2-m_{_{\rm w}}^2}
-{p^2\over q_2^2-m_{_{F_\beta}}^2}
\nonumber\\
&&\hspace{2.8cm}
+{4(q_1\cdot(p+k))^2+8(q_1\cdot p)(q_1\cdot(p+k))+12(q_1\cdot p)^2
\over(q_1^2-m_{_{\rm w}}^2)^2}+{4(q_2\cdot p)^2\over(q_2^2-m_{_{F_\beta}}^2)^2}
\nonumber\\
&&\hspace{2.8cm}
+{4(q_1\cdot(3p+k))(q_2\cdot p)\over(q_1^2-m_{_{\rm w}}^2)
(q_2^2-m_{_{F_\beta}}^2)}\Bigg\}
\overline{\psi}_{_f}\Big[\gamma^\mu/\!\!\!q_1\gamma^\nu\omega_-\Big]\psi_{_f}
\Big[-g_{\mu\sigma}(2p+k-2q_1)_\rho
\nonumber\\
&&\hspace{2.8cm}
+2(g_{\rho\mu}k_\sigma-g_{\rho\sigma}k_\mu)\Big]
{\bf Tr}\Bigg[\gamma^\sigma\Big\{\zeta^{L*}_{_{\alpha\beta}}\omega_-
+\zeta^{R*}_{_{\alpha\beta}}\omega_+\Big\}
(/\!\!\!q_2-/\!\!\!q_1+m_{_{F_\alpha}})
\nonumber\\
&&\hspace{2.8cm}\times
\gamma_\nu\Big\{\zeta^L_{_{\alpha\beta}}\omega_-+\zeta^R_{_{\alpha\beta}}\omega_+\Big\}
(/\!\!\!q_2-/\!\!\!\!p+m_{_{F_\beta}})\Bigg]
%%%%%%%%%%%%%%%%%%%%%%%%%%%%%%%%%%%%%%%%%%%%%%%%%%%%%%%%%%%%%%%%%%%%%%%%%%%%%%%%%%%
\label{eq-wa2}
\end{eqnarray}
since we only consider the corrections to lepton MDM and EDM from dimension 6 operators.

Because the denominators of all terms are invariant under the reversal
$q_1\rightarrow-q_1,q_2\rightarrow-q_2$, those terms in odd powers of
loop momenta can be abandoned, and the terms in even powers of loop
momenta can be simplified by
\begin{eqnarray}
%%%%%%%%%%%%%%%%%%%%%%%%%%%%%%%%%%%%%%%%%%%%%%%%%%%%%%%%%%%%%%%%%%%%%%%%%%%%%%%%%%%
&&\int{d^Dq_1\over(2\pi)^D}{d^Dq_2\over(2\pi)^D}{q_{1\mu}q_{1\nu}q_{1\rho}
q_{1\sigma}q_{1\alpha}q_{1\beta},\;q_{1\mu}q_{1\nu}q_{1\rho}
q_{1\sigma}q_{1\alpha}q_{2\beta}\over((q_2-q_1)^2-m_{_0}^2)(q_1^2-m_{_1}^2)(q_2^2-m_{_2}^2)}
\nonumber\\
&&\hspace{-1.0cm}
\longrightarrow{S_{_{\mu\nu\rho\sigma\alpha\beta}}\over D(D+2)(D+4)}
\int{d^Dq_1\over(2\pi)^D}{d^Dq_2\over(2\pi)^D}{(q_1)^3,\;(q_1)^2q_1\cdot q_2
\over((q_2-q_1)^2-m_{_0}^2)(q_1^2-m_{_1}^2)(q_2^2-m_{_2}^2)}
\;,\nonumber\\
%%%%%%%%%%%%%%%%%%%%%%%%%%%%%%%%%%%%%%%%%%%%%%%%%%%%%%%%%%%%%%%%%%%%%%%%%%%%%%%%%%%
&&\int{d^Dq_1\over(2\pi)^D}{d^Dq_2\over(2\pi)^D}{q_{1\mu}q_{1\nu}q_{1\rho}
q_{1\sigma}q_{2\alpha}q_{2\beta}\over((q_2-q_1)^2-m_{_0}^2)(q_1^2-m_{_1}^2)(q_2^2-m_{_2}^2)}
\nonumber\\
&&\hspace{-1.0cm}
\longrightarrow\int{d^Dq_1\over(2\pi)^D}{d^Dq_2\over(2\pi)^D}
{1\over((q_2-q_1)^2-m_{_0}^2)(q_1^2-m_{_1}^2)
(q_2^2-m_{_2}^2)}
\nonumber\\
&&\hspace{-0.2cm}\times
\Big[{Dq_{_1}^2(q_{_1}\cdot q_{_2})^2-(q_{_1}^2)^2q_{_2}^2\over D(D-1)(D+2)(D+4)}
S_{_{\mu\nu\rho\sigma\alpha\beta}}
-{q_{_1}^2(q_{_1}\cdot q_{_2})^2-(q_{_1}^2)^2q_{_2}^2
\over D(D-1)(D+2)}T_{_{\mu\nu\rho\sigma}}g_{_{\alpha\beta}}\Big]
\;,\nonumber\\
%%%%%%%%%%%%%%%%%%%%%%%%%%%%%%%%%%%%%%%%%%%%%%%%%%%%%%%%%%%%%%%%%%%%%%%%%%%%%%%%%%%
&&\int{d^Dq_1\over(2\pi)^D}{d^Dq_2\over(2\pi)^D}{q_{1\mu}q_{1\nu}q_{1\rho}
q_{2\alpha}q_{2\beta}q_{2\delta}\over((q_2-q_1)^2-m_{_0}^2)(q_1^2-m_{_1}^2)(q_2^2-m_{_2}^2)}
\nonumber\\
&&\hspace{-1.0cm}
\longrightarrow\int{d^Dq_1\over(2\pi)^D}{d^Dq_2\over(2\pi)^D}
{1\over((q_2-q_1)^2-m_{_0}^2)(q_1^2-m_{_1}^2)
(q_2^2-m_{_2}^2)}
\nonumber\\
&&\hspace{-0.2cm}\times
\Big[{(D+1)q_{_1}^2q_{_1}\cdot q_{_2}q_{_2}^2-2(q_{_1}\cdot q_{_2})^3
\over D(D-1)(D+2)(D+4)}S_{_{\mu\nu\rho\alpha\beta\delta}}
+{(q_{_1}\cdot q_{_2})^3-q_{_1}^2q_{_1}\cdot q_{_2}q_{_2}^2\over
D(D-1)(D+2)}\Big(g_{_{\mu\alpha}}(g_{_{\nu\beta}}g_{_{\rho\delta}}
\nonumber\\
&&\hspace{-0.2cm}
+g_{_{\nu\delta}}g_{_{\rho\beta}})+g_{_{\mu\beta}}(g_{_{\nu\alpha}}
g_{_{\rho\delta}}+g_{_{\nu\delta}}g_{_{\rho\alpha}})
+g_{_{\mu\delta}}(g_{_{\nu\alpha}}g_{_{\rho\beta}}
+g_{_{\nu\beta}}g_{_{\rho\alpha}})\Big)\Big]\;,
%%%%%%%%%%%%%%%%%%%%%%%%%%%%%%%%%%%%%%%%%%%%%%%%%%%%%%%%%%%%%%%%%%%%%%%%%%%%%%%%%%%
\label{eq-wa3}
\end{eqnarray}
and those similar formulae presented in Eq.(5) of Ref\cite{Feng1},
where the tensors are defined as
\begin{eqnarray}
%%%%%%%%%%%%%%%%%%%%%%%%%%%%%%%%%%%%%%%%%%%%%%%%%%%%%%%%%%%%%%%%%%%%%%%%%%%%%%%%%%%
&&T_{_{\mu\nu\rho\sigma}}=g_{_{\mu\nu}}g_{_{\rho\sigma}}+g_{_{\mu\rho}}g_{_{\nu\sigma}}
+g_{_{\mu\sigma}}g_{_{\nu\rho}}
\;,\nonumber\\
%%%%%%%%%%%%%%%%%%%%%%%%%%%%%%%%%%%%%%%%%%%%%%%%%%%%%%%%%%%%%%%%%%%%%%%%%%%%%%%%%%%
&&S_{_{\mu\nu\rho\sigma\alpha\beta}}=
g_{_{\mu\nu}}T_{_{\rho\sigma\alpha\beta}}+g_{_{\mu\rho}}T_{_{\nu\sigma\alpha\beta}}
+g_{_{\mu\sigma}}T_{_{\nu\rho\alpha\beta}}+g_{_{\mu\alpha}}T_{_{\nu\rho\sigma\beta}}
+g_{_{\mu\beta}}T_{_{\nu\rho\sigma\alpha}}\;.
%%%%%%%%%%%%%%%%%%%%%%%%%%%%%%%%%%%%%%%%%%%%%%%%%%%%%%%%%%%%%%%%%%%%%%%%%%%%%%%%%%%
\label{eq-wa4}
\end{eqnarray}
Summing over those indices which appear both as superscripts and subscripts,
we derive all possible dimension 6 operators in the momentum space together with their
coefficients which are expressed in the linear combinations of one and
two loop vacuum integrals. In a similar way, one obtains the amplitude
of other diagrams. Before integrating with the loop momenta, we apply the loop momentum
translating invariance to formulate the sum of those amplitude in explicitly
QED gauge invariant form, then extract the Wilson coefficients of those
dimension 6 operators listed in Eq.(\ref{ops}). Actually, we can easily
verify the equation
\begin{eqnarray}
&&\int\int{d^Dq_1\over(2\pi)^D}{d^Dq_2\over(2\pi)^D}{q_{1\mu}\over
(q_1^2-m_1^2)(q_2^2-m_2^2)((q_2-q_1)^2-m_0^2)}\equiv0\;.
\label{l-tran1}
\end{eqnarray}
Performing an infinitesimal translation $q_1\rightarrow q_1,\;q_2\rightarrow
q_2-a$ with $a_\rho\rightarrow0\;(\rho=0,1,\cdots,D)$, one can write the
left-handed side of above equation as
\begin{eqnarray}
&&\int\int{d^Dq_1\over(2\pi)^D}{d^Dq_2\over(2\pi)^D}{q_{1\mu}\over
(q_1^2-m_1^2)(q_2^2-m_2^2)((q_2-q_1)^2-m_0^2)}
\nonumber\\
&&\hspace{-0.6cm}=
\int\int{d^Dq_1\over(2\pi)^D}{d^Dq_2\over(2\pi)^D}{q_{1\mu}\over
(q_1^2-m_1^2)(q_2^2-m_2^2)((q_2-q_1)^2-m_0^2)}
\nonumber\\
&&\hspace{-0.2cm}\times
\Big\{1+{2q_2\cdot a\over q_2^2-m_2^2}+{2(q_2-q_1)\cdot a\over(q_2-q_1)^2-m_0^2}+\cdots\Big\}\;.
\label{l-tran2}
\end{eqnarray}
This result implies
\begin{eqnarray}
&&\int\int{d^Dq_1\over(2\pi)^D}{d^Dq_2\over(2\pi)^D}{q_1\cdot q_2\over
(q_1^2-m_1^2)(q_2^2-m_2^2)^2((q_2-q_1)^2-m_0^2)}
\nonumber\\
&&\hspace{-0.6cm}=
\int\int{d^Dq_1\over(2\pi)^D}{d^Dq_2\over(2\pi)^D}{q_1^2-q_1\cdot q_2\over
(q_1^2-m_1^2)(q_2^2-m_2^2)((q_2-q_1)^2-m_0^2)^2}\;.
\label{l-tran3}
\end{eqnarray}
In a similar way, other identities listed in Ref.\cite{Feng1} can be derived.
Using the expression of two loop vacuum integral\cite{2vac}
\begin{eqnarray}
&&\Lambda_{_{\rm RE}}^{4\epsilon}\int\int{d^Dq_1\over(2\pi)^D}{d^Dq_2\over(2\pi)^D}{1\over
(q_1^2-m_1^2)(q_2^2-m_2^2)((q_2-q_1)^2-m_0^2)}
\nonumber\\
&&\hspace{-0.6cm}=
{\Lambda^2\over2(4\pi)^4}{\Gamma^2(1+\epsilon)\over(1-\epsilon)^2}
\Big({4\pi x_{_R}}\Big)^{2\epsilon}
\Big\{-{1\over\epsilon^2}\Big(x_0+x_1+x_2\Big)
\nonumber\\&&
+{1\over\epsilon}\Big(2(x_0\ln x_0+x_1\ln x_1+x_2\ln x_2)-x_0-x_1-x_2\Big)
\nonumber\\&&
-2(x_0+x_1+x_2)+2(x_0\ln x_0+x_1\ln x_1+x_2\ln x_2)
\nonumber\\&&
-x_0\ln^2x_0-x_1\ln^2x_1-x_2\ln^2x_2-\Phi(x_0,x_1,x_2)\Big\}
\label{2l-vacuum}
\end{eqnarray}
and
\begin{eqnarray}
&&\Phi(x,y,z)=(x+y-z)\ln x\ln y+(x-y+z)\ln x\ln z
\nonumber\\
&&\hspace{2.2cm}
+(-x+y+z)\ln y\ln z+{\rm sign}(\lambda^2)\sqrt{|\lambda^2|}\Psi(x,y,z)\;,
\nonumber\\
&&{\partial\Phi\over\partial x}(x,y,z)=\ln x\ln y+\ln x\ln z
-\ln y\ln z+2\ln x+{x-y-z\over\sqrt{|\lambda^2|}}\Psi(x,y,z)\;,
\label{phi}
\end{eqnarray}
one obtains easily
\begin{eqnarray}
&&{\Lambda_{_{\rm RE}}^{4\epsilon}\over\Lambda^2}{\partial\over\partial x_0}
\bigg\{\int\int{d^Dq_1\over(2\pi)^D}{d^Dq_2\over(2\pi)^D}
{q_1^2\over(q_1^2-m_1^2)(q_2^2-m_2^2)((q_2-q_1)^2-m_0^2)}\bigg\}
\nonumber\\
%%%%%%%%%%%%%%%%%%%%%%%%%%%%%%%%%%%%%%%%%%%%%%%%%%%%%%%%%%%%%%%%%%%%%%%%%%%%
&&\hspace{-0.5cm}=
{\Lambda_{_{\rm RE}}^{4\epsilon}\over\Lambda^2}
\Big\{{\partial\over\partial x_0}+{\partial\over\partial x_2}\Big\}
\bigg\{\int\int{d^Dq_1\over(2\pi)^D}{d^Dq_2\over(2\pi)^D}{q_1\cdot q_2\over
(q_1^2-m_1^2)(q_2^2-m_2^2)((q_2-q_1)^2-m_0^2)}\bigg\}
\nonumber\\
&&\hspace{-0.5cm}=
{\Lambda^2\over2(4\pi)^4}{\Gamma^2(1+\epsilon)\over(1-\epsilon)^2}
\Big({4\pi x_{_R}}\Big)^{2\epsilon}\Big\{-{x_1+2x_2\over\epsilon^2}
+{1\over\epsilon}\Big(x_1(1+2\ln x_0)+2x_2(1+\ln x_0+\ln x_2)\Big)
\nonumber\\&&
-(x_1+x_2)\ln^2x_0-(x_1+2x_2)\ln x_0\ln x_2-x_2\ln^2x_2-x_1\ln x_0\ln x_1
+x_1\ln x_1\ln x_2
\nonumber\\&&
-2(x_1+x_2)\ln x_0-2x_2\ln x_2-{x_1(x_0-x_1-x_2)\over\sqrt{|\lambda^2|}}
\Psi(x_0,x_1,x_2)\Big\}\;,
\label{l-tran4}
\end{eqnarray}
which is equivalent to the identity Eq.(\ref{l-tran3}).
Here, $\varepsilon=2-{D/2}$ with $D$ denoting the dimension of space-time,
$\Lambda$ is a energy scale to define $x_i=m_i^2/\Lambda^2$,
and $x_{_R}=\Lambda_{_{\rm RE}}^2/\Lambda^2$. Additionally, $\lambda^2=x^2+y^2+z^2-2xy-2xz-2yz$,
and the concrete expression of $\Psi(x,y,z)$ can be found in the appendix.
Actually, the equation Eq.(\ref{l-tran4}) provides a crosscheck of Eq.(\ref{2l-vacuum})
and Eq.(\ref{phi}) rather than a verification of Eq.(\ref{l-tran3}).
In the limit $z\ll x,y$, we can expand $\Phi(x,y,z)$ according $z$ as
\begin{eqnarray}
%%%%%%%%%%%%%%%%%%%%%%%%%%%%%%%%%%%%%%%%%%%%%%%%%%%%%%%%%%%%%%%%%%%%%%
&&\Phi(x,y,z)=\varphi_0(x,y)+z\varphi_1(x,y)+{z^2\over2!}\varphi_2(x,y)
+{z^3\over3!}\varphi_3(x,y)
\nonumber\\
&&\hspace{2.2cm}
+2z\Big(\ln z-1\Big)\pi_{_1}(x,y)
+2z^2\Big({\ln z\over2!}-{3\over4}\Big)\pi_{_2}(x,y)
\nonumber\\
&&\hspace{2.2cm}
+2z^3\Big({\ln z\over3!}-{11\over36}\Big)\pi_{_3}(x,y)+\cdots
%%%%%%%%%%%%%%%%%%%%%%%%%%%%%%%%%%%%%%%%%%%%%%%%%%%%%%%%%%%%%%%%%%%%%%
\label{phi-expand}
\end{eqnarray}
with
\begin{eqnarray}
%%%%%%%%%%%%%%%%%%%%%%%%%%%%%%%%%%%%%%%%%%%%%%%%%%%%%%%%%%%%%%%%%%%%%%
&&\pi_{_1}(x,y)=1+\varrho_{_{1,1}}(x,y),
\nonumber\\
&&\pi_{_2}(x,y)=-{x+y\over(x-y)^2}-{2xy\over(x-y)^3}\ln{y\over x},
\nonumber\\
&&\pi_{_3}(x,y)=-{1\over(x-y)^2}-{12xy\over(x-y)^4}
-{6xy(x+y)\over(x-y)^5}\ln{y\over x}\;,
%%%%%%%%%%%%%%%%%%%%%%%%%%%%%%%%%%%%%%%%%%%%%%%%%%%%%%%%%%%%%%%%%%%%%%
\label{pi1}
\end{eqnarray}
and the concrete expressions of function $\varphi_i(x,y)\;(i=0,1,2,3)$
can be found in appendix.

After applying those identities derived from loop momentum translating invariance,
we formulate the sum of the amplitude for Fig.\ref{fig1}(a,b,c,d)
satisfying QED gauge invariance and
CPT symmetry explicitly, and extract the Wilson coefficients of those
operators in Eq.(\ref{ops}).

Since only the operators ${\cal O}_{_{2,3,6}}^\mp$ actually contribute
to the MDMs and EDMs of leptons when the equations of motion are applied
to the incoming and outgoing leptons separately, the relevant terms in the
effective Lagrangian are formulated as:
\begin{eqnarray}
%%%%%%%%%%%%%%%%%%%%%%%%%%%%%%%%%%%%%%%%%%%%%%%%%%%%%%%%%%%%%%%%%%%%%%
&&{\cal L}_{_{\rm ww}}^{eff}
=-{(4\pi)^2e^4\over s_{_{\rm w}}^4Q_{_f}}\cdot\Lambda_{_{\rm RE}}^{4\epsilon}
\int{d^D q_1\over(2\pi)^D}{d^D q_2\over(2\pi)^D}{1\over q_1^2(q_1^2-m_{_{\rm w}}^2)^2
((q_2-q_1)^2-m_{_{F_\alpha}}^2)(q_2^2-m_{_{F_\beta}}^2)}
\nonumber\\
&&\hspace{1.4cm}\times
\Bigg\{\Bigg[\Big(\zeta^{L*}_{_{\alpha\beta}}\zeta^R_{_{\alpha\beta}}
+\zeta^{R*}_{_{\alpha\beta}}\zeta^L_{_{\alpha\beta}}\Big)
{\cal N}_{_{\rm ww}}^{(1)}
%%%%%%%%%%%%%%%%%%%%%%%%%%%%%%%%%%%%%%%%%%%%%%%%%%%%%%%%%%%%%%%%%%%%%%%%%%%%%%%%%%%
+\Big(\zeta^{L*}_{_{\alpha\beta}}\zeta^L_{_{\alpha\beta}}
-\zeta^{R*}_{_{\alpha\beta}}\zeta^R_{_{\alpha\beta}}\Big)
{\cal N}_{_{\rm ww}}^{(2)}
\nonumber\\
%%%%%%%%%%%%%%%%%%%%%%%%%%%%%%%%%%%%%%%%%%%%%%%%%%%%%%%%%%%%%%%%%%%%%%%%%%%%%%%%%%%
&&\hspace{1.4cm}
+m_{_{\chi_\alpha^0}}m_{_{\chi_\beta^\pm}}\Big(\zeta^{L*}_{_{\alpha\beta}}
\zeta^R_{_{\alpha\beta}}+\zeta^{R*}_{_{\alpha\beta}}\zeta^L_{_{\alpha\beta}}\Big)
{\cal N}_{_{\rm ww}}^{(3)}\Bigg]\Big({\cal O}_{_2}^-+{\cal O}_{_3}^-\Big)
\nonumber\\
%%%%%%%%%%%%%%%%%%%%%%%%%%%%%%%%%%%%%%%%%%%%%%%%%%%%%%%%%%%%%%%%%%%%%%%%%%%%%%%%%%%
&&\hspace{1.4cm}
+m_{_{\chi_\alpha^0}}m_{_{\chi_\beta^\pm}}\Big(\zeta^{R*}_{_{\alpha\beta}}
\zeta^L_{_{\alpha\beta}}-\zeta^{L*}_{_{\alpha\beta}}\zeta^R_{_{\alpha\beta}}\Big)
{\cal N}_{_{\rm ww}}^{(4)}\Big({\cal O}_{_2}^--{\cal O}_{_3}^-\Big)\Bigg\}+\cdots\;,
%%%%%%%%%%%%%%%%%%%%%%%%%%%%%%%%%%%%%%%%%%%%%%%%%%%%%%%%%%%%%%%%%%%%%%%%%%%%%%%%%%%
\label{eff-wff0}
\end{eqnarray}
where $Q_{_f}=-1$ represents the charge of leptons, and the expressions of
form factors ${\cal N}_{_{\rm ww}}^{(i)}\;(i=1,\;2,\;3,\;4)$ are presented
in appendix.
%%%%%%%%%%%%%%%%%%%%%%%%%%%END MODIFICATION%%%%%%%%%%%%%%%%%%%%%%%%%%%%%

Integrating over loop momenta, one gets the following terms
in the effective Lagrangian:
\begin{eqnarray}
&&{\cal L}^{eff}_{_{\rm ww}}={\sqrt{2}G_{_F}\alpha_{_e}x_{_{\rm w}}\over\pi s_{_{\rm w}}^2Q_{_f}}
(4\pi x_{_{\rm R}})^{2\varepsilon}{\Gamma^2(1+\varepsilon)\over
(1-\varepsilon)^2}\Bigg\{\Big(\zeta^{L*}_{_{\alpha\beta}}
\zeta^L_{_{\alpha\beta}}+\zeta^{R*}_{_{\alpha\beta}}\zeta^R_{_{\alpha\beta}}\Big)
\nonumber\\
&&\hspace{1.4cm}\times
\Big[-{5\over24\varepsilon}{x_{_{F_\alpha}}+x_{_{F_\beta}}\over x_{_{\rm w}}^2}
+{5\over24x_{_{\rm w}}^2}\varrho_{_{2,1}}(x_{_{F_\alpha}},x_{_{F_\beta}})
+{x_{_{F_\alpha}}+x_{_{F_\beta}}\over x_{_{\rm w}}^2}\Big({7\over27}
+{5\over24}\ln x_{_{\rm w}}\Big)
\nonumber\\
&&\hspace{1.4cm}
-{1\over9x_{_{\rm w}}}
+T_1(x_{_{\rm w}},x_{_{F_\alpha}},x_{_{F_\beta}})\Big]({\cal O}_{_2}^-+{\cal O}_{_3}^-)
\nonumber\\
%---------------------------------------------------------------------
&&\hspace{1.4cm}
+\Big(\zeta^{L*}_{_{\alpha\beta}}\zeta^L_{_{\alpha\beta}}
-\zeta^{R*}_{_{\alpha\beta}}\zeta^R_{_{\alpha\beta}}\Big)
T_2(x_{_{\rm w}},x_{_{F_\alpha}},x_{_{F_\beta}})
({\cal O}_{_2}^-+{\cal O}_{_3}^-)
\nonumber\\
%---------------------------------------------------------------------
&&\hspace{1.4cm}
+\Big(\zeta^{L*}_{_{\alpha\beta}}\zeta^R_{_{\alpha\beta}}+\zeta^{R*}_{_{\alpha\beta}}
\zeta^L_{_{\alpha\beta}}\Big)(x_{_{F_\alpha}}x_{_{F_\beta}})^{1/2}
\Big[{5\over12\varepsilon x_{_{\rm w}}^2}-{5\over12x_{_{\rm w}}^2}
\varrho_{_{1,1}}(x_{_{F_\alpha}},x_{_{F_\beta}})
\nonumber\\
&&\hspace{1.4cm}
+{19\over72x_{_{\rm w}}^2}-{5\over12x_{_{\rm w}}^2}\ln x_{_{\rm w}}
+T_3(x_{_{\rm w}},x_{_{F_\alpha}},x_{_{F_\beta}})\Big]
({\cal O}_{_2}^-+{\cal O}_{_3}^-)
\nonumber\\
%---------------------------------------------------------------------
&&\hspace{1.4cm}
+\Big(\zeta^{L*}_{_{\alpha\beta}}\zeta^R_{_{\alpha\beta}}-\zeta^{R*}_{_{\alpha\beta}}
\zeta^L_{_{\alpha\beta}}\Big)(x_{_{F_\alpha}}x_{_{F_\beta}})^{1/2}
T_4(x_{_{\rm w}},x_{_{F_\alpha}},x_{_{F_\beta}})\Big]
({\cal O}_{_2}^--{\cal O}_{_3}^-)\Bigg\}+\cdots\;,
\label{eff-WFF}
\end{eqnarray}
where $G_{_F}=1.16639\times10^{-5}\;{\rm GeV}^{-2}$ is the 4-fermion coupling,
and $\alpha_{_e}=e^2/4\pi$.
Note that the above result does not depend on the concrete choice
of energy scale $\Lambda$, and the concrete expressions of
$T_i(x,y,z),\;\varrho_{_{i,j}}(x,y)\;(i,\;j=1,\;2\;\cdots)$ can be found in appendix.

The charged gauge boson self energy composed of a closed heavy fermion loop
induces the ultraviolet divergence in the Wilson coefficients of effective
Lagrangian, the unrenormalized $W^\pm$ self energy is generally expressed as
\begin{eqnarray}
%%%%%%%%%%%%%%%%%%%%%%%%%%%%%%%%%%%%%%%%%%%%%%%%%%%%%%%%%%%%%%%%%%%%%%%%%%%%%%%%%%%
&&\Sigma_{_{\mu\nu}}^{\rm w}(p,\Lambda_{_{\rm RE}})=\Lambda^2A_0^{\rm w}g_{\mu\nu}+\Big(A_1^{\rm w}
+{p^2\over\Lambda^2}A_2^{\rm w}+\cdots\Big)(p^2g_{\mu\nu}-p_\mu p_\nu)
\nonumber\\
&&\hspace{2.5cm}
+\Big(B_1^{\rm w}+{p^2\over\Lambda^2}B_2^{\rm w}+\cdots\Big)p_\mu p_\nu\;,
%%%%%%%%%%%%%%%%%%%%%%%%%%%%%%%%%%%%%%%%%%%%%%%%%%%%%%%%%%%%%%%%%%%%%%%%%%%%%%%%%%%
\label{eq-w1}
\end{eqnarray}
where the form factors $A_{0,1,2}^{\rm w}$ and $B_{1,2}^{\rm w}$ only depend on
the virtual field masses and renormalization scale.
Here, we omit those terms which are strongly suppressed at the limit
of heavy virtual fermion masses. The corresponding counter terms are given as
\begin{eqnarray}
%%%%%%%%%%%%%%%%%%%%%%%%%%%%%%%%%%%%%%%%%%%%%%%%%%%%%%%%%%%%%%%%%%%%%%%%%%%%%%%%%%%
&&\Sigma_{_{\mu\nu}}^{\rm wC}(p,\Lambda_{_{\rm RE}})=-\Big[\delta m_{_{\rm w}}^2(\Lambda_{_{\rm RE}})
+m_{_{\rm w}}^2\delta Z_{_{\rm w}}(\Lambda_{_{\rm RE}})\Big]g_{\mu\nu}
-\delta Z_{_{\rm w}}(\Lambda_{_{\rm RE}})\Big[p^2g_{\mu\nu}-p_\mu p_\nu\Big]\;.
%%%%%%%%%%%%%%%%%%%%%%%%%%%%%%%%%%%%%%%%%%%%%%%%%%%%%%%%%%%%%%%%%%%%%%%%%%%%%%%%%%%
\label{eq-w2}
\end{eqnarray}

The renormalized self energy is given by
\begin{eqnarray}
%%%%%%%%%%%%%%%%%%%%%%%%%%%%%%%%%%%%%%%%%%%%%%%%%%%%%%%%%%%%%%%%%%%%%%%%%%%%%%%%%%%
&&\hat{\Sigma}_{_{\mu\nu}}^{\rm w}(p,\Lambda_{_{\rm RE}})=
\Sigma_{_{\mu\nu}}^{\rm w}(p,\Lambda_{_{\rm RE}})
+\Sigma_{_{\mu\nu}}^{\rm wC}(p,\Lambda_{_{\rm RE}})\;.
%%%%%%%%%%%%%%%%%%%%%%%%%%%%%%%%%%%%%%%%%%%%%%%%%%%%%%%%%%%%%%%%%%%%%%%%%%%%%%%%%%%
\label{eq-w3}
\end{eqnarray}
For on-shell external gauge boson $W^\pm$, we have \cite{onshell}
\begin{eqnarray}
%%%%%%%%%%%%%%%%%%%%%%%%%%%%%%%%%%%%%%%%%%%%%%%%%%%%%%%%%%%%%%%%%%%%%%%%%%%%%%%%%%%
&&\hat{\Sigma}_{_{\mu\nu}}^{\rm w}(p,m_{_{\rm w}})\epsilon^\nu(p)\Big|_{p^2=m_{_{\rm w}}^2}=0
\;,\nonumber\\
&&\lim\limits_{p^2\rightarrow m_{_{\rm w}}^2}{1\over p^2-m_{_{\rm w}}^2}
\hat{\Sigma}_{_{\mu\nu}}^{\rm w}(p,m_{_{\rm w}})\epsilon^\nu(p)=\epsilon_{_\mu}(p)\;,
%%%%%%%%%%%%%%%%%%%%%%%%%%%%%%%%%%%%%%%%%%%%%%%%%%%%%%%%%%%%%%%%%%%%%%%%%%%%%%%%%%%
\label{eq-w4}
\end{eqnarray}
where $\epsilon(p)$ is the polarization vector of $W^\pm$ gauge boson.
Inserting Eq. (\ref{eq-w1}) and Eq. (\ref{eq-w2}) into Eq. (\ref{eq-w4}),
we derive the counter terms for the $W^\pm$ self energy in on-shell scheme as
\begin{eqnarray}
%%%%%%%%%%%%%%%%%%%%%%%%%%%%%%%%%%%%%%%%%%%%%%%%%%%%%%%%%%%%%%%%%%%%%%%%%%%%%%%%%%%
&&\delta Z_{_{\rm w}}^{os}(m_{_{\rm w}})=A_1^{\rm w}+{m_{_{\rm w}}^2\over\Lambda^2}A_2^{\rm w}
=A_1^{\rm w}+x_{_{\rm w}}A_2^{\rm w}\;,
\nonumber\\
&&\delta m_{_{\rm w}}^{2,os}(m_{_{\rm w}})=A_0^{\rm w}\Lambda^2
-m_{_{\rm w}}^2\delta Z_{_{\rm w}}^{os}\;.
%%%%%%%%%%%%%%%%%%%%%%%%%%%%%%%%%%%%%%%%%%%%%%%%%%%%%%%%%%%%%%%%%%%%%%%%%%%%%%%%%%%
\label{eq-w5}
\end{eqnarray}

We should derive the counter term for the vertex $\gamma W^+W^-$ here
since the corresponding coupling is not zero at tree level.
In the nonlinear $R_\xi$ gauge with $\xi=1$,
the counter term for the vertex $\gamma W^+W^-$ is
\begin{eqnarray}
%%%%%%%%%%%%%%%%%%%%%%%%%%%%%%%%%%%%%%%%%%%%%%%%%%%%%%%%%%%%%%%%%%%%%%%%%%%%%%%%%%%
&&i\delta C_{\gamma W^+W^-}=ie\cdot\delta Z_{_{\rm w}}(\Lambda_{_{\rm RE}})
\Big[g_{\mu\nu}(k_1-k_2)_\rho+g_{\nu\rho}(k_2-k_3)_\mu+g_{\rho\mu}(k_3-k_1)_\nu\Big]\;,
%%%%%%%%%%%%%%%%%%%%%%%%%%%%%%%%%%%%%%%%%%%%%%%%%%%%%%%%%%%%%%%%%%%%%%%%%%%%%%%%%%%
\label{eq-w6}
\end{eqnarray}
where $k_i\;(i=1,\;2,\;3)$ denote the incoming momenta of $W^\pm$ and photon,
and $\mu,\;\nu,\;\rho$ denote the corresponding Lorentz indices respectively.

We present the counter term diagrams to cancel the ultraviolet divergence
contained by the bare effective Lagrangian in Fig.\ref{fig1}(e,f,g),
and we can verify that the sum of corresponding amplitude
satisfies the Ward identity required by the QED gauge invariance obviously.

Accordingly, the effective Lagrangian from the counter term diagrams is written as
\begin{eqnarray}
%%%%%%%%%%%%%%%%%%%%%%%%%%%%%%%%%%%%%%%%%%%%%%%%%%%%%%%%%%%%%%%%%%%%%%%
&&\delta{\cal L}_{_{\rm ww}}^C=i{e^2\over2s_{_{\rm w}}^2\Lambda^2Q_{_f}}
(4\pi x_{_{\rm R}})^{\varepsilon}{\Gamma(1+\varepsilon)\over(1-\varepsilon)}
\Big\{A_0^{\rm w}\Big[{5\over12x_{_{\rm w}}^2}+{19\varepsilon\over
72x_{_{\rm w}}^2}-{5\varepsilon\over12x_{_{\rm w}}^2}\ln x_{_{\rm w}}\Big]
\nonumber\\
&&\hspace{1.6cm}
+{5\varepsilon\over12x_{_{\rm w}}}(A_1^{\rm w}+x_{_{\rm w}}A_2^{\rm w})\Big\}
({\cal O}_{_2}^-+{\cal O}_{_3}^-)
\nonumber\\
%---------------------------------------------------------------------
&&\hspace{1.2cm}=
{\sqrt{2}G_{_F}\alpha_{_e}x_{_{\rm w}}\over\pi s_{_{\rm w}}^2Q_{_f}}
(4\pi x_{_{\rm R}})^{2\varepsilon}{\Gamma^2(1+\varepsilon)\over
(1-\varepsilon)^2}\Bigg\{\Big(\zeta^{L*}_{_{\alpha\beta}}
\zeta^L_{_{\alpha\beta}}+\zeta^{R*}_{_{\alpha\beta}}\zeta^R_{_{\alpha\beta}}\Big)
\Big[{5\over24\varepsilon}{x_{_{F_\alpha}}+x_{_{F_\beta}}\over x_{_{\rm w}}^2}
\nonumber\\
&&\hspace{1.6cm}
-{5\over24x_{_{\rm w}}^2}\varrho_{_{2,1}}(x_{_{F_\alpha}},x_{_{F_\beta}})
-{x_{_{F_\alpha}}+x_{_{F_\beta}}\over x_{_{\rm w}}^2}\Big({7\over27}
+{5\over24}\ln x_{_{\rm w}}\Big)
+{1\over9x_{_{\rm w}}}\Big]({\cal O}_{_2}^-+{\cal O}_{_3}^-)
\nonumber\\
%---------------------------------------------------------------------
&&\hspace{1.6cm}
+\Big(\zeta^{L*}_{_{\alpha\beta}}\zeta^R_{_{\alpha\beta}}+\zeta^{R*}_{_{\alpha\beta}}
\zeta^L_{_{\alpha\beta}}\Big)(x_{_{F_\alpha}}x_{_{F_\beta}})^{1/2}
\Big[-{5\over12\varepsilon x_{_{\rm w}}^2}+{5\over12x_{_{\rm w}}^2}
\varrho_{_{1,1}}(x_{_{F_\alpha}},x_{_{F_\beta}})
\nonumber\\
&&\hspace{1.6cm}
-{19\over72x_{_{\rm w}}^2}+{5\over12x_{_{\rm w}}^2}\ln x_{_{\rm w}}\Big]
({\cal O}_{_2}^-+{\cal O}_{_3}^-)\Bigg\}+\cdots\;.
%%%%%%%%%%%%%%%%%%%%%%%%%%%%%%%%%%%%%%%%%%%%%%%%%%%%%%%%%%%%%%%%%%%%%%%%%%%%%%%%%%%
\label{w-counter}
\end{eqnarray}

%%%%%%%%%%%%%%%%%%%%%%%%%BEGIN MODIFICATION%%%%%%%%%%%%%%%%%%%%%%%%%%%%%
Adding the counter terms to bare Lagrangian Eq.(\ref{eff-WFF}),
we cancel the ultraviolet divergence there. However, the diagrams in
Fig.\ref{fig1} include the virtual neutrino which belongs to
light freedoms contained by the effective Lagrangian.
It is unreasonable obviously in the above analysis
that the propagators of virtual neutrino are expanded according
to the external momenta. In order to obtain the corrections to lepton
MDMs and EDMs from light freedoms properly, we match the sum of amplitude
from full theory to that from effective theory\cite{matching} at first:
\begin{eqnarray}
%%%%%%%%%%%%%%%%%%%%%%%%%%%%%%%%%%%%%%%%%%%%%%%%%%%%%%%%%%%%%%%%%%%%%%%%%%%%%%%%%%%
&&\sum\limits_if_{_{0,i}}^\mp(m_{_{Vh}},m_{_{Vl}}){\cal O}_{_i}^\mp
=\sum\limits_i\Big[f_{_{h,i}}^\mp(m_{_{Vh}},\Lambda_{_{\rm MA}})
+f_{_{l,i}}^\mp(m_{_{Vl}},\Lambda_{_{\rm MA}})\Big]{\cal O}_{_i}^\mp\;,
%%%%%%%%%%%%%%%%%%%%%%%%%%%%%%%%%%%%%%%%%%%%%%%%%%%%%%%%%%%%%%%%%%%%%%%%%%%%%%%%%%%
\label{ww-match}
\end{eqnarray}
where $\Lambda_{_{\rm MA}}$ represents the matching scale, and $m_{_{Vh}},\;m_{_{Vl}}$
denote the masses of virtual heavy and light freedoms respectively. The left-handed
side of above equation denotes the amplitude from Fig.\ref{fig1} derived through
the above steps, the first term of right-handed
side is the corrections to effective Lagrangian from heavy freedoms only,
and the second term of right-handed side is the corrections obtained unsuitably
to effective Lagrangian from light freedoms.
%%%%%%%%%%%%%%%%%%%%%%%%%%%%%%%%%%%%%%%%%%%%%%%%%%%%%%%%%%%%%%%%%%%
\begin{figure}[h]
\setlength{\unitlength}{1mm}
\begin{center}
\begin{picture}(0,20)(0,0)
\put(-62,-110){\includegraphics{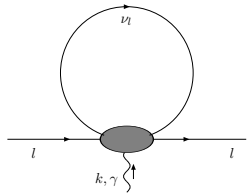}}
\end{picture}
\caption[]{The diagram of effective theory corresponds to those
diagrams in Fig.\ref{fig1}.}
\label{fig2}
\end{center}
\end{figure}
%%%%%%%%%%%%%%%%%%%%%%%%%%%%%%%%%%%%%%%%%%%%%%%%%%%%%%%%%%%%%%%%%%%
Shrinking those heavy freedoms of Fig.\ref{fig1} in a point, one
obtains the corresponding Feynman diagrams of effective theory in Fig.\ref{fig2}.
Expanding the amplitude of diagram for effective theory in powers of external
momenta, we can derive the corrections $\sum\limits_if_{_{l,i}}^\mp(m_{_{Vl}},
\Lambda_{_{\rm MA}}){\cal O}_{_i}^\mp$ which are originated from light freedoms
purely. Inserting the concrete expressions of $f_{_{l,i}}^\mp(m_{_{vl}},
\Lambda_{_{\rm MA}})$ into Eq.(\ref{ww-match}), one gets the corrections
$f_{_{h,i}}^\mp(m_{_{Vh}},\Lambda_{_{\rm MA}})$
which are originated from heavy freedoms only. Finally, we analyze the amplitude
of Fig.\ref{fig2} strictly to get the corrections from virtual light freedoms
to lepton MDMs and EDMs. Because the effective
coupling among leptons and photon in Fig.\ref{fig2} is induced
by the dimension 8 operators at least, the corrections from Fig.\ref{fig2}
to lepton MDMs and EDMs contain the additional suppression factor
$m_{_l}^2/\Lambda_{_{\rm EW}}^2$ comparing with that only from the heavy freedoms.
Under our approximation,  the resulted lepton MDMs and EDMs are respectively
formulated as
\begin{eqnarray}
%%%%%%%%%%%%%%%%%%%%%%%%%%%%%%%%%%%%%%%%%%%%%%%%%%%%%%%%%%%%%%%%%%%%%%
&&a_{l,F}^{\rm ww}={G_{_F}\alpha_{_e}m_{_l}^2\over2\sqrt{2}\pi^3s_{_{\rm w}}^2}
x_{_{\rm w}}\Big\{\Big(|\zeta^L_{_{\alpha\beta}}|^2+|\zeta^R_{_{\alpha\beta}}|^2\Big)
T_1(x_{_{\rm w}},x_{_{F_\alpha}},x_{_{F_\beta}})
\nonumber\\
%---------------------------------------------------------------------
&&\hspace{1.6cm}
+\Big(|\zeta^L_{_{\alpha\beta}}|^2-|\zeta^R_{_{\alpha\beta}}|^2\Big)
T_2(x_{_{\rm w}},x_{_{F_\alpha}},x_{_{F_\beta}})
\nonumber\\
%---------------------------------------------------------------------
&&\hspace{1.6cm}
+2(x_{_{F_\alpha}}x_{_{F_\beta}})^{1/2}\Re(\zeta^{R*}_{_{\alpha\beta}}
\zeta^L_{_{\alpha\beta}})T_3(x_{_{\rm w}},x_{_{F_\alpha}},x_{_{F_\beta}})\Big\}\;,
\nonumber\\
%%%%%%%%%%%%%%%%%%%%%%%%%%%%%%%%%%%%%%%%%%%%%%%%%%%%%%%%%%%%%%%%%%%%%%
&&d_{l,F}^{\rm ww}=-{G_{_F}\alpha_{_e}em_{_l}\over2\sqrt{2}\pi^3s_{_{\rm w}}^2}
x_{_{\rm w}}(x_{_{F_\alpha}}x_{_{F_\beta}})^{1/2}
\Im(\zeta^{R*}_{_{\alpha\beta}}\zeta^L_{_{\alpha\beta}})
T_4(x_{_{\rm w}},x_{_{F_\alpha}},x_{_{F_\beta}})\;,
%%%%%%%%%%%%%%%%%%%%%%%%%%%%%%%%%%%%%%%%%%%%%%%%%%%%%%%%%%%%%%%%%%%%%%
\label{MEDM-ww}
\end{eqnarray}
which only depend on the masses of virtual fields. It should be
clarified that the corrections to
lepton EDMs from the diagrams (a,b,c,d) in Fig.\ref{fig1} do not depend on
the concrete renormalization scheme adopted here since the relevant
terms from bare Lagrangian do not contain the ultraviolet divergence.
Using the expansion of $\Phi(x,y,z)$ in Eq(.\ref{phi-expand}),
we get the asymptotic expressions of $T_i(x,z,u),\;(i=1,\cdots,4)$
at the limit $z,\;u\gg x$ as
\begin{eqnarray}
%%%%%%%%%%%%%%%%%%%%%%%%%%%%%%%%%%%%%%%%%%%%%%%%%%%%%%%%%%%%%%%%%%%%%%
&&T_1(x,z,u)\simeq
-{3-3Q_\beta\over8x}+{3-2Q_\beta\over8x}\ln u
-{3-2Q_\beta\over8x}\pi_{_1}(z,u)
\nonumber\\
%---------------------------------------------------------------------
&&\hspace{2.8cm}
+{(3-2Q_\beta)z+6u\over8x}{\partial\pi_{_1}\over\partial u}(z,u)
-{(3-2Q_\beta)zu+(3+2Q_\beta)u^2\over8x}{\partial^2\pi_{_1}\over\partial u^2}(z,u)
\nonumber\\
%---------------------------------------------------------------------
&&\hspace{2.8cm}
+{1\over12x}u^2(z-u){\partial^3\pi_{_1}\over\partial u^3}(z,u)
-{2(4-Q_\beta)z-(1-2Q_\beta)u\over16x}\pi_{_2}(z,u)
\nonumber\\
%---------------------------------------------------------------------
&&\hspace{2.8cm}
+{7\over16x}u(z-u){\partial\pi_{_2}\over\partial u}(z,u)
+{(z-u)^2\over24}\pi_{_3}(z,u)+\cdots\;,
\nonumber\\
%%%%%%%%%%%%%%%%%%%%%%%%%%%%%%%%%%%%%%%%%%%%%%%%%%%%%%%%%%%%%%%%%%%%%%
&&T_2(x,z,u)\simeq
-{\ln u\over8x}+{1\over8x}\pi_{_1}(z,u)
-{(5-Q_\beta)u\over4x}{\partial\pi_{_1}\over\partial u}(z,u)
+{u(z-u)\over8x}{\partial^2\pi_{_1}\over\partial u^2}(z,u)
+\cdots\;,
\nonumber\\
%%%%%%%%%%%%%%%%%%%%%%%%%%%%%%%%%%%%%%%%%%%%%%%%%%%%%%%%%%%%%%%%%%%%%%
&&T_3(x,z,u)\simeq
{1-3Q_\beta\over24xu}+{1+Q_\beta\over2x}{\partial\pi_{_1}\over\partial u}(z,u)
+{1-Q_\beta\over4x}{\partial\pi_{_1}\over\partial z}(z,u)
-{u(z-u)\over16x}{\partial^3\pi_{_1}\over\partial u^3}(z,u)
\nonumber\\
%---------------------------------------------------------------------
&&\hspace{2.8cm}
-{Q_\beta z-(3+Q_\beta)u\over8x}{\partial^2\pi_{_1}\over\partial u^2}(z,u)
-{1-Q_\beta\over8x}(z-u){\partial^2\pi_{_1}\over\partial z\partial u}(z,u)
\nonumber\\
&&\hspace{2.8cm}
+{3\over4x}\pi_{_2}(z,u)
-{1-Q_\beta\over16x}(z-u)\Big[{\partial\pi_{_2}\over\partial u}
+{\partial\pi_{_2}\over\partial z}\Big](z,u)+\cdots\;,
\nonumber\\
%%%%%%%%%%%%%%%%%%%%%%%%%%%%%%%%%%%%%%%%%%%%%%%%%%%%%%%%%%%%%%%%%%%%%%
&&T_4(x,z,u)\simeq
-{Q_\beta\over8xu}+{4Q_\beta-Q_\alpha\over8x}{\partial\pi_{_1}\over\partial u}(z,u)
-{-Q_\alpha\over8x}{\partial\pi_{_1}\over\partial z}(z,u)
\nonumber\\
&&\hspace{2.8cm}
-{Q_\beta\over8x}(z-u){\partial^2\pi_{_1}\over\partial u^2}(z,u)
-{Q_\alpha\over8x}(z-u){\partial^2\pi_{_1}\over\partial z\partial u}(z,u)
+\cdots\;.
%%%%%%%%%%%%%%%%%%%%%%%%%%%%%%%%%%%%%%%%%%%%%%%%%%%%%%%%%%%%%%%%%%%%%%
\label{T-asymp}
\end{eqnarray}
This implies that the leading contributions contained in the asymptotic
form of Eq.\ref{MEDM-ww} under the assumption $m_{_F}=m_{_{F_\alpha}}=m_{_{F_\beta}}
\gg m_{_{\rm w}}$ can be written as:
\begin{eqnarray}
%%%%%%%%%%%%%%%%%%%%%%%%%%%%%%%%%%%%%%%%%%%%%%%%%%%%%%%%%%%%%%%%%%%%%%
&&a_{l,F}^{\rm ww}\approx
{G_{_F}\alpha_{_e}m_{_l}^2\over48\sqrt{2}\pi^3s_{_{\rm w}}^2}
\Big\{(18Q_\beta-13)\Big(|\zeta^L_{_{\alpha\beta}}|^2
+|\zeta^R_{_{\alpha\beta}}|^2\Big)
\nonumber\\
%---------------------------------------------------------------------
&&\hspace{1.4cm}
+3(Q_\beta-3)\Big(|\zeta^L_{_{\alpha\beta}}|^2-|\zeta^R_{_{\alpha\beta}}|^2\Big)
+11\Re(\zeta^{R*}_{_{\alpha\beta}}\zeta^L_{_{\alpha\beta}})
\Big\}+\cdots\;,
\nonumber\\
%%%%%%%%%%%%%%%%%%%%%%%%%%%%%%%%%%%%%%%%%%%%%%%%%%%%%%%%%%%%%%%%%%%%%%
&&d_{l,F}^{\rm ww}\approx
-{G_{_F}\alpha_{_e}em_{_l}(2+Q_\beta)\over16\sqrt{2}\pi^3s_{_{\rm w}}^2}
\Im(\zeta^{R*}_{_{\alpha\beta}}\zeta^L_{_{\alpha\beta}})
+\cdots\;,
%%%%%%%%%%%%%%%%%%%%%%%%%%%%%%%%%%%%%%%%%%%%%%%%%%%%%%%%%%%%%%%%%%%%%%
\label{asyHF-MDM-W}
\end{eqnarray}
where ellipses represent those relatively unimportant corrections.

Comparing the result in Eq.(\ref{MEDM-ww}), the contributions from
the corresponding diagrams contain
the additional suppressed factor $m_{_l}^2/\Lambda_{_{\rm EW}}^2$
when both of virtual charged gauge bosons in Fig.\ref{fig1}(a,b,c,d)
are replaced with the charged Goldstone $G^\pm$. However, we should consider
the corrections from those two loop diagrams in which one of virtual
charged gauge bosons is replaced with the charged Goldstone $G^\pm$ since it
represents the longitudinal component of charged gauge boson
in nonlinear $R_\xi$ gauge. For many extensions of the SM contain
the charged Higgs, we also generalize the result directly to the diagrams
in which a closed heavy loop is attached to the virtual $H^\pm$
and $W^\pm$ fields simultaneously.

%%%%%%%%%%%%%%%%%%%%%%%%%%%%%%%%%%%%%%%%%%%%%%%%%%%%%%%%%%%%%%%%%%%
\begin{figure}[t]
\setlength{\unitlength}{1mm}
\begin{center}
\begin{picture}(0,60)(0,0)
\put(-62,-70){\includegraphics{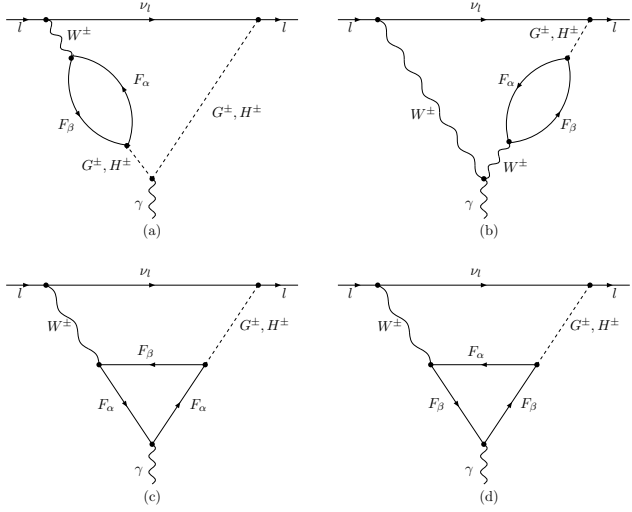}}
\end{picture}
\caption[]{The two-loop diagrams in which a closed heavy
fermion loop is attached to the virtual $W^\pm,\;G^\pm$ or $H^\pm$ bosons.
In concrete calculation, the contributions from those mirror diagrams
should be included also.}
\label{fig3}
\end{center}
\end{figure}
%%%%%%%%%%%%%%%%%%%%%%%%%%%%%%%%%%%%%%%%%%%%%%%%%%%%%%%%%%%%%%%%%%%

\subsection{The corrections from the diagrams where a closed heavy
fermion loop is attached to the virtual $W^\pm,\;G^\pm\;(H^\pm)$ bosons}
\indent\indent
Similarly, the renormalizable interaction among the electroweak charged
Goldstone/Higgs $G^\pm\;(H^\pm)$ and the heavy fermions $F_{\alpha,\beta}$
can be expressed in a more universal form as
\begin{eqnarray}
&&{\cal L}_{_{S^\pm FF}}={e\over s_{_{\rm w}}}\Big[G^{-}\bar{F}_\alpha
({\cal G}^{c,L}_{_{\alpha\beta}}\omega_-+{\cal G}^{c,R}_{_{\alpha\beta}}\omega_+)F_\beta
+H^{-}\bar{F}_\alpha({\cal H}^{c,L}_{_{\alpha\beta}}\omega_-
+{\cal H}^{c,R}_{_{\alpha\beta}}\omega_+)F_\beta\Big]+h.c.\;,
\label{charged-G-H-FF}
\end{eqnarray}
where the concrete expressions of ${\cal G}^{c,L,R}_{_{\alpha\beta}},\;
{\cal H}^{c,L,R}_{_{\alpha\beta}}$ depend on the
models employed in our calculation, the conservation of electric charge
requires $Q_\beta-Q_\alpha=1$.
Generally, the couplings among the charged Goldstone/Higgs
and leptons are written as
\begin{eqnarray}
&&{\cal L}_{_{S^\pm l\nu_{_l}}}={em_{_l}\over\sqrt{2}m_{_{\rm w}}
s_{_{\rm w}}}\Big[G^{-}\bar{l}\omega_-\nu_{_l}
+{\cal B}_{c}H^{-}\bar{l}\omega_-\nu_{_l}\Big]+h.c.\;,
\label{charged-G-H-lepton}
\end{eqnarray}
where the parameter ${\cal B}_{c}$ also depends on the concrete
models adopted in our analysis. In full theory,
the couplings in Eq.(\ref{charged-G-H-FF}) induce the corrections
to lepton MDMs and EDMs through the diagrams in Fig.\ref{fig3},
and the corresponding diagram of effective theory
is same as that presented in Fig.\ref{fig2}.

After the steps taken in $WW$ sector, the corresponding corrections
from the diagrams in Fig.\ref{fig3} to the lepton MDMs and EDMs are
formulated as
\begin{eqnarray}
%%%%%%%%%%%%%%%%%%%%%%%%%%%%%%%%%%%%%%%%%%%%%%%%%%%%%%%%%%%%%%%%%%%%%%
&&a_l^{WG}={G_{_F}\alpha_{_e}m_{_l}^2\over32\sqrt{2}\pi^3s_{_{\rm w}}^2}
x_{_{\rm w}}\Bigg\{\Big({x_{_{F_\beta}}\over x_{_{\rm w}}}\Big)^{1/2}
F_1(x_{_{\rm w}},x_{_{\rm w}},x_{_{F_\alpha}},x_{_{F_\beta}})
\Re\Big({\cal G}^{c,L}_{_{\beta\alpha}}\zeta^L_{_{\alpha\beta}}
+{\cal G}^{c,R}_{_{\beta\alpha}}\zeta^R_{_{\alpha\beta}}\Big)
\nonumber\\
%---------------------------------------------------------------------
&&\hspace{1.2cm}
+\Big({x_{_{F_\alpha}}\over x_{_{\rm w}}}\Big)^{1/2}
F_2(x_{_{\rm w}},x_{_{\rm w}},x_{_{F_\alpha}},x_{_{F_\beta}})
\Re\Big({\cal G}^{c,L}_{_{\beta\alpha}}\zeta^R_{_{\alpha\beta}}
+{\cal G}^{c,R}_{_{\beta\alpha}}\zeta^L_{_{\alpha\beta}}\Big)
\nonumber\\
%---------------------------------------------------------------------
&&\hspace{1.2cm}
+\Big({x_{_{F_\beta}}\over x_{_{\rm w}}}\Big)^{1/2}
F_3(x_{_{\rm w}},x_{_{\rm w}},x_{_{F_\alpha}},x_{_{F_\beta}})
\Re\Big({\cal G}^{c,L}_{_{\beta\alpha}}\zeta^L_{_{\alpha\beta}}
-{\cal G}^{c,R}_{_{\beta\alpha}}\zeta^R_{_{\alpha\beta}}\Big)
\nonumber\\
%---------------------------------------------------------------------
&&\hspace{1.2cm}
+\Big({x_{_{F_\alpha}}\over x_{_{\rm w}}}\Big)^{1/2}
F_4(x_{_{\rm w}},x_{_{\rm w}},x_{_{F_\alpha}},x_{_{F_\beta}})
\Re\Big({\cal G}^{c,L}_{_{\beta\alpha}}\zeta^R_{_{\alpha\beta}}
-{\cal G}^{c,R}_{_{\beta\alpha}}\zeta^L_{_{\alpha\beta}}\Big)\Bigg\}\;,
\nonumber\\
%%%%%%%%%%%%%%%%%%%%%%%%%%%%%%%%%%%%%%%%%%%%%%%%%%%%%%%%%%%%%%%%%%%%%%
&&d_l^{WG}={G_{_F}\alpha_{_e}em_{_l}\over64\sqrt{2}\pi^3s_{_{\rm w}}^2}
x_{_{\rm w}}\Bigg\{\Big({x_{_{F_\beta}}\over x_{_{\rm w}}}\Big)^{1/2}
F_1(x_{_{\rm w}},x_{_{\rm w}},x_{_{F_\alpha}},x_{_{F_\beta}})
\Im\Big({\cal G}^{c,L}_{_{\beta\alpha}}\zeta^L_{_{\alpha\beta}}
+{\cal G}^{c,R}_{_{\beta\alpha}}\zeta^R_{_{\alpha\beta}}\Big)
\nonumber\\
%---------------------------------------------------------------------
&&\hspace{1.2cm}
+\Big({x_{_{F_\alpha}}\over x_{_{\rm w}}}\Big)^{1/2}
F_2(x_{_{\rm w}},x_{_{\rm w}},x_{_{F_\alpha}},x_{_{F_\beta}})
\Im\Big({\cal G}^{c,L}_{_{\beta\alpha}}\zeta^R_{_{\alpha\beta}}
+{\cal G}^{c,R}_{_{\beta\alpha}}\zeta^L_{_{\alpha\beta}}\Big)
\nonumber\\
%---------------------------------------------------------------------
&&\hspace{1.2cm}
+\Big({x_{_{F_\beta}}\over x_{_{\rm w}}}\Big)^{1/2}
F_3(x_{_{\rm w}},x_{_{\rm w}},x_{_{F_\alpha}},x_{_{F_\beta}})
\Im\Big({\cal G}^{c,L}_{_{\beta\alpha}}\zeta^L_{_{\alpha\beta}}
-{\cal G}^{c,R}_{_{\beta\alpha}}\zeta^R_{_{\alpha\beta}}\Big)
\nonumber\\
%---------------------------------------------------------------------
&&\hspace{1.2cm}
+\Big({x_{_{F_\alpha}}\over x_{_{\rm w}}}\Big)^{1/2}
F_4(x_{_{\rm w}},x_{_{\rm w}},x_{_{F_\alpha}},x_{_{F_\beta}})
\Im\Big({\cal G}^{c,L}_{_{\beta\alpha}}\zeta^R_{_{\alpha\beta}}
-{\cal G}^{c,R}_{_{\beta\alpha}}\zeta^L_{_{\alpha\beta}}\Big)\Bigg\}\;,
\nonumber\\
%%%%%%%%%%%%%%%%%%%%%%%%%%%%%%%%%%%%%%%%%%%%%%%%%%%%%%%%%%%%%%%%%%%%%%
&&a_l^{WH}={G_{_F}\alpha_{_e}m_{_l}^2{\cal B}_{c}\over32\sqrt{2}\pi^3s_{_{\rm w}}^2}
x_{_{\rm w}}\Bigg\{\Big({x_{_{F_\beta}}\over x_{_{\rm w}}}\Big)^{1/2}
F_1(x_{_{\rm w}},x_{_{H^\pm}},x_{_{F_\alpha}},x_{_{F_\beta}})
\Re\Big({\cal H}^{c,L}_{_{\beta\alpha}}\zeta^L_{_{\alpha\beta}}
+{\cal H}^{c,R}_{_{\beta\alpha}}\zeta^R_{_{\alpha\beta}}\Big)
\nonumber\\
%---------------------------------------------------------------------
&&\hspace{1.2cm}
+\Big({x_{_{F_\alpha}}\over x_{_{\rm w}}}\Big)^{1/2}
F_2(x_{_{\rm w}},x_{_{H^\pm}},x_{_{F_\alpha}},x_{_{F_\beta}})
\Re\Big({\cal H}^{c,L}_{_{\beta\alpha}}\zeta^R_{_{\alpha\beta}}
+{\cal H}^{c,R}_{_{\beta\alpha}}\zeta^L_{_{\alpha\beta}}\Big)
\nonumber\\
%---------------------------------------------------------------------
&&\hspace{1.2cm}
+\Big({x_{_{F_\beta}}\over x_{_{\rm w}}}\Big)^{1/2}
F_3(x_{_{\rm w}},x_{_{H^\pm}},x_{_{F_\alpha}},x_{_{F_\beta}})
\Re\Big({\cal H}^{c,L}_{_{\beta\alpha}}\zeta^L_{_{\alpha\beta}}
-{\cal H}^{c,R}_{_{\beta\alpha}}\zeta^R_{_{\alpha\beta}}\Big)
\nonumber\\
%---------------------------------------------------------------------
&&\hspace{1.2cm}
+\Big({x_{_{F_\alpha}}\over x_{_{\rm w}}}\Big)^{1/2}
F_4(x_{_{\rm w}},x_{_{H^\pm}},x_{_{F_\alpha}},x_{_{F_\beta}})
\Re\Big({\cal H}^{c,L}_{_{\beta\alpha}}\zeta^R_{_{\alpha\beta}}
-{\cal H}^{c,R}_{_{\beta\alpha}}\zeta^L_{_{\alpha\beta}}\Big)\Bigg\}\;,
\nonumber\\
%%%%%%%%%%%%%%%%%%%%%%%%%%%%%%%%%%%%%%%%%%%%%%%%%%%%%%%%%%%%%%%%%%%%%%
&&d_l^{WH}={G_{_F}\alpha_{_e}em_{_l}{\cal B}_{c}\over64\sqrt{2}\pi^3s_{_{\rm w}}^2}
x_{_{\rm w}}\Bigg\{\Big({x_{_{F_\beta}}\over x_{_{\rm w}}}\Big)^{1/2}
F_1(x_{_{\rm w}},x_{_{H^\pm}},x_{_{F_\alpha}},x_{_{F_\beta}})
\Im\Big({\cal H}^{c,L}_{_{\beta\alpha}}\zeta^L_{_{\alpha\beta}}
+{\cal H}^{c,R}_{_{\beta\alpha}}\zeta^R_{_{\alpha\beta}}\Big)
\nonumber\\
%---------------------------------------------------------------------
&&\hspace{1.2cm}
+\Big({x_{_{F_\alpha}}\over x_{_{\rm w}}}\Big)^{1/2}
F_2(x_{_{\rm w}},x_{_{H^\pm}},x_{_{F_\alpha}},x_{_{F_\beta}})
\Im\Big({\cal H}^{c,L}_{_{\beta\alpha}}\zeta^R_{_{\alpha\beta}}
+{\cal H}^{c,R}_{_{\beta\alpha}}\zeta^L_{_{\alpha\beta}}\Big)
\nonumber\\
%---------------------------------------------------------------------
&&\hspace{1.2cm}
+\Big({x_{_{F_\beta}}\over x_{_{\rm w}}}\Big)^{1/2}
F_3(x_{_{\rm w}},x_{_{H^\pm}},x_{_{F_\alpha}},x_{_{F_\beta}})
\Im\Big({\cal H}^{c,L}_{_{\beta\alpha}}\zeta^L_{_{\alpha\beta}}
-{\cal H}^{c,R}_{_{\beta\alpha}}\zeta^R_{_{\alpha\beta}}\Big)
\nonumber\\
%---------------------------------------------------------------------
&&\hspace{1.2cm}
+\Big({x_{_{F_\alpha}}\over x_{_{\rm w}}}\Big)^{1/2}
F_4(x_{_{\rm w}},x_{_{H^\pm}},x_{_{F_\alpha}},x_{_{F_\beta}})
\Im\Big({\cal H}^{c,L}_{_{\beta\alpha}}\zeta^R_{_{\alpha\beta}}
-{\cal H}^{c,R}_{_{\beta\alpha}}\zeta^L_{_{\alpha\beta}}\Big)\Bigg\}\;.
%%%%%%%%%%%%%%%%%%%%%%%%%%%%%%%%%%%%%%%%%%%%%%%%%%%%%%%%%%%%%%%%%%%%%%
\label{MED-W-G-H}
\end{eqnarray}
The expressions of form factors $F_i(x,y,z,u)\;(i=1,\cdots,4)$ can
be found in appendix.

Using the asymptotic formulae of form factors $F_i,\;(i=1,\cdots,4)$
under the condition $z,\;u\gg x,\;y$,
\begin{eqnarray}
%%%%%%%%%%%%%%%%%%%%%%%%%%%%%%%%%%%%%%%%%%%%%%%%%%%%%%%%%%%%%%%%%%%%%%
&&F_1(x,y,z,u)\simeq-{3(2-3Q_\beta)z^3-32z^2u-(20-9Q_\beta)zu^2\over
3(z-u)^4}\ln{z\over u}
\nonumber\\
&&\hspace{2.8cm}
+{(11-54Q_\beta)z^2-(151-54Q_\beta)zu+2u^2\over9(z-u)^3}
\nonumber\\
&&\hspace{2.8cm}
-\Big[5{\partial\over\partial u}+u{\partial^2\over\partial u^2}\Big]
\Big[2\varrho_{_{1,1}}(x,y)\pi_{_1}(z,u)+\varphi_1(z,u)\Big]
\nonumber\\
&&\hspace{2.8cm}
+\Big[(Q_\beta-6)+\Big((4-Q_\beta)z+(6+Q_\beta)u\Big){\partial\over\partial u}
\nonumber\\
&&\hspace{2.8cm}
-u(z-u){\partial^2\over\partial u^2}\Big]
\Big[\varrho_{_{1,1}}(x,y)\pi_{_2}(z,u)+{1\over2}\varphi_2(z,u)\Big]
\nonumber\\
&&\hspace{2.8cm}
-{10(z-u)\over3}
\Big[\varrho_{_{1,1}}(x,y)\pi_{_3}(z,u)-{1\over2}\varphi_3(z,u)\Big]
+\cdots\;,
\nonumber\\
%%%%%%%%%%%%%%%%%%%%%%%%%%%%%%%%%%%%%%%%%%%%%%%%%%%%%%%%%%%%%%%%%%%%%%
&&F_2(x,y,z,u)\simeq-{3z^3-(35-3Q_\beta)z^2u-(29-12Q_\beta)zu^2
+15(1-Q_\beta)u^3\over3(z-u)^4}\ln{z\over u}
\nonumber\\
&&\hspace{2.8cm}
+{(11-18Q_\beta)z^2-(223-90Q_\beta)zu+(74-72Q_\beta)u^2\over9(z-u)^3}
\nonumber\\
&&\hspace{2.8cm}
-\Big[{\partial\over\partial u}+(1-Q_\beta){\partial\over\partial z}
-u{\partial^2\over\partial u^2}\Big]
\Big[2\varrho_{_{1,1}}(x,y)\pi_{_1}(z,u)+\varphi_1(z,u)\Big]
\nonumber\\
&&\hspace{2.8cm}
+\Big[(2-Q_\beta)+(z+9u){\partial\over\partial u}
+(1-Q_\beta)(z-u){\partial\over\partial z}
\nonumber\\
&&\hspace{2.8cm}
-u(z-u){\partial^2\over\partial u^2}\Big]
\Big[\varrho_{_{1,1}}(x,y)\pi_{_2}(z,u)+{1\over2}\varphi_2(z,u)\Big]
\nonumber\\
&&\hspace{2.8cm}
-{10(z-u)\over3}
\Big[\varrho_{_{1,1}}(x,y)\pi_{_3}(z,u)+{1\over2}\varphi_3(z,u)\Big]
+\cdots\;,
\nonumber\\
%%%%%%%%%%%%%%%%%%%%%%%%%%%%%%%%%%%%%%%%%%%%%%%%%%%%%%%%%%%%%%%%%%%%%%
&&F_3(x,y,z,u)\simeq{(1+4Q_\beta)z^2+(5-4Q_\beta)zu\over(z-u)^3}\ln{z\over u}
\nonumber\\
&&\hspace{2.8cm}
-{4(1+Q_\beta)z+2(1-2Q_\beta)u\over(z-u)^2}
\nonumber\\
&&\hspace{2.8cm}
+(1+2Q_\beta){\partial\over\partial u}
\Big[2\varrho_{_{1,1}}(x,y)\pi_{_1}(z,u)+\varphi_1(z,u)\Big]
\nonumber\\
&&\hspace{2.8cm}
+\Big[1-(z-u){\partial\over\partial u}\Big]
\Big[\varrho_{_{1,1}}(x,y)\pi_{_2}(z,u)+{1\over2}\varphi_2(z,u)\Big]
+\cdots\;,
\nonumber\\
%%%%%%%%%%%%%%%%%%%%%%%%%%%%%%%%%%%%%%%%%%%%%%%%%%%%%%%%%%%%%%%%%%%%%%
&&F_4(x,y,z,u)\simeq-{z^2+(4+Q_\beta)zu-5(1-Q_\beta)u^2\over(z-u)^3}\ln{z\over u}
\nonumber\\
&&\hspace{2.8cm}
+{(6-2Q_\beta)z-(6-8Q_\beta)u\over(z-u)^2}
\nonumber\\
&&\hspace{2.8cm}
+\Big[{\partial\over\partial u}+(1-Q_\beta){\partial\over\partial z}
\Big]\Big[2\varrho_{_{1,1}}(x,y)\pi_{_1}(z,u)+\varphi_1(z,u)\Big]
\nonumber\\
&&\hspace{2.8cm}
-\Big[Q_\beta-(z-u){\partial\over\partial u}
-(1-Q_\beta)(z-u){\partial\over\partial z}\Big]
\Big[\varrho_{_{1,1}}(x,y)\pi_{_2}(z,u)
\nonumber\\
&&\hspace{2.8cm}
+{1\over2}\varphi_2(z,u)\Big]
+\cdots\;,
%%%%%%%%%%%%%%%%%%%%%%%%%%%%%%%%%%%%%%%%%%%%%%%%%%%%%%%%%%%%%%%%%%%%%
\label{F-asymp}
\end{eqnarray}
we simplify the expressions of Eq.(\ref{MED-W-G-H})
in the limit $m_{_F}=m_{_{F_\alpha}}=m_{_{F_\beta}}\gg m_{_{\rm w}}$ as:
\begin{eqnarray}
%%%%%%%%%%%%%%%%%%%%%%%%%%%%%%%%%%%%%%%%%%%%%%%%%%%%%%%%%%%%%%%%%%%%%%
&&a_l^{WG}={G_{_F}\alpha_{_e}m_{_l}^2m_{_{\rm w}}\over32\sqrt{2}\pi^3s_{_{\rm w}}^2m_{_F}}
\Bigg\{\Big[{9\over4}-{11\over18}Q_\beta
+(3+{Q_\beta\over3})\ln{m_{_F}^2\over m_{_{\rm w}}^2}\Big]
\Re\Big({\cal G}^{c,L}_{_{\beta\alpha}}\zeta^L_{_{\alpha\beta}}
+{\cal G}^{c,R}_{_{\beta\alpha}}\zeta^R_{_{\alpha\beta}}\Big)
\nonumber\\
%---------------------------------------------------------------------
&&\hspace{1.4cm}
+\Big[{13-8Q_\beta\over9}+{2-4Q_\beta\over3}
\ln{m_{_F}^2\over m_{_{\rm w}}^2}\Big]
\Re\Big({\cal G}^{c,L}_{_{\beta\alpha}}\zeta^R_{_{\alpha\beta}}
+{\cal G}^{c,R}_{_{\beta\alpha}}\zeta^L_{_{\alpha\beta}}\Big)
\nonumber\\
%---------------------------------------------------------------------
&&\hspace{1.4cm}
+\Big[-{2(5-9Q_\beta)\over9}-{2(1+3Q_\beta)\over3}
\ln{m_{_F}^2\over m_{_{\rm w}}^2}\Big]
\Re\Big({\cal G}^{c,L}_{_{\beta\alpha}}\zeta^L_{_{\alpha\beta}}
-{\cal G}^{c,R}_{_{\beta\alpha}}\zeta^R_{_{\alpha\beta}}\Big)
\nonumber\\
%---------------------------------------------------------------------
&&\hspace{1.4cm}
+\Big[{2(9-4Q_\beta)\over9}
-{2(3-Q_\beta)\over3}\ln{m_{_F}^2\over m_{_{\rm w}}^2}\Big]
\Re\Big({\cal G}^{c,L}_{_{\beta\alpha}}\zeta^R_{_{\alpha\beta}}
-{\cal G}^{c,R}_{_{\beta\alpha}}\zeta^L_{_{\alpha\beta}}\Big)\Bigg\}\;,
\nonumber\\
%%%%%%%%%%%%%%%%%%%%%%%%%%%%%%%%%%%%%%%%%%%%%%%%%%%%%%%%%%%%%%%%%%%%%%
&&d_l^{WG}={G_{_F}\alpha_{_e}em_{_l}m_{_{\rm w}}\over64\sqrt{2}\pi^3s_{_{\rm w}}^2m_{_F}}
\Bigg\{\Big[{9\over4}-{11\over18}Q_\beta
+(3+{Q_\beta\over3})\ln{m_{_F}^2\over m_{_{\rm w}}^2}\Big]
\Im\Big({\cal G}^{c,L}_{_{\beta\alpha}}\zeta^L_{_{\alpha\beta}}
+{\cal G}^{c,R}_{_{\beta\alpha}}\zeta^R_{_{\alpha\beta}}\Big)
\nonumber\\
%---------------------------------------------------------------------
&&\hspace{1.4cm}
+\Big[{13-8Q_\beta\over9}
+{2-4Q_\beta\over3}\ln{m_{_F}^2\over m_{_{\rm w}}^2}\Big]
\Im\Big({\cal G}^{c,L}_{_{\beta\alpha}}\zeta^R_{_{\alpha\beta}}
+{\cal G}^{c,R}_{_{\beta\alpha}}\zeta^L_{_{\alpha\beta}}\Big)
\nonumber\\
%---------------------------------------------------------------------
&&\hspace{1.4cm}
+\Big[-{2(5-9Q_\beta)\over9}
-{2(1+3Q_\beta)\over3}\ln{m_{_F}^2\over m_{_{\rm w}}^2}\Big]
\Im\Big({\cal G}^{c,L}_{_{\beta\alpha}}\zeta^L_{_{\alpha\beta}}
-{\cal G}^{c,R}_{_{\beta\alpha}}\zeta^R_{_{\alpha\beta}}\Big)
\nonumber\\
%---------------------------------------------------------------------
&&\hspace{1.4cm}
+\Big[{2(9-4Q_\beta)\over9}
-{2(3-Q_\beta)\over3}\ln{m_{_F}^2\over m_{_{\rm w}}^2}\Big]
\Im\Big({\cal G}^{c,L}_{_{\beta\alpha}}\zeta^R_{_{\alpha\beta}}
-{\cal G}^{c,R}_{_{\beta\alpha}}\zeta^L_{_{\alpha\beta}}\Big)\Bigg\}
\;,\nonumber\\
%%%%%%%%%%%%%%%%%%%%%%%%%%%%%%%%%%%%%%%%%%%%%%%%%%%%%%%%%%%%%%%%%%%%%%
%%%%%%%%%%%%%%%%%%%%%%%%%%%%%%%%%%%%%%%%%%%%%%%%%%%%%%%%%%%%%%%%%%%%%%
&&a_l^{WH}={G_{_F}\alpha_{_e}m_{_l}^2m_{_{\rm w}}{\cal B}_{c}
\over32\sqrt{2}\pi^3s_{_{\rm w}}^2m_{_F}}
\Bigg\{\Big[{21\over4}-{5\over18}Q_\beta+(3+{Q_\beta\over3})\Big(\ln m_{_F}^2
-\varrho_{_{1,1}}(m_{_{\rm w}}^2,m_{_{H^\pm}}^2)\Big)\Big]
\nonumber\\
&&\hspace{1.4cm}\times
\Re\Big({\cal H}^{c,L}_{_{\beta\alpha}}\zeta^L_{_{\alpha\beta}}
+{\cal H}^{c,R}_{_{\beta\alpha}}\zeta^R_{_{\alpha\beta}}\Big)
\nonumber\\
%---------------------------------------------------------------------
&&\hspace{1.4cm}
+\Big[{19-20Q_\beta\over9}+{2-4Q_\beta\over3}\Big(\ln m_{_F}^2
-\varrho_{_{1,1}}(m_{_{\rm w}}^2,m_{_{H^\pm}}^2)\Big)\Big]
\Re\Big({\cal H}^{c,L}_{_{\beta\alpha}}\zeta^R_{_{\alpha\beta}}
+{\cal H}^{c,R}_{_{\beta\alpha}}\zeta^L_{_{\alpha\beta}}\Big)
\nonumber\\
%---------------------------------------------------------------------
&&\hspace{1.4cm}
+\Big[-{16\over9}-{2+6Q_\beta\over3}\Big(\ln m_{_F}^2
-\varrho_{_{1,1}}(m_{_{\rm w}}^2,m_{_{H^\pm}}^2)\Big)\Big]
\Re\Big({\cal H}^{c,L}_{_{\beta\alpha}}\zeta^L_{_{\alpha\beta}}
-{\cal H}^{c,R}_{_{\beta\alpha}}\zeta^R_{_{\alpha\beta}}\Big)
\nonumber\\
%---------------------------------------------------------------------
&&\hspace{1.4cm}
+\Big[-{2Q_\beta\over9}-{6-2Q_\beta\over3}\Big(\ln m_{_F}^2
-\varrho_{_{1,1}}(m_{_{\rm w}}^2,m_{_{H^\pm}}^2)\Big)\Big]
\Re\Big({\cal H}^{c,L}_{_{\beta\alpha}}\zeta^R_{_{\alpha\beta}}
-{\cal H}^{c,R}_{_{\beta\alpha}}\zeta^L_{_{\alpha\beta}}\Big)\Bigg\}\;,
\nonumber\\
%%%%%%%%%%%%%%%%%%%%%%%%%%%%%%%%%%%%%%%%%%%%%%%%%%%%%%%%%%%%%%%%%%%%%%
&&d_l^{WH}={G_{_F}\alpha_{_e}em_{_l}m_{_{\rm w}}{\cal B}_{c}
\over64\sqrt{2}\pi^3s_{_{\rm w}}^2m_{_F}}
\Bigg\{\Big[{21\over4}-{5\over18}Q_\beta+(3+{Q_\beta\over3})\Big(\ln m_{_F}^2
-\varrho_{_{1,1}}(m_{_{\rm w}}^2,m_{_{H^\pm}}^2)\Big)\Big]
\nonumber\\
&&\hspace{1.4cm}\times
\Im\Big({\cal H}^{c,L}_{_{\beta\alpha}}\zeta^L_{_{\alpha\beta}}
+{\cal H}^{c,R}_{_{\beta\alpha}}\zeta^R_{_{\alpha\beta}}\Big)
\nonumber\\
%---------------------------------------------------------------------
&&\hspace{1.4cm}
+\Big[{19-20Q_\beta\over9}+{2-4Q_\beta\over3}\Big(\ln m_{_F}^2
-\varrho_{_{1,1}}(m_{_{\rm w}}^2,m_{_{H^\pm}}^2)\Big)\Big]
\Im\Big({\cal H}^{c,L}_{_{\beta\alpha}}\zeta^R_{_{\alpha\beta}}
+{\cal H}^{c,R}_{_{\beta\alpha}}\zeta^L_{_{\alpha\beta}}\Big)
\nonumber\\
%---------------------------------------------------------------------
&&\hspace{1.4cm}
+\Big[-{16\over9}-{2+6Q_\beta\over3}\Big(\ln m_{_F}^2
-\varrho_{_{1,1}}(m_{_{\rm w}}^2,m_{_{H^\pm}}^2)\Big)\Big]
\Im\Big({\cal H}^{c,L}_{_{\beta\alpha}}\zeta^L_{_{\alpha\beta}}
-{\cal H}^{c,R}_{_{\beta\alpha}}\zeta^R_{_{\alpha\beta}}\Big)
\nonumber\\
%---------------------------------------------------------------------
&&\hspace{1.4cm}
+\Big[-{2Q_\beta\over9}-{6-2Q_\beta\over3}\Big(\ln m_{_F}^2
-\varrho_{_{1,1}}(m_{_{\rm w}}^2,m_{_{H^\pm}}^2)\Big)\Big]
\Im\Big({\cal H}^{c,L}_{_{\beta\alpha}}\zeta^R_{_{\alpha\beta}}
-{\cal H}^{c,R}_{_{\beta\alpha}}\zeta^L_{_{\alpha\beta}}\Big)\Bigg\}\;.
%%%%%%%%%%%%%%%%%%%%%%%%%%%%%%%%%%%%%%%%%%%%%%%%%%%%%%%%%%%%%%%%%%%%%%
\label{ASY-MED-W-GH}
\end{eqnarray}
The results indicate that the corrections to $a_l,\;d_l$ from
the diagrams in Fig.\ref{fig3} are suppressed in the limit
$m_{_F}=m_{_{F_\alpha}}=m_{_{F_\beta}}\gg m_{_{\rm w}}$
unless the couplings ${\cal H}^{c,L,R}_{_{\beta\alpha}}$ violate the
decoupling theorem.

%%%%%%%%%%%%%%%%%%%%%%%%%%%%%%%%%%%%%%%%%%%%%%%%%%%%%%%%%%%%%%%%%%%
\begin{figure}[h]
\setlength{\unitlength}{1mm}
\begin{center}
\begin{picture}(0,60)(0,0)
\put(-62,-65){\includegraphics{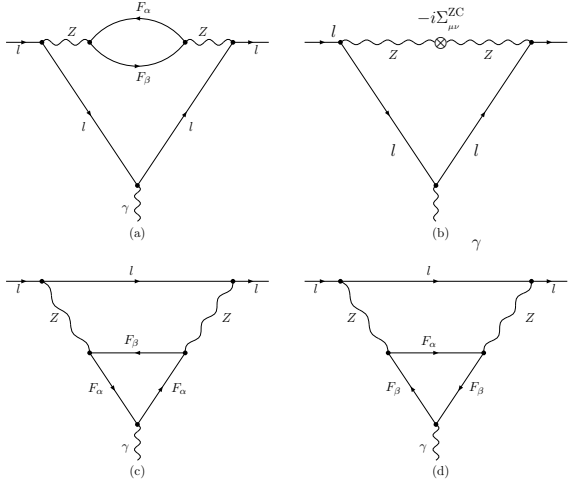}}
\end{picture}
\caption[]{The two loop diagrams in which a closed heavy
fermion loop is inserted into the propagator of virtual neutral gauge boson,
where diagram (b) contributes the counter terms to cancel the ultraviolet
divergence arisen by divergent subdiagram in (a).}
\label{fig4}
\end{center}
\end{figure}
%%%%%%%%%%%%%%%%%%%%%%%%%%%%%%%%%%%%%%%%%%%%%%%%%%%%%%%%%%%%%%%%%%%

\subsection{The corrections from the diagram where a closed heavy fermion
loop is inserted into the self energy of $Z$ gauge boson}
\indent\indent
In order to get the amplitude of the diagram in Fig.\ref{fig4}, we write
the renormalizable interaction among the electroweak neutral gauge boson $Z$ and
the heavy fermions $F_{\alpha,\beta}$ in a more universal form as
\begin{eqnarray}
&&{\cal L}_{_{ZFF}}={e\over2s_{_{\rm w}}c_{_{\rm w}}}Z^\mu\bar{F}_\alpha\gamma_\mu
(\xi^L_{_{\alpha\beta}}\omega_-+\xi^R_{_{\alpha\beta}}\omega_+)F_\beta\;,
\label{ZFF}
\end{eqnarray}
where the expressions of $\xi^{L,R}_{_{\alpha\beta}}$ depend on the
concrete models employed in our calculation, and the CPT symmetry
requires $\xi^L_{_{\alpha\beta}}=\xi^{L*}_{_{\beta\alpha}},
\;\xi^R_{_{\alpha\beta}}=\xi^{R*}_{_{\beta\alpha}}$.
%%%%%%%%%%%%%%%%%%%%%%%%%%%%%%%%%%%%%%%%%%%%%%%%%%%%%%%%%%%%%%%%%%%
\begin{figure}[h]
\setlength{\unitlength}{1mm}
\begin{center}
\begin{picture}(0,30)(0,0)
\put(-60,-80){\includegraphics{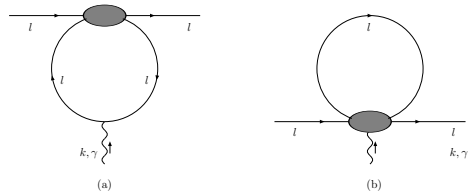}}
\end{picture}
\caption[]{The diagrams of effective theory correspond to those
diagrams in Fig.\ref{fig4}. Where the diagram (a) corresponds
to Fig.\ref{fig4}(a) and (b) in full theory, and the diagram (b) corresponds
to Fig.\ref{fig4}(c) and (d) in full theory respectively.}
\label{fig5}
\end{center}
\end{figure}
%%%%%%%%%%%%%%%%%%%%%%%%%%%%%%%%%%%%%%%%%%%%%%%%%%%%%%%%%%%%%%%%%%%

It is easy to check that the divergent amplitude of Fig.\ref{fig4}(a)
satisfies the Ward identity:
\begin{eqnarray}
&&k^\rho{\cal A}_{{\rm zz},\rho}^{(a)}(p,k)=e[\Sigma_{_{\rm zz}}(p+k)
-\Sigma_{_{\rm zz}}(p)]\;,
\label{WTI-zz(a)}
\end{eqnarray}
where $p,\;k$ are the incoming momenta of lepton and photon fields,
$\rho$ is the Lorentz index of photon, ${\cal A}_{{\rm zz},\rho}^{(a)}$
denotes the sum of amplitude for the diagrams in Fig.\ref{fig5}(a),
and $\Sigma_{\rm zz}$ denotes the amplitude of corresponding self energy
diagram, respectively. After a tedious calculation, the bare Lagrangian
from the diagram Fig.\ref{fig4}(a) can be written as
\begin{eqnarray}
%%%%%%%%%%%%%%%%%%%%%%%%%%%%%%%%%%%%%%%%%%%%%%%%%%%%%%%%%%%%%%%%%%%%%%
&&{\cal L}_{_{Z(a)}}^{eff}=
-{G_{_F}\alpha_{_e}x_{_{\rm w}}\over6\sqrt{2}\pi s_{_{\rm w}}^2c_{_{\rm w}}^4}
{\Gamma^2(1+\epsilon)\over(1-\epsilon)^2}(4\pi x_{_R})^{2\epsilon}
\Bigg\{\Big(\xi^L_{_{\beta\alpha}}\xi^L_{_{\alpha\beta}}
+\xi^R_{_{\beta\alpha}}\xi^R_{_{\alpha\beta}}\Big)
\Big[{1\over\varepsilon}{x_{_{F_\alpha}}+x_{_{F_\beta}}\over x_{_{\rm z}}^2}
\nonumber\\
&&\hspace{1.6cm}
+{x_{_{F_\alpha}}+x_{_{F_\beta}}\over x_{_{\rm z}}^2}\Big({4\over3}
-\ln x_{_{\rm z}}\Big)
-{\varrho_{_{2,1}}(x_{_{F_\alpha}},x_{_{F_\beta}})\over x_{_{\rm z}}^2}
-{2\over3x_{_{\rm z}}}
+T_5(x_{_{\rm z}},x_{_{F_\alpha}},x_{_{F_\beta}})\Big]
\nonumber\\
&&\hspace{1.6cm}\times
\Big[\Big(T_{_f}^Z-Q_{_f}s_{_{\rm w}}^2\Big)^2({\cal O}_{_2}^-+{\cal O}_{_3}^-)
+Q_{_f}^2s_{_{\rm w}}^4({\cal O}_{_2}^++{\cal O}_{_3}^+)\Big]
\nonumber\\
%---------------------------------------------------------------------
&&\hspace{1.6cm}
+\Big(\xi^R_{_{\beta\alpha}}\xi^L_{_{\alpha\beta}}+\xi^L_{_{\beta\alpha}}
\xi^R_{_{\alpha\beta}}\Big)(x_{_{F_\alpha}}x_{_{F_\beta}})^{1/2}
\Big[-{2\over\varepsilon x_{_{\rm z}}^2}
+{2\varrho_{_{1,1}}(x_{_{F_\alpha}},x_{_{F_\beta}})\over x_{_{\rm z}}^2}
+{2\over x_{_{\rm z}}^2}\ln x_{_{\rm z}}
\nonumber\\
&&\hspace{1.6cm}
-{11\over3x_{_{\rm z}}^2}\Big]
\Big[\Big(T_{_f}^Z-Q_{_f}s_{_{\rm w}}^2\Big)^2({\cal O}_{_2}^-+{\cal O}_{_3}^-)
+Q_{_f}^2s_{_{\rm w}}^4({\cal O}_{_2}^++{\cal O}_{_3}^+)\Big]
\nonumber\\
%---------------------------------------------------------------------
&&\hspace{1.6cm}
+3Q_{_f}\Big(\xi^L_{_{\alpha\beta}}\xi^L_{_{\beta\alpha}}
+\xi^R_{_{\alpha\beta}}\xi^R_{_{\beta\alpha}}\Big)s_{_{\rm w}}^2
\Big(T_{_f}^Z-Q_{_f}s_{_{\rm w}}^2\Big)
\Big[{1\over\varepsilon}{x_{_{F_\alpha}}+x_{_{F_\beta}}\over x_{_{\rm z}}^2}
\nonumber\\
&&\hspace{1.6cm}
-{\varrho_{_{2,1}}(x_{_{F_\alpha}},x_{_{F_\beta}})\over x_{_{\rm z}}^2}
-{x_{_{F_\alpha}}+x_{_{F_\beta}}\over x_{_{\rm z}}^2}({3\over2}+\ln x_{_l})
\Big]({\cal O}_{_6}^++{\cal O}_{_6}^-)
\nonumber\\
%---------------------------------------------------------------------
&&\hspace{1.6cm}
-6Q_{_f}\Big(\xi^L_{_{\alpha\beta}}\xi^R_{_{\beta\alpha}}
+\xi^R_{_{\alpha\beta}}\xi^L_{_{\beta\alpha}}\Big)
s_{_{\rm w}}^2\Big(T_{_f}^Z-Q_{_f}s_{_{\rm w}}^2\Big)
(x_{_{F_\alpha}}x_{_{F_\beta}})^{1/2}
\Big[{1\over\varepsilon x_{_{\rm z}}^2}
\nonumber\\
&&\hspace{1.6cm}
-{\varrho_{_{1,1}}(x_{_{F_\alpha}},x_{_{F_\beta}})\over x_{_{\rm z}}^2}
-{1+\ln x_{_l}\over x_{_{\rm z}}^2}\Big]({\cal O}_{_6}^++{\cal O}_{_6}^-)\Bigg\}
+\cdots\;,
%%%%%%%%%%%%%%%%%%%%%%%%%%%%%%%%%%%%%%%%%%%%%%%%%%%%%%%%%%%%%%%%%%%%%%
\label{zz}
\end{eqnarray}
where $T_{_f}^Z=-1/2$ represents the isospin of charged leptons.

Generally, the unrenormalized self energy of the weak gauge boson
$Z$ can be written as
\begin{eqnarray}
%%%%%%%%%%%%%%%%%%%%%%%%%%%%%%%%%%%%%%%%%%%%%%%%%%%%%%%%%%%%%%%%%%%%%%%%%%%%%%%%%%%
&&\Sigma_{_{\mu\nu}}^{\rm z}(p,\Lambda_{_{\rm RE}})=\Lambda^2A_0^zg_{\mu\nu}+\Big(A_1^z
+{p^2\over\Lambda^2}A_2^z+\cdots\Big)(p^2g_{\mu\nu}-p_\mu p_\nu)
\nonumber\\
&&\hspace{2.5cm}
+\Big(B_1^z+{p^2\over\Lambda^2}B_2^z+\cdots\Big)p_\mu p_\nu\;.
%%%%%%%%%%%%%%%%%%%%%%%%%%%%%%%%%%%%%%%%%%%%%%%%%%%%%%%%%%%%%%%%%%%%%%%%%%%%%%%%%%%
\label{eq-z1}
\end{eqnarray}
Here, we omit those terms which are strongly suppressed at the limit
of the large virtual fermion masses, and those form factors are actually decided
by the masses of virtual fields and the renormalization scale.
Correspondingly, the counter terms are given as
\begin{eqnarray}
%%%%%%%%%%%%%%%%%%%%%%%%%%%%%%%%%%%%%%%%%%%%%%%%%%%%%%%%%%%%%%%%%%%%%%%%%%%%%%%%%%%
&&\Sigma_{_{\mu\nu}}^{\rm zC}(p,\Lambda_{_{\rm RE}})=
-\Big[\delta m_{_{\rm z}}^2(\Lambda_{_{\rm RE}})
+m_{_{\rm z}}^2\delta Z_{_{\rm z}}(\Lambda_{_{\rm RE}})\Big]g_{\mu\nu}
-\delta Z_{_{\rm z}}(\Lambda_{_{\rm RE}})\Big[p^2g_{\mu\nu}-p_\mu p_\nu\Big]\;.
%%%%%%%%%%%%%%%%%%%%%%%%%%%%%%%%%%%%%%%%%%%%%%%%%%%%%%%%%%%%%%%%%%%%%%%%%%%%%%%%%%%
\label{eq-z2}
\end{eqnarray}
The renormalized self energy is expressed as
\begin{eqnarray}
%%%%%%%%%%%%%%%%%%%%%%%%%%%%%%%%%%%%%%%%%%%%%%%%%%%%%%%%%%%%%%%%%%%%%%%%%%%%%%%%%%%
&&\hat{\Sigma}_{_{\mu\nu}}^{\rm z}(p,\Lambda_{_{\rm RE}})
=\Sigma_{_{\mu\nu}}^{\rm z}(p,\Lambda_{_{\rm RE}})
+\Sigma_{_{\mu\nu}}^{\rm zC}(p,\Lambda_{_{\rm RE}})\;.
%%%%%%%%%%%%%%%%%%%%%%%%%%%%%%%%%%%%%%%%%%%%%%%%%%%%%%%%%%%%%%%%%%%%%%%%%%%%%%%%%%%
\label{eq-z3}
\end{eqnarray}
For on-shell external gauge boson $Z$, we have \cite{onshell}
\begin{eqnarray}
%%%%%%%%%%%%%%%%%%%%%%%%%%%%%%%%%%%%%%%%%%%%%%%%%%%%%%%%%%%%%%%%%%%%%%%%%%%%%%%%%%%
&&\hat{\Sigma}_{_{\mu\nu}}^{\rm z}(p,m_{_{\rm z}})\epsilon^\nu(p)\Big|_{p^2=m_{_{\rm z}}^2}=0
\;,\nonumber\\
&&\lim\limits_{p^2\rightarrow m_{_{\rm z}}^2}{1\over p^2-m_{_{\rm z}}^2}
\hat{\Sigma}_{_{\mu\nu}}^{\rm z}(p,m_{_{\rm z}})\epsilon^\nu(p)=\epsilon_{_\mu}(p)\;,
%%%%%%%%%%%%%%%%%%%%%%%%%%%%%%%%%%%%%%%%%%%%%%%%%%%%%%%%%%%%%%%%%%%%%%%%%%%%%%%%%%%
\label{eq-z4}
\end{eqnarray}
where $\epsilon(p)$ is the polarization vector of neutral gauge boson.
From Eq.(\ref{eq-z4}), we get the counter terms at electroweak scale
in on-shell scheme:
\begin{eqnarray}
%%%%%%%%%%%%%%%%%%%%%%%%%%%%%%%%%%%%%%%%%%%%%%%%%%%%%%%%%%%%%%%%%%%%%%%%%%%%%%%%%%%
&&\delta Z_{_{\rm z}}^{os}(m_{_{\rm z}})=A_1^z+{m_{_{\rm z}}^2\over\Lambda^2}A_2^z
=A_1^z+x_{_{\rm z}}A_2^z\;,
\nonumber\\
&&\delta m_{_{\rm z}}^{2,os}(m_{_{\rm z}})=A_0^z\Lambda^2
-m_{_{\rm z}}^2\delta Z_{_{\rm z}}^{os}\;.
%%%%%%%%%%%%%%%%%%%%%%%%%%%%%%%%%%%%%%%%%%%%%%%%%%%%%%%%%%%%%%%%%%%%%%%%%%%%%%%%%%%
\label{eq-z5}
\end{eqnarray}

Accordingly, the effective Lagrangian from the counter term diagrams is written as
\begin{eqnarray}
%%%%%%%%%%%%%%%%%%%%%%%%%%%%%%%%%%%%%%%%%%%%%%%%%%%%%%%%%%%%%%%%%%%%%%%
&&\delta{\cal L}_{\rm zz}^C=i{e^2\over s_{_{\rm w}}^2c_{_{\rm w}}^2\Lambda^2}
(4\pi x_{_{\rm R}})^{\varepsilon}{\Gamma(1+\varepsilon)\over(1-\varepsilon)}
\Bigg\{\Big[A_0^{\rm z}\Big({1\over3x_{_{\rm z}}^2}+{11\varepsilon\over
18x_{_{\rm z}}^2}-{\varepsilon\over3x_{_{\rm z}}^2}\ln x_{_{\rm z}}\Big)
\nonumber\\
&&\hspace{1.6cm}
+{\varepsilon\over3x_{_{\rm z}}}(A_1^{\rm z}+x_{_{\rm z}}A_2^{\rm z})\Big]
\Big[\Big(T_{_f}^Z-Q_{_f}s_{_{\rm w}}^2\Big)^2({\cal O}_{_2}^-+{\cal O}_{_3}^-)
+Q_{_f}^2s_{_{\rm w}}^4({\cal O}_{_2}^++{\cal O}_{_3}^+)\Big]
\nonumber\\
%---------------------------------------------------------------------
&&\hspace{1.6cm}
-Q_{_f}s_{_{\rm w}}^2\Big(T_{_f}^Z-Q_{_f}s_{_{\rm w}}^2\Big)
A_0^{\rm z}\Big[-{1\over x_{_{\rm z}}^2}+{\varepsilon\over x_{_{\rm z}}^2}
(1+\ln x_{_l})\Big]({\cal O}_{_6}^++{\cal O}_{_6}^-)\Bigg\}
\nonumber\\
%---------------------------------------------------------------------
&&\hspace{1.2cm}=
{G_{_F}\alpha_{_e}x_{_{\rm w}}\over6\sqrt{2}\pi s_{_{\rm w}}^2c_{_{\rm w}}^4}
{\Gamma^2(1+\epsilon)\over(1-\epsilon)^2}(4\pi x_{_R})^{2\epsilon}
\Bigg\{\Big(\xi^L_{_{\beta\alpha}}\xi^L_{_{\alpha\beta}}
+\xi^R_{_{\beta\alpha}}\xi^R_{_{\alpha\beta}}\Big)
\Big[{1\over\varepsilon}{x_{_{F_\alpha}}+x_{_{F_\beta}}\over x_{_{\rm z}}^2}
\nonumber\\
&&\hspace{1.6cm}
+{x_{_{F_\alpha}}+x_{_{F_\beta}}\over x_{_{\rm z}}^2}\Big({4\over3}
-\ln x_{_{\rm z}}\Big)
-{\varrho_{_{2,1}}(x_{_{F_\alpha}},x_{_{F_\beta}})\over x_{_{\rm z}}^2}
-{2\over3x_{_{\rm z}}}\Big]
\nonumber\\
&&\hspace{1.6cm}\times
\Big[\Big(T_{_f}^Z-Q_{_f}s_{_{\rm w}}^2\Big)^2({\cal O}_{_2}^-+{\cal O}_{_3}^-)
+Q_{_f}^2s_{_{\rm w}}^4({\cal O}_{_2}^++{\cal O}_{_3}^+)\Big]
\nonumber\\
%---------------------------------------------------------------------
&&\hspace{1.6cm}
+\Big(\xi^R_{_{\beta\alpha}}\xi^L_{_{\alpha\beta}}+\xi^L_{_{\beta\alpha}}
\xi^R_{_{\alpha\beta}}\Big)(x_{_{F_\alpha}}x_{_{F_\beta}})^{1/2}
\Big[-{2\over\varepsilon x_{_{\rm z}}^2}
+{2\varrho_{_{1,1}}(x_{_{F_\alpha}},x_{_{F_\beta}})\over x_{_{\rm z}}^2}
+{2\over x_{_{\rm z}}^2}\ln x_{_{\rm z}}
\nonumber\\
&&\hspace{1.6cm}
-{11\over3x_{_{\rm z}}^2}\Big]
\Big[\Big(T_{_f}^Z-Q_{_f}s_{_{\rm w}}^2\Big)^2({\cal O}_{_2}^-+{\cal O}_{_3}^-)
+Q_{_f}^2s_{_{\rm w}}^4({\cal O}_{_2}^++{\cal O}_{_3}^+)\Big]
\nonumber\\
%---------------------------------------------------------------------
&&\hspace{1.6cm}
+3Q_{_f}\Big(\xi^L_{_{\alpha\beta}}\xi^L_{_{\beta\alpha}}
+\xi^R_{_{\alpha\beta}}\xi^R_{_{\beta\alpha}}\Big)s_{_{\rm w}}^2
\Big(T_{_f}^Z-Q_{_f}s_{_{\rm w}}^2\Big)
\Big[{1\over\varepsilon}{x_{_{F_\alpha}}+x_{_{F_\beta}}\over x_{_{\rm z}}^2}
\nonumber\\
&&\hspace{1.6cm}
-{\varrho_{_{2,1}}(x_{_{F_\alpha}},x_{_{F_\beta}})\over x_{_{\rm z}}^2}
-{x_{_{F_\alpha}}+x_{_{F_\beta}}\over x_{_{\rm z}}^2}({3\over2}+\ln x_{_l})
\Big]({\cal O}_{_6}^++{\cal O}_{_6}^-)
\nonumber\\
%---------------------------------------------------------------------
&&\hspace{1.6cm}
-6Q_{_f}\Big(\xi^L_{_{\alpha\beta}}\xi^R_{_{\beta\alpha}}
+\xi^R_{_{\alpha\beta}}\xi^L_{_{\beta\alpha}}\Big)
s_{_{\rm w}}^2\Big(T_{_f}^Z-Q_{_f}s_{_{\rm w}}^2\Big)
(x_{_{F_\alpha}}x_{_{F_\beta}})^{1/2}
\Big[{1\over\varepsilon x_{_{\rm z}}^2}
\nonumber\\
&&\hspace{1.6cm}
-{\varrho_{_{1,1}}(x_{_{F_\alpha}},x_{_{F_\beta}})\over x_{_{\rm z}}^2}
-{1+\ln x_{_l}\over x_{_{\rm z}}^2}\Big]({\cal O}_{_6}^++{\cal O}_{_6}^-)\Bigg\}
+\cdots\;,
%%%%%%%%%%%%%%%%%%%%%%%%%%%%%%%%%%%%%%%%%%%%%%%%%%%%%%%%%%%%%%%%%%%%%%%%%%%%%%%%%%%
\label{z-counter}
\end{eqnarray}

As a result of the preparation mentioned above, we use the contributions from
the counter term diagram in  Fig.\ref{fig4}(b) to cancel the ultraviolet divergence
in Eq.(\ref{zz}). After matching the amplitude from Fig.\ref{fig4}(a, b) to that
from the diagram of effective theory (Fig.\ref{fig5}(a)) and analyzing
the corrections from light freedoms properly, we get the theoretical
prediction on the lepton MDMs as
\begin{eqnarray}
%%%%%%%%%%%%%%%%%%%%%%%%%%%%%%%%%%%%%%%%%%%%%%%%%%%%%%%%%%%%%%%%%%%%%%
&&a_{l(a+b)}^{\rm zz}=
-{G_{_F}\alpha_{_e}Q_{_f}m_{_l}^2x_{_{\rm w}}\over24\sqrt{2}\pi^3
s_{_{\rm w}}^2c_{_{\rm w}}^4}\Big(|\xi^L_{_{\alpha\beta}}|^2
+|\xi^R_{_{\alpha\beta}}|^2\Big)
\Big[\Big(T_{_f}^Z-Q_{_f}s_{_{\rm w}}^2\Big)^2
+Q_{_f}^2s_{_{\rm w}}^4\Big]
T_5(x_{_{\rm z}},x_{_{F_\alpha}},x_{_{F_\beta}})\;.
%%%%%%%%%%%%%%%%%%%%%%%%%%%%%%%%%%%%%%%%%%%%%%%%%%%%%%%%%%%%%%%%%%%%%%
\label{HF-MDM}
\end{eqnarray}
In fact, the four-fermion interaction does not induce the corrections
to lepton MDMs and EDMs in Fig.\ref{fig5}(a).
Using the asymptotic expression of $T_5(x,z,u)$ at the limit $z,\;u\gg x$:
\begin{eqnarray}
%%%%%%%%%%%%%%%%%%%%%%%%%%%%%%%%%%%%%%%%%%%%%%%%%%%%%%%%%%%%%%%%%%%%%%
&&T_5(x,z,u)\simeq-{1\over9x}-{z+u\over6x}\pi_{_2}(z,u)
+{(z-u)^2\over18x}\pi_{_3}(z,u)+\cdots\;,
%%%%%%%%%%%%%%%%%%%%%%%%%%%%%%%%%%%%%%%%%%%%%%%%%%%%%%%%%%%%%%%%%%%%%%
\label{T5-asymp}
\end{eqnarray}
one finally obtains the leading corrections contained in the asymptotic
form of Eq.(\ref{HF-MDM}) under the assumption $m_{_F}=m_{_{F_\alpha}}=m_{_{F_\beta}}
\gg m_{_{\rm w}}$:
\begin{eqnarray}
%%%%%%%%%%%%%%%%%%%%%%%%%%%%%%%%%%%%%%%%%%%%%%%%%%%%%%%%%%%%%%%%%%%%%%
&&a_{l(a+b)}^{\rm zz}\propto{\cal O}(m_{_{\rm w}}^2/m_{_F}^2)+\cdots\;,
%%%%%%%%%%%%%%%%%%%%%%%%%%%%%%%%%%%%%%%%%%%%%%%%%%%%%%%%%%%%%%%%%%%%%%
\label{asyHF-MDM-Z}
\end{eqnarray}
where ellipsis represents those relatively unimportant corrections.

In a similar way, we can verify the identity
\begin{eqnarray}
&&k^\rho{\cal A}_{_{{\rm zz},\rho}}^{(c+d)}(p,k)=0\;.
\label{WTI-zz(b+c)}
\end{eqnarray}
${\cal A}_{_{{\rm zz},\rho}}^{(c+d)}$
denotes the sum of amplitude for the diagrams in Fig.\ref{fig4}(c,d).
After the steps adopted above, the corrections from Fig.\ref{fig4}(c,d)
to lepton MDMs and EDMs are formulated as
\begin{eqnarray}
%%%%%%%%%%%%%%%%%%%%%%%%%%%%%%%%%%%%%%%%%%%%%%%%%%%%%%%%%%%%%%%%%%%%%%
&&a_{l(c+d)}^{\rm zz}=
-{Q_\beta G_{_F}\alpha_{_e}m_{_l}^2\over64\sqrt{2}\pi^3s_{_{\rm w}}^2c_{_{\rm w}}^4}
x_{_{\rm w}}\Bigg\{\Big(|\xi^L_{_{\alpha\beta}}|^2+|\xi^R_{_{\alpha\beta}}|^2\Big)
\Big[\Big(T_{_f}^Z-Q_{_f}s_{_{\rm w}}^2\Big)^2+Q_{_f}^2s_{_{\rm w}}^4\Big]
\Bigg[T_6(x_{_{\rm z}},x_{_{F_\alpha}},x_{_{F_\beta}})\Bigg]
\nonumber\\
%---------------------------------------------------------------------
&&\hspace{1.6cm}
+\Big(|\xi^L_{_{\alpha\beta}}|^2-|\xi^R_{_{\alpha\beta}}|^2\Big)
\Big[\Big(T_{_f}^Z-Q_{_f}s_{_{\rm w}}^2\Big)^2-Q_{_f}^2s_{_{\rm w}}^4\Big]
T_{7}(x_{_{\rm z}},x_{_{F_\alpha}},x_{_{F_\beta}})
\nonumber\\
%---------------------------------------------------------------------
&&\hspace{1.6cm}
-\Re(\xi^L_{_{\alpha\beta}}\xi^R_{_{\beta\alpha}})
\Big[\Big(T_{_f}^Z-Q_{_f}s_{_{\rm w}}^2\Big)^2+Q_{_f}^2s_{_{\rm w}}^4\Big]
(x_{_{F_\alpha}}x_{_{F_\beta}})^{1/2}
T_{8}(x_{_{\rm z}},x_{_{F_\alpha}},x_{_{F_\beta}})\Bigg]
\nonumber\\
%---------------------------------------------------------------------
&&\hspace{1.6cm}
-Q_{_f}\Big(|\xi^L_{_{\alpha\beta}}|^2+|\xi^R_{_{\alpha\beta}}|^2\Big)s_{_{\rm w}}^2
\Big(T_{_f}^Z-Q_{_f}s_{_{\rm w}}^2\Big)
T_{10}(x_{_{\rm z}},x_{_{F_\alpha}},x_{_{F_\beta}})
\Bigg\}\;,
\nonumber\\
%---------------------------------------------------------------------
&&d_{l(c+d)}^{\rm zz}=
{Q_\beta G_{_F}\alpha_{_e}em_{_l}\over8\sqrt{2}\pi^3s_{_{\rm w}}^2c_{_{\rm w}}^4}
x_{_{\rm w}}\cdot\Im(\xi^L_{_{\alpha\beta}}\xi^R_{_{\beta\alpha}})
(x_{_{F_\alpha}}x_{_{F_\beta}})^{1/2}\Bigg\{
Q_{_f}s_{_{\rm w}}^2\Big(T_{_f}^Z-Q_{_f}s_{_{\rm w}}^2\Big)
\nonumber\\
&&\hspace{1.6cm}\times
\Big({\partial^2\over\partial x_{_{\rm z}}\partial x_{_{F_\beta}}}
-{\partial^2\over\partial x_{_{\rm z}}\partial x_{_{F_\alpha}}}\Big)
\Big({\Phi(x_{_{\rm z}},x_{_{F_\alpha}},x_{_{F_\beta}})
-\varphi_0(x_{_{F_\alpha}},x_{_{F_\beta}})
\over x_{_{\rm z}}}\Big)
\nonumber\\
&&\hspace{1.6cm}
-{1\over16}\Big[\Big(T_{_f}^Z-Q_{_f}s_{_{\rm w}}^2\Big)^2+Q_{_f}^2s_{_{\rm w}}^4\Big]
T_{9}(x_{_{\rm z}},x_{_{F_\alpha}},x_{_{F_\beta}})\Bigg\}\;.
%%%%%%%%%%%%%%%%%%%%%%%%%%%%%%%%%%%%%%%%%%%%%%%%%%%%%%%%%%%%%%%%%%%%%%
\label{HF-MDM(b+c)}
\end{eqnarray}

%%%%%%%%%%%%%%%%%%%%%%%%BEGIN NEW ADDING%%%%%%%%%%%%%%%%%%%%%%%%%%%
As a closed heavy fermion loop is inserted into neutral or charged
gauge boson propagator, the counter terms to self energy diagrams of
$Z$ or $W$ gauge bosons induce the renormalization for Weinberg angle
\begin{eqnarray}
%%%%%%%%%%%%%%%%%%%%%%%%%%%%%%%%%%%%%%%%%%%%%%%%%%%%%%%%%%%%%%%%%%%%%%%%%%%%%%%%%%%
&&{\delta s_{_{\rm w}}\over s_{_{\rm w}}}=-{c_{_{\rm w}}^2\over2s_{_{\rm w}}^2}
\Big[{\delta m_{_{\rm w}}^{2,os}\over m_{_{\rm w}}^2}
-{\delta m_{_{\rm z}}^{2,os}\over m_{_{\rm z}}^2}\Big]
\simeq{c_{_{\rm w}}^2\over2s_{_{\rm w}}^2}\Delta \rho_{_F}+\cdots\;,
%%%%%%%%%%%%%%%%%%%%%%%%%%%%%%%%%%%%%%%%%%%%%%%%%%%%%%%%%%%%%%%%%%%%%%%%%%%%%%%%%%%
\label{dsw}
\end{eqnarray}
where the dots indicate again nonleading contributions.
Furthermore, $\Delta \rho_{_F}$ denotes the 1-loop corrections from the heavy fermions
to the $\rho$ parameter, which appears in the ratio of weak neutral to charged
current amplitudes and comparisons of the $Z$ and $W^\pm$ masses.
Using the above equation, one can express the corresponding counter terms for
the vertices $Z\bar{l}l$ as
\begin{eqnarray}
%%%%%%%%%%%%%%%%%%%%%%%%%%%%%%%%%%%%%%%%%%%%%%%%%%%%%%%%%%%%%%%%%%%%%%%%%%%%%%%%%%%
&&i\delta C_{{\rm z}\overline{l}l}={ie\over s_{_{\rm w}}c_{_{\rm w}}^3}
\Big[{\delta s_{_{\rm w}}\over2s_{_{\rm w}}}\omega_-+s_{_{\rm w}}
\delta s_{_{\rm w}}\omega_+\Big]\;.
%%%%%%%%%%%%%%%%%%%%%%%%%%%%%%%%%%%%%%%%%%%%%%%%%%%%%%%%%%%%%%%%%%%%%%%%%%%%%%%%%%%
\label{z-mu-mu}
\end{eqnarray}
Inserting the counter terms into one loop standard model diagrams, we finally
obtain
\begin{eqnarray}
%%%%%%%%%%%%%%%%%%%%%%%%%%%%%%%%%%%%%%%%%%%%%%%%%%%%%%%%%%%%%%%%%%%%%%%%%%%%%%%%%%%
&&a_l^{\rho}=-{e^2m_{_l}^2\over3(4\pi)^2m_{_{\rm w}}^2s_{_{\rm w}}^2c_{_{\rm w}}^2}\Big(
1-2s_{_{\rm w}}^2+2s_{_{\rm w}}^4\Big){\delta s_{_{\rm w}}\over s_{_{\rm w}}}
\nonumber\\
&&\hspace{0.5cm}
=-{G_{_F}m_{_l}^2\over12\sqrt{2}\pi^2s_{_{\rm w}}^2}\Big(
1-2s_{_{\rm w}}^2+2s_{_{\rm w}}^4\Big)\Delta \rho_{_F}+\cdots\;.
%%%%%%%%%%%%%%%%%%%%%%%%%%%%%%%%%%%%%%%%%%%%%%%%%%%%%%%%%%%%%%%%%%%%%%%%%%%%%%%%%%%
\label{amu-sw}
\end{eqnarray}
In principle, we should consider the corrections from the counter terms for
$W^+\bar{\nu}_ll$:
\begin{eqnarray}
%%%%%%%%%%%%%%%%%%%%%%%%%%%%%%%%%%%%%%%%%%%%%%%%%%%%%%%%%%%%%%%%%%%%%%%%%%%%%%%%%%%
&&i\delta C_{{\rm w}^+\overline{\nu}_ll}=-{ie\over\sqrt{2}s_{_{\rm w}}^2}
\delta s_{_{\rm w}}\omega_-\;.
%%%%%%%%%%%%%%%%%%%%%%%%%%%%%%%%%%%%%%%%%%%%%%%%%%%%%%%%%%%%%%%%%%%%%%%%%%%%%%%%%%%
\label{w-mu-nu}
\end{eqnarray}
However, the corrections can be absorbed in the one-loop
result as we parametrize the final result in terms of $G_{_F}$ determined from
the muon's lifetime.
%%%%%%%%%%%%%%%%%%%%%%%%%END NEW ADDING%%%%%%%%%%%%%%%%%%%%%%%%%%%%

The contributions from the corresponding diagrams contain
the additional suppressed factor $m_{_l}^2/\Lambda_{_{\rm EW}}^2$
when both of virtual neutral gauge bosons in Fig.\ref{fig4}
are replaced with the neutral Goldstone $G^0$. Nevertheless, we should consider
the corrections from those two loop diagrams in which one of virtual
neutral gauge bosons is replaced with the neutral Goldstone $G^0$ since it
represents the longitudinal component of charged gauge boson
in nonlinear $R_\xi$ gauge. For many electroweak theories contain
the neutral CP-even and CP-odd Higgs, we also generalize the result
directly to the diagrams in which a closed heavy loop is attached on
the virtual neutral gauge boson $Z$ and neutral scalars $h^0,\;A^0$.

%%%%%%%%%%%%%%%%%%%%%%%%%%%%%%%%%%%%%%%%%%%%%%%%%%%%%%%%%%%%%%%%%%%
\begin{figure}[t]
\setlength{\unitlength}{1mm}
\begin{center}
\begin{picture}(0,30)(0,0)
\put(-62,-100){\includegraphics{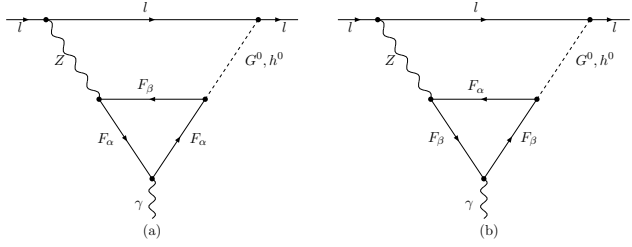}}
\end{picture}
\caption[]{The two-loop diagrams in which a closed heavy
fermion loop is attached to the virtual $Z,\;G^0$ or $h^0$ bosons.
In concrete calculation, the contributions from those mirror diagrams
should be included also.}
\label{fig6}
\end{center}
\end{figure}
%%%%%%%%%%%%%%%%%%%%%%%%%%%%%%%%%%%%%%%%%%%%%%%%%%%%%%%%%%%%%%%%%%%

\subsection{The corrections from the diagrams where a closed heavy
fermion loop is attached to the virtual $Z,\;G^0\;(h^0,\;A^0)$ bosons}
\indent\indent
Similarly, the renormalizable interaction among the electroweak neutral
Goldstone/Higgs $G^0\;(h^0,\;A^0)$ and the heavy fermions $F_{\alpha,\beta}$
can be expressed in a more universal form as
\begin{eqnarray}
&&{\cal L}_{_{S^0FF}}={e\over s_{_{\rm w}}}\Big[-iG^{0}\bar{F}_\alpha
({\cal G}^{\cal N}_{_{\alpha\beta}}\omega_-
-{\cal G}^{{\cal N}\dagger}_{_{\alpha\beta}}\omega_+)F_\beta
+h^0\bar{F}_\alpha({\cal H}^{\cal E}_{_{\alpha\beta}}\omega_-
+{\cal H}^{{\cal E}\dagger}_{_{\alpha\beta}}\omega_+)F_\beta
\nonumber\\
&&\hspace{1.6cm}
-iA^0\bar{F}_\alpha({\cal H}^{\cal O}_{_{\alpha\beta}}\omega_-
-{\cal H}^{{\cal O}\dagger}_{_{\alpha\beta}}\omega_+)F_\beta\Big]+h.c.\;,
\label{neutral-G-H-FF}
\end{eqnarray}
where the concrete expressions of ${\cal G}^{\cal N}_{_{\alpha\beta}},\;
{\cal H}^{\cal E}_{_{\alpha\beta}}$  and ${\cal H}^{\cal O}_{_{\alpha\beta}}$
depend on the models employed in our calculation, and the conservation of electric charge
requires $Q_\beta=Q_\alpha$. Generally, the couplings among the neutral Goldstone/Higgs
and leptons are written as
\begin{eqnarray}
&&{\cal L}_{_{S^0 ll}}={em_{_l}\over2m_{_{\rm w}}s_{_{\rm w}}}
\Big[-iG^0\bar{l}\gamma_5l+{\cal B}_{\cal E}h^0\bar{l}l
-i{\cal B}_{\cal O}A^0\bar{l}\gamma_5l\Big]\;,
\label{neutral-G-H-lepton}
\end{eqnarray}
and the parameters ${\cal B}_{\cal E},\;{\cal B}_{\cal O}$ depend
on the concrete models adopted in our analysis. In full theory,
the couplings in Eq.(\ref{neutral-G-H-FF}) induce the corrections
to lepton MDMs and EDMs from the diagrams in Fig.\ref{fig6}.
Accordingly, the corresponding diagram of effective theory
is presented in Fig.\ref{fig5}(b).
In principle, we should also consider the corrections from those
two loop diagrams in which one of virtual $Z$ gauge bosons is replaced
with the neutral Goldstone/Higgs $G^0\;(h^0,\;A^0)$ in Fig.\ref{fig4}(a). However, the sum
of amplitude from those two loop diagrams does not contribute
to the lepton MDMs and EDMs when one of virtual
neutral gauge bosons in Fig.\ref{fig4}(a) is replaced with the neutral
Goldstone/Higgs $G^0\;(h^0,\;A^0)$\cite{Feng4}.

After the steps taken in $WW$ sector, we formulate
the corresponding corrections  to the lepton MDMs and EDMs
from the diagrams in Fig.\ref{fig6} as
\begin{eqnarray}
%%%%%%%%%%%%%%%%%%%%%%%%%%%%%%%%%%%%%%%%%%%%%%%%%%%%%%%%%%%%%%%%%%%%%%
&&a_l^{Zh^0}=-{G_{_F}\alpha_{_e}m_{_l}^2{\cal B}_{\cal E}\over64\pi^3s_{_{\rm w}}^2c_{_{\rm w}}^2}
(T_{_f}^Z-2Q_{_f}s_{_{\rm w}}^2)(x_{_{\rm w}}x_{_{F_\beta}})^{1/2}
\Big[2(2+\ln x_{_{F_\beta}})\varrho_{_{0,1}}(x_{_{\rm z}},x_{_h})
\nonumber\\
&&\hspace{1.4cm}
+F_5(x_{_{\rm z}},x_{_h},x_{_{F_\alpha}},x_{_{F_\beta}})\Big]
\Re\Big({\cal H}^{\cal E}\xi^L_{_{\alpha\beta}}
+{\cal H}^{{\cal E}\dagger}_{_{\beta\alpha}}\xi^R_{_{\alpha\beta}}\Big)
\;,\nonumber\\
%---------------------------------------------------------------------
&&d_l^{Zh^0}={G_{_F}\alpha_{_e}em_{_l}{\cal B}_{\cal E}\over128\pi^3s_{_{\rm w}}^2c_{_{\rm w}}^2}
(T_{_f}^Z-2Q_{_f}s_{_{\rm w}}^2)(x_{_{\rm w}}x_{_{F_\beta}})^{1/2}
\Big[2(\ln x_{_{F_\alpha}}-\ln x_{_{F_\beta}})
\varrho_{_{0,1}}(x_{_{\rm z}},x_{_h})
\nonumber\\
&&\hspace{1.4cm}
-F_5(x_{_{\rm z}},x_{_h},x_{_{F_\alpha}},x_{_{F_\beta}})
-F_6(x_{_{\rm z}},x_{_h},x_{_{F_\beta}},x_{_{F_\alpha}})\Big]
\Im\Big({\cal H}^{\cal E}_{_{\beta\alpha}}\xi^L_{_{\alpha\beta}}
-{\cal H}^{{\cal E}\dagger}_{_{\beta\alpha}}\xi^R_{_{\alpha\beta}}\Big)\;,
\nonumber\\
%%%%%%%%%%%%%%%%%%%%%%%%%%%%%%%%%%%%%%%%%%%%%%%%%%%%%%%%%%%%%%%%%%%%%%
&&a_l^{ZA^0}=-{G_{_F}\alpha_{_e}m_{_l}^2{\cal B}_{\cal O}\over64\pi^3s_{_{\rm w}}^2c_{_{\rm w}}^2}
(T_{_f}^Z-2Q_{_f}s_{_{\rm w}}^2)(x_{_{\rm w}}x_{_{F_\beta}})^{1/2}
\Big[-2(\ln x_{_{F_\alpha}}-\ln x_{_{F_\beta}})
\varrho_{_{0,1}}(x_{_{\rm z}},x_{_A})
\nonumber\\
&&\hspace{1.4cm}
+F_5(x_{_{\rm z}},x_{_A},x_{_{F_\alpha}},x_{_{F_\beta}})
+F_6(x_{_{\rm z}},x_{_A},x_{_{F_\beta}},x_{_{F_\alpha}})\Big]
\Re\Big({\cal H}^{\cal O}_{_{\beta\alpha}}\xi^L_{_{\alpha\beta}}
-{\cal H}^{{\cal O}\dagger}_{_{\beta\alpha}}\xi^R_{_{\alpha\beta}}\Big)\;,
\nonumber\\
%---------------------------------------------------------------------
&&d_l^{ZA^0}={G_{_F}\alpha_{_e}em_{_l}{\cal B}_{\cal O}\over128\pi^3s_{_{\rm w}}^2c_{_{\rm w}}^2}
(T_{_f}^Z-2Q_{_f}s_{_{\rm w}}^2)(x_{_{\rm w}}x_{_{F_\beta}})^{1/2}
\Big[2(2+\ln x_{_{F_\beta}})\varrho_{_{0,1}}(x_{_{\rm z}},x_{_A})
\nonumber\\
&&\hspace{1.2cm}
+F_5(x_{_{\rm z}},x_{_A},x_{_{F_\alpha}},x_{_{F_\beta}})\Big]
\Im\Big({\cal H}^{\cal O}_{_{\beta\alpha}}\xi^L_{_{\alpha\beta}}
+{\cal H}^{{\cal O}\dagger}_{_{\beta\alpha}}\xi^R_{_{\alpha\beta}}\Big)
\;,\nonumber\\
%%%%%%%%%%%%%%%%%%%%%%%%%%%%%%%%%%%%%%%%%%%%%%%%%%%%%%%%%%%%%%%%%%%%%%
&&a_l^{ZG}=-{G_{_F}\alpha_{_e}m_{_l}^2\over64\pi^3s_{_{\rm w}}^2c_{_{\rm w}}^2}
(T_{_f}^Z-2Q_{_f}s_{_{\rm w}}^2)(x_{_{\rm w}}x_{_{F_\beta}})^{1/2}
\Big[-{2\over  x_{_{\rm z}}}(\ln x_{_{F_\alpha}}-\ln x_{_{F_\beta}})
\nonumber\\
&&\hspace{1.2cm}
+F_5(x_{_{\rm z}},x_{_{\rm z}},x_{_{F_\alpha}},x_{_{F_\beta}})
+F_6(x_{_{\rm z}},x_{_{\rm z}},x_{_{F_\beta}},x_{_{F_\alpha}})\Big]
\Re\Big({\cal G}^{\cal N}_{_{\beta\alpha}}\xi^L_{_{\alpha\beta}}
-{\cal G}^{{\cal N}\dagger}_{_{\beta\alpha}}\xi^R_{_{\alpha\beta}}\Big)\;,
\nonumber\\
&&d_l^{ZG}={G_{_F}\alpha_{_e}em_{_l}\over128\pi^3s_{_{\rm w}}^2c_{_{\rm w}}^2}
(T_{_f}^Z-2Q_{_f}s_{_{\rm w}}^2)(x_{_{\rm w}}x_{_{F_\beta}})^{1/2}
\Big[{2\over x_{_{\rm z}}}(2+\ln x_{_{F_\beta}})
\nonumber\\
&&\hspace{1.2cm}
+F_5(x_{_{\rm z}}, x_{_{\rm z}},x_{_{F_\alpha}},x_{_{F_\beta}})\Big]
\Im\Big({\cal G}^{\cal N}_{_{\beta\alpha}}\xi^L_{_{\alpha\beta}}
+{\cal G}^{{\cal N}\dagger}_{_{\beta\alpha}}\xi^R_{_{\alpha\beta}}\Big)\;.
%%%%%%%%%%%%%%%%%%%%%%%%%%%%%%%%%%%%%%%%%%%%%%%%%%%%%%%%%%%%%%%%%%%%%%
\label{MED-Z-HAG}
\end{eqnarray}
The expressions of form factors $F_i(x,y,z,u)\;(i=5,\;6)$ can
be found in appendix. In the heavy mass limit $m_{_F}=m_{_{F_\alpha}}=m_{_{F_\beta}}
\gg m_{_h},\;m_{_A},\;m_{_{\rm w}}$, we have
\begin{eqnarray}
%%%%%%%%%%%%%%%%%%%%%%%%%%%%%%%%%%%%%%%%%%%%%%%%%%%%%%%%%%%%%%%%%%%%%%
&&a_l^{Zh^0}=-{G_{_F}\alpha_{_e}m_{_l}^2m_{_{\rm w}}{\cal B}_{\cal E}
\over64\pi^3s_{_{\rm w}}^2c_{_{\rm w}}^2m_{_F}}
\Big[\varrho_{_{1,1}}(m_{_{\rm z}}^2,m_{_h}^2)-\ln m_{_F}^2-1\Big]
\nonumber\\
&&\hspace{1.4cm}\times
(T_{_f}^Z-2Q_{_f}s_{_{\rm w}}^2)
\Re\Big({\cal H}^{\cal E}_{_{\beta\alpha}}\xi^L_{_{\alpha\beta}}
+{\cal H}^{{\cal E}\dagger}_{_{\beta\alpha}}\xi^R_{_{\alpha\beta}}\Big)\;,
\nonumber\\
%%%%%%%%%%%%%%%%%%%%%%%%%%%%%%%%%%%%%%%%%%%%%%%%%%%%%%%%%%%%%%%%%%%%%%
&&a_l^{ZA^0}=-{G_{_F}\alpha_{_e}m_{_l}^2m_{_{\rm w}}{\cal B}_{\cal O}
\over32\pi^3s_{_{\rm w}}^2c_{_{\rm w}}^2m_{_F}}\Big[\varrho_{_{1,1}}(m_{_{\rm z}}^2,m_{_A}^2)
-\ln m_{_F}^2-1\Big]
\nonumber\\
&&\hspace{1.4cm}\times
(T_{_f}^Z-2Q_{_f}s_{_{\rm w}}^2)
\Re\Big({\cal H}^{\cal O}_{_{\beta\alpha}}\xi^L_{_{\alpha\beta}}
-{\cal H}^{{\cal O}\dagger}_{_{\beta\alpha}}\xi^R_{_{\alpha\beta}}\Big)\;,
\nonumber\\
%%%%%%%%%%%%%%%%%%%%%%%%%%%%%%%%%%%%%%%%%%%%%%%%%%%%%%%%%%%%%%%%%%%%%%
&&a_l^{ZG}={G_{_F}\alpha_{_e}m_{_l}^2m_{_{\rm w}}
\over32\pi^3s_{_{\rm w}}^2c_{_{\rm w}}^2m_{_F}}\ln\Big({m_{_F}^2\over m_{_{\rm z}}^2}\Big)
(T_{_f}^Z-2Q_{_f}s_{_{\rm w}}^2)
\Re\Big({\cal G}^{\cal N}_{_{\beta\alpha}}\xi^L_{_{\alpha\beta}}
-{\cal G}^{{\cal N}\dagger}_{_{\beta\alpha}}\xi^R_{_{\alpha\beta}}\Big)\;,
\nonumber\\
%---------------------------------------------------------------------
&&d_l^{Zh^0}=-{G_{_F}\alpha_{_e}em_{_l}m_{_{\rm w}}{\cal B}_{\cal E}
\over64\pi^3s_{_{\rm w}}^2c_{_{\rm w}}^2m_{_F}}
\Big[\varrho_{_{1,1}}(m_{_{\rm z}}^2,m_{_h}^2)-\ln m_{_F}^2-1\Big]
\nonumber\\
&&\hspace{1.4cm}\times
(T_{_f}^Z-2Q_{_f}s_{_{\rm w}}^2)
\Im\Big({\cal H}^{\cal E}_{_{\beta\alpha}}\xi^L_{_{\alpha\beta}}
-{\cal H}^{{\cal E}\dagger}_{_{\beta\alpha}}\xi^R_{_{\alpha\beta}}\Big)
\;,\nonumber\\
%%%%%%%%%%%%%%%%%%%%%%%%%%%%%%%%%%%%%%%%%%%%%%%%%%%%%%%%%%%%%%%%%%%%%%
&&d_l^{ZA^0}={G_{_F}\alpha_{_e}em_{_l}m_{_{\rm w}}{\cal B}_{\cal O}
\over128\pi^3s_{_{\rm w}}^2c_{_{\rm w}}^2m_{_F}}
\Big[\varrho_{_{1,1}}(m_{_{\rm z}}^2,m_{_A}^2)-\ln m_{_F}^2-1\Big]
\nonumber\\
&&\hspace{1.4cm}\times
(T_{_f}^Z-2Q_{_f}s_{_{\rm w}}^2)
\Im\Big({\cal H}^{\cal O}_{_{\beta\alpha}}\xi^L_{_{\alpha\beta}}
+{\cal H}^{{\cal O}\dagger}_{_{\beta\alpha}}\xi^R_{_{\alpha\beta}}\Big)
\;,\nonumber\\
%%%%%%%%%%%%%%%%%%%%%%%%%%%%%%%%%%%%%%%%%%%%%%%%%%%%%%%%%%%%%%%%%%%%%%
&&d_l^{ZG}={G_{_F}\alpha_{_e}em_{_l}m_{_{\rm w}}
\over128\pi^3s_{_{\rm w}}^2c_{_{\rm w}}^2m_{_F}}
\ln\Big({m_{_F}^2\over m_{_{\rm z}}^2}\Big)(T_{_f}^Z-2Q_{_f}s_{_{\rm w}}^2)
\Im\Big({\cal G}^{\cal N}_{_{\beta\alpha}}\xi^L_{_{\alpha\beta}}
+{\cal G}^{{\cal N}\dagger}_{_{\beta\alpha}}\xi^R_{_{\alpha\beta}}\Big)\;.
%%%%%%%%%%%%%%%%%%%%%%%%%%%%%%%%%%%%%%%%%%%%%%%%%%%%%%%%%%%%%%%%%%%%%%
\label{ASYMED-Z-H}
\end{eqnarray}

\subsection{The corrections from the diagrams where a closed heavy
fermion loop is attached to the virtual $\gamma$ and $Z$ bosons}
\begin{figure}[h]
\setlength{\unitlength}{1mm}
\begin{center}
\begin{picture}(0,40)(0,0)
\put(-62,-100){\includegraphics{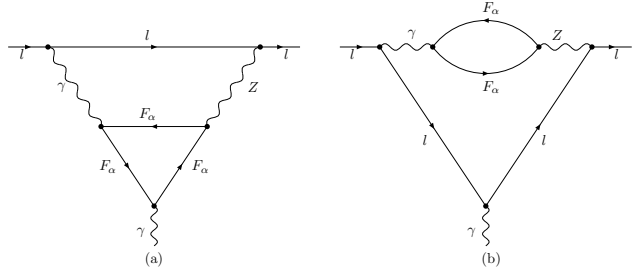}}
\end{picture}
\caption[]{The two-loop diagrams in which a closed heavy
fermion loop is attached to the virtual $\gamma$ and $Z$ bosons.
In concrete calculation, the contributions from those mirror diagrams
should be included also.}
\label{fig7}
\end{center}
\end{figure}

\begin{figure}[h]
\setlength{\unitlength}{1mm}
\begin{center}
\begin{picture}(0,30)(0,0)
\put(-60,-80){\includegraphics{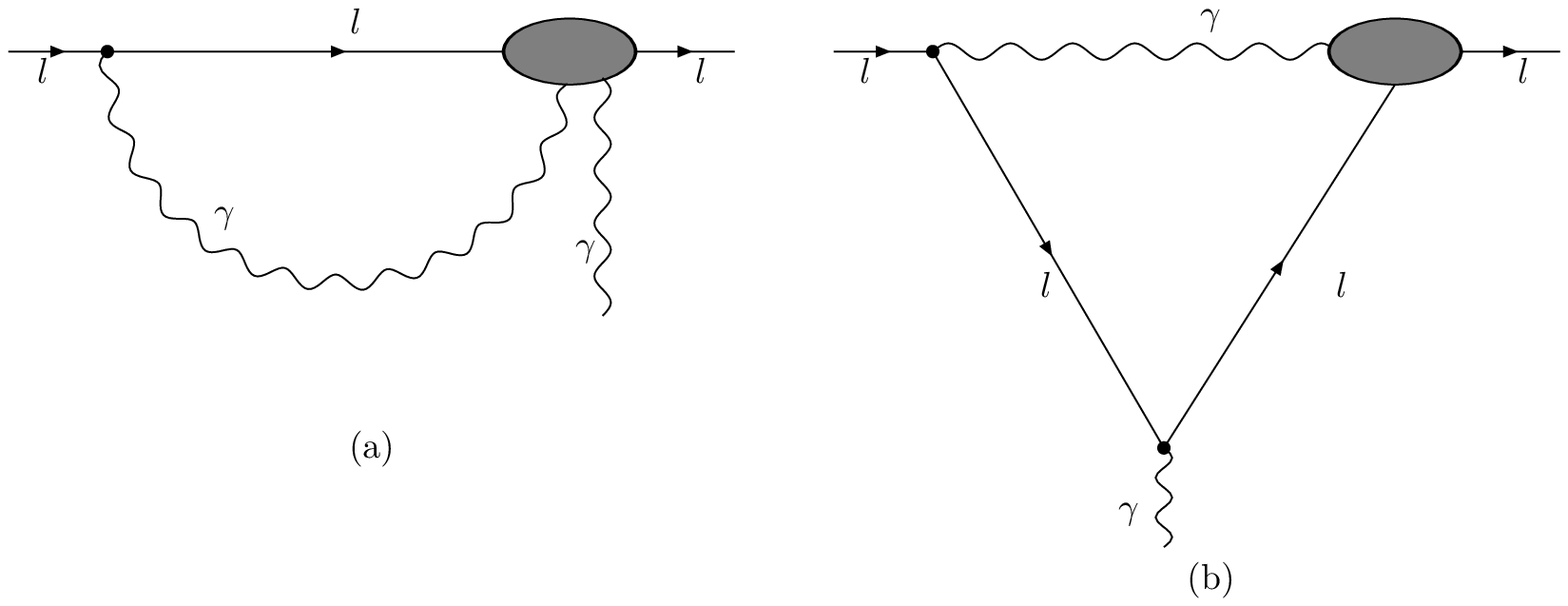}}
\end{picture}
\caption[]{The diagrams of effective theory corresponds to those
diagrams in Fig.\ref{fig7}. Where the diagram (a) corresponds
to Fig.\ref{fig7}(a) in full theory, and the diagram (b) corresponds
to Fig.\ref{fig7}(b) in full theory respectively.}
\label{fig8}
\end{center}
\end{figure}

\indent\indent
When a closed fermion loop is attached to the virtual $\gamma$ and $Z$
gauge bosons (Fig.\ref{fig7}), the corresponding diagrams of effective
theory are presented in Fig.\ref{fig8}. Taking the steps above,
one can get the tedious correction to the effective Lagrangian. If we ignore the terms
which are proportional to the suppression factor $1-4s_{_{\rm w}}^2$, the correction
from this sector to the lepton MDMs from this sector is drastically simplified as
%%%%%%%%%%%%%%%%%%%%%%%%%%%%%%%%%%%%%%%%%%%%%%%%%%%%%%%%%%%%%%%%%%%
%%%%%%%%%%%%%%%%%%%%%%%%%%%%%%%%%%%%%%%%%%%%%%%%%%%%%%%%%%%%%%%%%%%

%%%%%%%%%%%%%%%%%%%%%%%%%%%%%%%%%%%%%%%%%%%%%%%%%%%%%%%%%%%%%%%%%%%
%%%%%%%%%%%%%%%%%%%%%%%%%%%%%%%%%%%%%%%%%%%%%%%%%%%%%%%%%%%%%%%%%%%

\begin{eqnarray}
%%%%%%%%%%%%%%%%%%%%%%%%%%%%%%%%%%%%%%%%%%%%%%%%%%%%%%%%%%%%%%%%%%%%%%
&&a_l^{\gamma Z}={G_{_F}\alpha_{_e}Q_{_f}Q_\alpha m_{_l}^2\over32\sqrt{2}
\pi^3c_{_{\rm w}}^2}\Big(\xi^L_{_{\alpha\alpha}}-\xi^R_{_{\alpha\alpha}}\Big)
x_{_{\rm w}}T_7(x_{_{\rm z}},x_{_{F_\alpha}},x_{_{F_\alpha}})\;,
%%%%%%%%%%%%%%%%%%%%%%%%%%%%%%%%%%%%%%%%%%%%%%%%%%%%%%%%%%%%%%%%%%%%%%
\label{MD-gamma-z}
\end{eqnarray}
and the correction to the lepton EDMs is zero. In the limit
$m_{_F}=m_{_{F_\alpha}}\gg m_{_{\rm z}}$, we approximate the correction
to the lepton MDMs from this sector as
\begin{eqnarray}
%%%%%%%%%%%%%%%%%%%%%%%%%%%%%%%%%%%%%%%%%%%%%%%%%%%%%%%%%%%%%%%%%%%%%%
&&a_l^{\gamma Z}={G_{_F}\alpha_{_e}Q_{_f}Q_\alpha m_{_l}^2m_{_{\rm w}}^2\over32\sqrt{2}
\pi^3c_{_{\rm w}}^2m_{_F}^2}
\Big(\xi^L_{_{\alpha\alpha}}-\xi^R_{_{\alpha\alpha}}\Big)\Big[
35+\ln{m_{_F}^2\over m_{_{\rm z}}^2}\Big]\;.
%%%%%%%%%%%%%%%%%%%%%%%%%%%%%%%%%%%%%%%%%%%%%%%%%%%%%%%%%%%%%%%%%%%%%%
\label{MD-gamma-z1}
\end{eqnarray}

\subsection{The corrections from the diagrams where a closed heavy
fermion loop is attached to the virtual $\gamma,\;G^0\;(h^0,\;A^0)$ bosons}
\indent\indent
As a closed fermion loop is attached to the virtual neutral Higgs and photon
fields, a real photon can be emitted from either the virtual lepton or the virtual charginos
in the self energy diagram. When a real photon is emitted
from the virtual charginos, the corresponding "triangle" diagrams
belong to the typical two-loop Bar-Zee-type diagrams \cite{Barr-Zee}.
Within the framework of CP violating MSSM,
the contributions from two-loop Bar-Zee-type diagrams to the EDMs of those light fermions
are discussed extensively in literature \cite{Pilaftsis}.
When a real photon is attached to the internal standard fermion,
the correction from corresponding triangle diagram to
the effective Lagrangian is zero because of the Furry theorem,
this point is also verified through a strict analysis.
Replacing the virtual neutral gauge boson with photon in Fig.\ref{fig6},
one obtains the relevant diagrams in full theory. Meanwhile, the
diagram of effective theory is same as that presented in Fig.\ref{fig8}(a).
After the steps adopted above, the corresponding corrections to lepton
MDMs and EDMs from this sector are expressed as
\begin{eqnarray}
%%%%%%%%%%%%%%%%%%%%%%%%%%%%%%%%%%%%%%%%%%%%%%%%%%%%%%%%%%%%%%%%%%%%%%
&&a_l^{\gamma h^0}={G_{_F}\alpha_{_e}Q_{_f}Q_\alpha m_{_l}^2{\cal B}_{\cal E}
\over8\pi^3}\Re({\cal H}^{\cal E}_{_{\alpha\alpha}})(x_{_{\rm w}}x_{_{F_\alpha}}
)^{1/2}T_{11}(x_{_{h}},x_{_{F_\alpha}},x_{_{F_\alpha}})\;,
\nonumber\\
%---------------------------------------------------------------------
&&a_l^{\gamma A^0}=
-{G_{_F}\alpha_{_e}Q_{_f}Q_\alpha m_{_l}^2{\cal B}_{\cal O}
\over8\pi^3}\Re({\cal H}^{\cal O}_{_{\alpha\alpha}})
(x_{_{\rm w}}x_{_{F_\alpha}})^{1/2}T_{12}(x_{_{A}},x_{_{F_\alpha}},x_{_{F_\alpha}})
\;,\nonumber\\
%---------------------------------------------------------------------
&&d_l^{\gamma h^0}=-{G_{_F}\alpha_{_e}eQ_{_f}Q_\alpha m_{_l}{\cal B}_{\cal E}
\over16\pi^3}\Im({\cal H}^{\cal E}_{_{\alpha\alpha}})
(x_{_{\rm w}}x_{_{F_\alpha}})^{1/2}T_{12}(x_{_{h}},x_{_{F_\alpha}},x_{_{F_\alpha}})\;,
\nonumber\\
%---------------------------------------------------------------------
&&d_l^{\gamma A^0}=-{G_{_F}\alpha_{_e}eQ_{_f}Q_\alpha m_{_l}{\cal B}_{\cal O}
\over16\pi^3}\Im({\cal H}^{\cal O}_{_{\alpha\alpha}})
(x_{_{\rm w}}x_{_{F_\alpha}})^{1/2}T_{11}(x_{_{A}},x_{_{F_\alpha}},x_{_{F_\alpha}})\;.
%%%%%%%%%%%%%%%%%%%%%%%%%%%%%%%%%%%%%%%%%%%%%%%%%%%%%%%%%%%%%%%%%%%%%%
\label{MD-gamma-h}
\end{eqnarray}
Note here that the corrections from this sector to the MDM of lepton
depend on real parts of the effective couplings
${\cal H}^{\cal E}_{_{\alpha\alpha}}$ and  ${\cal H}^{\cal O}_{_{\alpha\alpha}}$,
and the corrections from this sector to the EDM of lepton depend on
imaginary parts of the effective couplings
${\cal H}^{\cal E}_{_{\alpha\alpha}}$ and ${\cal H}^{\cal O}_{_{\alpha\alpha}}$.
In the limit $m_{_F}=m_{_{F_\alpha}}\gg m_{_{h}},\;m_{_{A}}$, the above expressions are
simplified as
\begin{eqnarray}
%%%%%%%%%%%%%%%%%%%%%%%%%%%%%%%%%%%%%%%%%%%%%%%%%%%%%%%%%%%%%%%%%%%%%%
&&a_l^{\gamma h^0}={G_{_F}\alpha_{_e}Q_{_f}Q_\alpha m_{_l}^2m_{_{\rm w}}{\cal B}_{\cal E}
\over8\pi^3m_{_F}}\Re({\cal H}^{\cal E}_{_{\alpha\alpha}})
\Big[1+\ln{m_{_F}^2\over m_{_h}^2}\Big]\;,
\nonumber\\
%---------------------------------------------------------------------
&&a_l^{\gamma A^0}=
{G_{_F}\alpha_{_e}Q_{_f}Q_\alpha m_{_l}^2m_{_{\rm w}}{\cal B}_{\cal O}
\over8\pi^3m_{_F}}\Re({\cal H}^{\cal O}_{_{\alpha\alpha}})
\Big[1+\ln{m_{_F}^2\over m_{_A}^2}\Big]
\;,\nonumber\\
%---------------------------------------------------------------------
&&d_l^{\gamma h^0}={G_{_F}\alpha_{_e}eQ_{_f}Q_\alpha m_{_l}m_{_{\rm w}}{\cal B}_{\cal E}
\over16\pi^3m_{_F}}\Im({\cal H}^{\cal E}_{_{\alpha\alpha}})
\Big[1+\ln{m_{_F}^2\over m_{_h}^2}\Big]\;,
\nonumber\\
%%---------------------------------------------------------------------
&&d_l^{\gamma A^0}=-{G_{_F}\alpha_{_e}eQ_{_f}Q_\alpha m_{_l}m_{_{\rm w}}{\cal B}_{\cal O}
\over16\pi^3m_{_F}}\Im({\cal H}^{\cal O}_{_{\alpha\alpha}})
\Big[1+\ln{m_{_F}^2\over m_{_A}^2}\Big]\;.
%%%%%%%%%%%%%%%%%%%%%%%%%%%%%%%%%%%%%%%%%%%%%%%%%%%%%%%%%%%%%%%%%%%%%%
\label{MD-gamma-h1}
\end{eqnarray}

Similarly, the corrections to the lepton MDMs and EDMs from the $\gamma G$
sector are:
\begin{eqnarray}
%%%%%%%%%%%%%%%%%%%%%%%%%%%%%%%%%%%%%%%%%%%%%%%%%%%%%%%%%%%%%%%%%%%%%%
&&a_l^{\gamma G}={G_{_F}\alpha_{_e}Q_{_f}Q_\alpha m_{_l}^2\over8\pi^3}
\Re({\cal G}^{{\cal N}}_{_{\alpha\alpha}})
(x_{_{\rm w}}x_{_{F_\alpha}})^{1/2}
T_{12}(x_{_{\rm z}},x_{_{F_\alpha}},x_{_{F_\alpha}})\;,
\nonumber\\
&&d_l^{\gamma G}={G_{_F}\alpha_{_e}eQ_{_f}Q_\alpha m_{_l}\over16\pi^3}
\Im({\cal G}^{{\cal N}}_{_{\alpha\alpha}})
(x_{_{\rm w}}x_{_{F_\alpha}})^{1/2}
T_{11}(x_{_{\rm z}},x_{_{F_\alpha}},x_{_{F_\alpha}})\;.
%%%%%%%%%%%%%%%%%%%%%%%%%%%%%%%%%%%%%%%%%%%%%%%%%%%%%%%%%%%%%%%%%%%%%%
\label{MD-gamma-G}
\end{eqnarray}
The corrections from this sector to the MDM of lepton are proportional
to real parts of the effective couplings ${\cal G}^{{\cal N}}_{_{\alpha\alpha}}$, and the corrections
from this sector to  the EDM of lepton are proportional to imaginary parts
of the effective couplings ${\cal G}^{{\cal N}}_{_{\alpha\alpha}}$, separately.
In the limit $m_{_F}=m_{_{F_\alpha}}\gg m_{_{\rm z}}$, we have
\begin{eqnarray}
%%%%%%%%%%%%%%%%%%%%%%%%%%%%%%%%%%%%%%%%%%%%%%%%%%%%%%%%%%%%%%%%%%%%%%
&&a_l^{\gamma G}={G_{_F}\alpha_{_e}Q_{_f}Q_\alpha m_{_l}^2m_{_{\rm w}}
\over8\pi^3m_{_F}}\Re({\cal G}^{{\cal N}}_{_{\alpha\alpha}})
\Big[1+\ln{m_{_F}^2\over m_{_{\rm z}}^2}\Big]\;,
\nonumber\\
&&d_l^{\gamma G}={G_{_F}\alpha_{_e}eQ_{_f}Q_\alpha m_{_l}m_{_{\rm w}}
\over16\pi^3m_{_F}}\Im({\cal G}^{{\cal N}}_{_{\alpha\alpha}})
\Big[1+\ln{m_{_F}^2\over m_{_{\rm z}}^2}\Big]\;,
%%%%%%%%%%%%%%%%%%%%%%%%%%%%%%%%%%%%%%%%%%%%%%%%%%%%%%%%%%%%%%%%%%%%%%
\label{MD-gamma-G1}
\end{eqnarray}
which are suppressed by the masses of heavy fermions.

\subsection{The corrections from the diagrams where a closed heavy
fermion loop is attached to the virtual $\gamma$ bosons}
\begin{figure}[h]
\setlength{\unitlength}{1mm}
\begin{center}
\begin{picture}(0,70)(0,0)
\put(-62,-70){\includegraphics{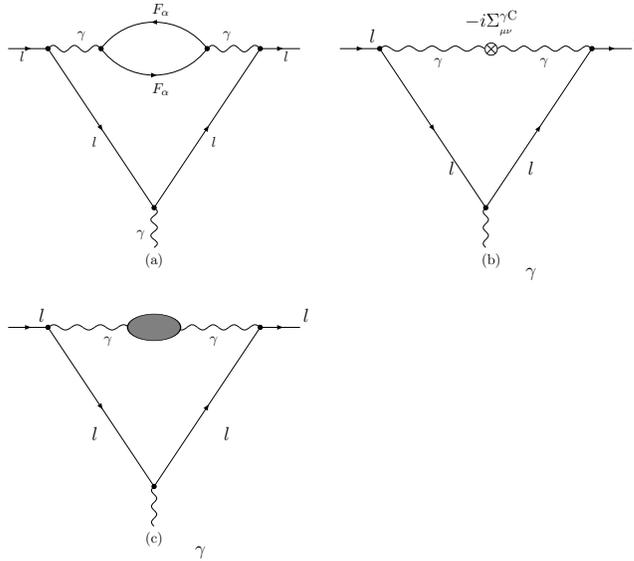}}
\end{picture}
\caption[]{The two-loop diagrams in which a closed heavy
fermion loop is attached to the virtual $\gamma$ bosons,
where subdiagram (a) represents the two loop diagram in full
theory, (b) represents the corresponding counter diagram,
and (c) is the corresponding diagram of effective theory.}
\label{fig9}
\end{center}
\end{figure}

\indent\indent
At the two loop level, there are QED diagrams involving a photon vacuum
polarization subdiagrams. For leptons or quarks, these contributions are
of course known\cite{Rafael1}. If heavy fermions contribute, the two loop
QED contributions are modified by the photon vacuum polarization (Fig.\ref{fig9}).
Adopting the same steps in $WW$ sector, we formulate the corrections from
Fig.\ref{fig9} to lepton MDM and EDM respectively as:
\begin{eqnarray}
%%%%%%%%%%%%%%%%%%%%%%%%%%%%%%%%%%%%%%%%%%%%%%%%%%%%%%%%%%%%%%%%%%%%%%
&&a_l^{\gamma\gamma}={\sqrt{2}G_{_F}\alpha_{_e}Q_\alpha^2m_{_l}^2\over45\pi^3}
{m_{_{\rm w}}^2\over m_{_{F_\alpha}}^2}
\;,\nonumber\\
&&d_l^{\gamma\gamma}=0\;,
%%%%%%%%%%%%%%%%%%%%%%%%%%%%%%%%%%%%%%%%%%%%%%%%%%%%%%%%%%%%%%%%%%%%%%
\label{gamma-vacuum}
\end{eqnarray}
which coincide with  the well known results in Ref.\cite{gamma-polarization}.
For masses $m_{_{F_\alpha}}\ge100{\rm GeV}$, these corrections from
this sector to the muon MDM $a_\mu$ are below $10^{-13}$ and hence negligible.
%%%%%%%%%%%%%%%%%%%%%%%%%%%%%%%%%%%%%%%%%%%%%%%%%%%%%%%%%%%%%%%%%%%
%%%%%%%%%%%%%%%%%%%%%%%%%%%%%%%%%%%%%%%%%%%%%%%%%%%%%%%%%%%%%%%%%%%

\section{The leading terms in two loop corrections to $a_l$ in concrete
electroweak theories\label{sec3}}
\indent\indent
In this section, we will present the leading terms from two loop corrections to
$a_l$ within some electroweak models discussed extensively in literature.

\subsection{the SM}
\indent\indent
Within the framework of SM, only top quark belongs to the fermion which
mass is heavier than that of weak gauge bosons. The couplings in Eq.(\ref{WFF})
and Eq.(\ref{ZFF}) are respectively expressed as
\begin{eqnarray}
%%%%%%%%%%%%%%%%%%%%%%%%%%%%%%%%%%%%%%%%%%%%%%%%%%%%%%%%%%%%%%%%%%%%%%
&&\xi_{_{tt}}^L=-1+{4\over3}s_{_{\rm w}}^2\;,
\;\;\xi_{_{tt}}^R=-{4\over3}s_{_{\rm w}}^2\;,
\nonumber\\
&&\zeta_{_{td_i}}^L=-{V_{_{td_i}}\over\sqrt{2}}\;,
\;\;\zeta_{_{td_i}}^R=0\;.
%%%%%%%%%%%%%%%%%%%%%%%%%%%%%%%%%%%%%%%%%%%%%%%%%%%%%%%%%%%%%%%%%%%%%%
\label{SM-couplings}
\end{eqnarray}
Using the unitary property of Cabibbo-Kabayashi-Maskawa (CKM) matrix
$V$ and $m_{_{\rm z}}=m_{_{\rm w}}/c_{_{\rm w}}$, one
formulates the leading corrections from top quark to lepton MDMs as
\begin{eqnarray}
%%%%%%%%%%%%%%%%%%%%%%%%%%%%%%%%%%%%%%%%%%%%%%%%%%%%%%%%%%%%%%%%%%%%%%
&&a_{2L}^{SM}=
{G_{_F}\alpha_{_e}m_{_l}^2\over8\sqrt{2}\pi^3s_{_{\rm w}}^2}
\Bigg\{3-{104\over9}s_{_{\rm w}}^2
-{16\over9}s_{_{\rm w}}^2\ln{m_{_t}^2\over m_{_h}^2}\Bigg\}
\nonumber\\
&&\hspace{1.3cm}
-{G_{_F}m_{_l}^2\over12\sqrt{2}\pi^2s_{_{\rm w}}^2}\Big(
1-2s_{_{\rm w}}^2+2s_{_{\rm w}}^4\Big)
\Delta \rho_{_{\rm SM}}\;.
%%%%%%%%%%%%%%%%%%%%%%%%%%%%%%%%%%%%%%%%%%%%%%%%%%%%%%%%%%%%%%%%%%%%%%
\label{2L-SM}
\end{eqnarray}
In order to produce the terms $\propto m_{_t}^2$, we take the limit
$s_{_{\rm w}}^2\rightarrow1/4$ used in Ref\cite{czarnecki}, and approximate
the last term as $-5G_{_F}m_{_l}^2\Delta \rho_{_{\rm SM}}/24\sqrt{2}\pi^2.$
Using the leading contributions $\Delta\rho_{_{\rm SM}}=3e^2m_{_t}^2/4(4\pi)^2
s_{_{\rm w}}^2m_{_{\rm w}}^2+\cdots$ in the limit $m_{_t}\gg m_{_{\rm w}}$,
we recover the terms $\propto m_{_t}^2$ in Ref\cite{czarnecki} perfectly.

\subsection{the Extension of SM with the Fourth Generation}
\indent\indent
Besides top quark, the extension of SM with the fourth generation
also includes additional three heavy fermions: $t',\;b',\;\tau'$. Correspondingly,
the couplings in Eq.(\ref{WFF}) and Eq.(\ref{ZFF}) are separately written as
\begin{eqnarray}
%%%%%%%%%%%%%%%%%%%%%%%%%%%%%%%%%%%%%%%%%%%%%%%%%%%%%%%%%%%%%%%%%%%%%%
&&\xi_{_{t't'}}^L=-1+{4\over3}s_{_{\rm w}}^2\;,
\;\;\xi_{_{t't'}}^R=-{4\over3}s_{_{\rm w}}^2\;,
\nonumber\\
&&\xi_{_{b'b'}}^L=-1+{2\over3}s_{_{\rm w}}^2\;,
\;\;\xi_{_{b'b'}}^R={2\over3}s_{_{\rm w}}^2\;,
\nonumber\\
&&\xi_{_{\tau'\tau'}}^L=1-2s_{_{\rm w}}^2\;,
\;\;\xi_{_{\tau'\tau'}}^R=-2s_{_{\rm w}}^2\;,
\nonumber\\
&&\zeta_{_{t'b'}}^L=-{V_{_{t'b'}}\over\sqrt{2}}\;,
\;\;\zeta_{_{t'b'}}^R=0\;,
\nonumber\\
&&\zeta_{_{\nu_{\tau'}\tau'}}^L=-{1\over\sqrt{2}}\;,
\;\;\zeta_{_{\nu_{\tau'}\tau'}}^R=0\;.
%%%%%%%%%%%%%%%%%%%%%%%%%%%%%%%%%%%%%%%%%%%%%%%%%%%%%%%%%%%%%%%%%%%%%%
\label{4G-couplings}
\end{eqnarray}
Assuming $m_{_{t'}}=m_{_{b'}}=m_{_{\tau'}}=m_{_F}\gg m_{_{\rm w}}$ and applying
the unitary property of $4\times4$ CKM matrix $V$, we finally obtain
\begin{eqnarray}
%%%%%%%%%%%%%%%%%%%%%%%%%%%%%%%%%%%%%%%%%%%%%%%%%%%%%%%%%%%%%%%%%%%%%%
&&a_{2L}^{4G}=a_{2L}^{SM}
+{G_{_F}\alpha_{_e}m_{_l}^2\over48\sqrt{2}\pi^3s_{_{\rm w}}^2}
\Big[31+12|V_{_{t'b'}}|^2\Big]
\nonumber\\
&&\hspace{1.3cm}
-{G_{_F}m_{_l}^2\over12\sqrt{2}\pi^2s_{_{\rm w}}^2}\Big(
1-2s_{_{\rm w}}^2+2s_{_{\rm w}}^4\Big)
\Delta \rho_{_{4G}}\;.
%%%%%%%%%%%%%%%%%%%%%%%%%%%%%%%%%%%%%%%%%%%%%%%%%%%%%%%%%%%%%%%%%%%%%%
\label{2L-4G}
\end{eqnarray}
Here, the 1-loop corrections to $\rho$-parameter from the heavy fermions of
4th generation can be written as
\begin{eqnarray}
%%%%%%%%%%%%%%%%%%%%%%%%%%%%%%%%%%%%%%%%%%%%%%%%%%%%%%%%%%%%%%%%%%%%%%
&&\Delta \rho_{_{4G}}={\alpha_{_e}\over16\pi s_{_{\rm w}}^2}{m_{_F}^2\over
m_{_{\rm w}}^2}\Big[19+12\ln{m_{_F}^2\over m_{_{\rm w}}^2}\Big]\;.
%%%%%%%%%%%%%%%%%%%%%%%%%%%%%%%%%%%%%%%%%%%%%%%%%%%%%%%%%%%%%%%%%%%%%%
\label{rho-4G}
\end{eqnarray}
In this model, the dominant contributions from Higgs sector are originated
from the two-loop $\gamma h$ diagrams. Under the assumption
$m_{_{t'}}=m_{_{b'}}=m_{_{\tau'}}=m_{_F}$, those contributions are zero
since the anomalous cancelation.
There is no correction from those heavy fermions
to lepton EDMs also in the extension of SM with the fourth generation.

\subsection{the minimal supersymmetric extension of SM}
\indent\indent
In this extension of SM, the additional possible heavy fermions include
two charginos $\chi_{1,2}^\pm$ and four neutralinos $\chi_i^0\;(i=1,\cdots,4)$
\cite{Su}. The couplings among weak gauge bosons and heavy fermions are
given as
\begin{eqnarray}
%%%%%%%%%%%%%%%%%%%%%%%%%%%%%%%%%%%%%%%%%%%%%%%%%%%%%%%%%%%%%%%%%%%%%%
&&\xi^L_{_{\chi_\alpha^\pm\chi_\beta^\pm}}=2\delta_{\alpha\beta}\cos2\theta_{_{\rm w}}
+(U_{_L}^\dagger)_{_{\alpha1}}(U_{_L})_{_{1\beta}}
\;,\nonumber\\
&&\xi^R_{_{\chi_\alpha^\pm\chi_\beta^\pm}}=2\delta_{\alpha\beta}\cos2\theta_{_{\rm w}}
+(U_{_R}^\dagger)_{_{\alpha1}}(U_{_R})_{_{1\beta}}\;,\;(\alpha,\beta=1,2)\;,
\nonumber\\
&&\xi^L_{_{\chi_\alpha^0\chi_\beta^0}}={\cal N}_{\alpha4}^\dagger{\cal N}_{4\beta}
\;,\nonumber\\
&&\xi^R_{_{\chi_\alpha^0\chi_\beta^0}}={\cal N}_{\beta3}^\dagger{\cal N}_{3\alpha}
\;,\;(\alpha,\beta=1,\cdots,4)\;,
\nonumber\\
&&\zeta^L_{_{\chi_\alpha^0\chi_\beta^\pm}}={\cal N}^\dagger_{\alpha2}(U_{_L})_{_{1\beta}}
-{1\over\sqrt{2}}{\cal N}^\dagger_{\alpha4}(U_{_L})_{_{2\beta}}
\;,\nonumber\\
&&\zeta^R_{_{\chi_\alpha^0\chi_\beta^\pm}}={\cal N}_{2\alpha}(U_{_R}^\dagger)_{_{\beta1}}
+{1\over\sqrt{2}}{\cal N}_{3\alpha}(U_{_R}^\dagger)_{_{\beta2}}
\;,\;(\alpha=1,\cdots,4;\;\;\beta=1,\;2)\;,
%%%%%%%%%%%%%%%%%%%%%%%%%%%%%%%%%%%%%%%%%%%%%%%%%%%%%%%%%%%%%%%%%%%%%%
\label{MSSM-couplings}
\end{eqnarray}
Here, the $4\times4$ matrix ${\cal N}$ denotes the mixing matrix of neutralinos,
two $2\times2$ matrices $U_{_L},\;U_{_R}$ denote the left- and right-handed
mixing matrices of charginos, respectively. In the limit of heavy masses,
the mass spectra of charginos and neutralinos are respectively approached as
\begin{eqnarray}
%%%%%%%%%%%%%%%%%%%%%%%%%%%%%%%%%%%%%%%%%%%%%%%%%%%%%%%%%%%%%%%%%%%%%%
&&m_{_{\chi^\pm}}\approx{\rm diag}(|m_2|,\;|\mu_{_H}|)\;,
\nonumber\\
&&m_{_{\chi^0}}\approx{\rm diag}(|m_1|,\;|m_2|,\;|\mu_{_H}|,\;|\mu_{_H}|)\;.
%%%%%%%%%%%%%%%%%%%%%%%%%%%%%%%%%%%%%%%%%%%%%%%%%%%%%%%%%%%%%%%%%%%%%%
\label{mass-matrices}
\end{eqnarray}
Here, $\mu_{_H}$ represents the $\mu$ parameter in superpotential,
and $m_2,\;m_1$ denote the soft breaking masses of $SU(2)\times U(1)$
gauginos,respectively.

Applying Eq.(\ref{MSSM-couplings}) and Eq.(\ref{mass-matrices}),
we get the two loop corrections to lepton MDMs in the heavy mass
limit $|m_1|=|m_2|=|\mu_{_H}|=m_{_F}
\gg m_{_{\rm w}}$ as
\begin{eqnarray}
%%%%%%%%%%%%%%%%%%%%%%%%%%%%%%%%%%%%%%%%%%%%%%%%%%%%%%%%%%%%%%%%%%%%%%
&&a_{2L}^{MSSM}=a_{2L}^{SM}
+{41G_{_F}\alpha_{_e}m_{_l}^2\over96\sqrt{2}\pi^3s_{_{\rm w}}^2}
-{G_{_F}m_{_l}^2\over12\sqrt{2}\pi^2s_{_{\rm w}}^2}\Big(
1-2s_{_{\rm w}}^2+2s_{_{\rm w}}^4\Big)\Delta \rho_{_{SUSY}}\;,
%%%%%%%%%%%%%%%%%%%%%%%%%%%%%%%%%%%%%%%%%%%%%%%%%%%%%%%%%%%%%%%%%%%%%%
\label{2L-MSSM}
\end{eqnarray}
the 1-loop corrections to $\rho$-parameter from the heavy supersymmetric fermions
can be written as \cite{rho-susy}
\begin{eqnarray}
%%%%%%%%%%%%%%%%%%%%%%%%%%%%%%%%%%%%%%%%%%%%%%%%%%%%%%%%%%%%%%%%%%%%%%
&&\Delta \rho_{_{SUSY}}={\alpha_{_e}\over16\pi s_{_{\rm w}}^2}\Big[-6
+{(1-2s_{_\beta}^2)^2\over c_{_{\rm w}}^2}
+{8c_{_\beta}^2c_{_{\rm w}}^2+4s_{_{\rm w}}^2+2s_{_\beta}^2\over
c_{_{\rm w}}^2-1/(2c_{_\beta}^2)}\ln(2c_{_\beta}^2c_{_{\rm w}}^2)
\nonumber\\
&&\hspace{1.8cm}
+{8s_{_\beta}^2c_{_{\rm w}}^2+4s_{_{\rm w}}^2+2c_{_\beta}^2\over
c_{_{\rm w}}^2-1/(2s_{_\beta}^2)}\ln(2s_{_\beta}^2c_{_{\rm w}}^2)\Big]\;.
%%%%%%%%%%%%%%%%%%%%%%%%%%%%%%%%%%%%%%%%%%%%%%%%%%%%%%%%%%%%%%%%%%%%%%
\label{rho-SUSY}
\end{eqnarray}
The abbreviation symbols $c_\beta=\cos\beta,\;s_\beta=\sin\beta$
with $\tan\beta=\upsilon_2/\upsilon_1$ denoting the ratio between
the absolute values of two vacuum expectations: $\upsilon_1,\;
\upsilon_2$. As $1\le\tan\beta\le60$ and $m_{_F}\sim1{\rm TeV}$,
the contributions from this sector to muon MDMs  is well below $10^{-11}$
which can be ignored safely.

At large $\tan\beta$, the dominant two-loop supersymmetric
corrections to lepton anomalous dipole moments are originated from those
Bar-Zee type diagrams which are analyzed extensively.
To obtain the corrections from those sectors, we formulate the relevant
couplings as
\begin{eqnarray}
%%%%%%%%%%%%%%%%%%%%%%%%%%%%%%%%%%%%%%%%%%%%%%%%%%%%%%%%%%%%%%%%%%%%%%
&&{\cal H}^{c,L}_{_{\chi_\beta^\pm\chi_\alpha^0}}=-c_\beta\Big[{1\over\sqrt{2}}(U_{_R})_{_{2\beta}}
\Big({s_{_{\rm w}}\over c_{_{\rm w}}}{\cal N}_{1\alpha}+{\cal N}_{2\alpha}\Big)
+(U_{_R})_{_{1\beta}}{\cal N}_{4\alpha}\Big]\;,
\nonumber\\
&&{\cal H}^{c,R}_{_{\chi_\beta^\pm\chi_\alpha^0}}=s_\beta\Big[{1\over\sqrt{2}}(U_{_L}^\dagger)_{_{\beta2}}
\Big({s_{_{\rm w}}\over c_{_{\rm w}}}{\cal N}^\dagger_{\alpha1}+{\cal N}^\dagger_{\alpha2}\Big)
-(U_{_L}^\dagger)_{_{1\beta}}{\cal N}^\dagger_{\alpha3}\Big]\;,
\nonumber\\
&&{\cal H}^{\cal E}_{_{\chi_\alpha^\pm\chi_\beta^\pm}}(h_k^0)={\cal Z}_{_{1k}}^{\cal E}
(U_{_L})_{_{2\alpha}}(U_{_R})_{_{1\beta}}+{\cal Z}_{_{2k}}^{\cal E}
(U_{_L})_{_{1\alpha}}(U_{_R})_{_{2\beta}}\;,\;(k=1,2)\;,
\nonumber\\
&&{\cal H}^{\cal O}_{_{\chi_\alpha^\pm\chi_\beta^\pm}}(h_k^0)=-s_\beta
(U_{_L})_{_{2\alpha}}(U_{_R})_{_{1\beta}}\;,
%%%%%%%%%%%%%%%%%%%%%%%%%%%%%%%%%%%%%%%%%%%%%%%%%%%%%%%%%%%%%%%%%%%%%%
\label{tan-beta1}
\end{eqnarray}
with ${\cal Z}^{\cal E}$ is the mixing matrix of two CP even Higgs.
Assuming $|\mu_{_{\rm H}}|=|m_2|=|m_1|=m_{_{F}}$ and $\theta_1=\theta_2=\theta_\mu$,
we expand the effective couplings in Eq.\ref{ASY-MED-W-GH} and
Eq.\ref{MD-gamma-h1} in powers of $m_{_{\rm w}}/m_{_F}$ and get
\begin{eqnarray}
%%%%%%%%%%%%%%%%%%%%%%%%%%%%%%%%%%%%%%%%%%%%%%%%%%%%%%%%%%%%%%%%%%%%%%
&&\sum\limits_{\alpha,\beta}{\cal H}^{c,L}_{_{\chi_\beta^\pm\chi_\alpha^0}}
\zeta^L_{_{\chi_\alpha^0\chi_\beta^\pm}}
={\cal O}({m_{_{\rm w}}^2\over m_{_F}^2})\;,
\nonumber\\
&&\sum\limits_{\alpha,\beta}{\cal H}^{c,L}_{_{\chi_\beta^\pm\chi_\alpha^0}}
\zeta^R_{_{\chi_\alpha^0\chi_\beta^\pm}}
={\cal O}({m_{_{\rm w}}^2\over m_{_F}^2})\;,
\nonumber\\
&&\sum\limits_{\alpha,\beta}{\cal H}^{c,R}_{_{\chi_\beta^\pm\chi_\alpha^0}}
\zeta^L_{_{\chi_\alpha^0\chi_\beta^\pm}}
={\cal O}({m_{_{\rm w}}^2\over m_{_F}^2})\;,
\nonumber\\
&&\sum\limits_{\alpha,\beta}{\cal H}^{c,R}_{_{\chi_\beta^\pm\chi_\alpha^0}}
\zeta^R_{_{\chi_\alpha^0\chi_\beta^\pm}}
\propto{m_{_{\rm w}}\over m_{_F}}e^{i\theta_\mu}\;,
\nonumber\\
&&\sum\limits_{\alpha}{\cal H}^{\cal E}_{_{\chi_\alpha^\pm\chi_\alpha^\pm}}(h_k^0)
\propto{m_{_{\rm w}}\over m_{_F}}e^{i\theta_\mu}\;,
\nonumber\\
&&\sum\limits_{\alpha}{\cal H}^{\cal O}_{_{\chi_\alpha^\pm\chi_\alpha^\pm}}
\propto{m_{_{\rm w}}\over m_{_F}}e^{i\theta_\mu}\;,
%%%%%%%%%%%%%%%%%%%%%%%%%%%%%%%%%%%%%%%%%%%%%%%%%%%%%%%%%%%%%%%%%%%%%%
\label{tan-beta2}
\end{eqnarray}
where $\theta_{1,2}=\arg(m_{1,2}),\; \theta_\mu=\arg(\mu_{_H})$
are the corresponding CP violating phases.
Applying the above equations and ${\cal B}_{\cal E}={\cal B}_{\cal O}
={\cal B}_{\cal C}=\tan\beta$, we find
\begin{eqnarray}
%%%%%%%%%%%%%%%%%%%%%%%%%%%%%%%%%%%%%%%%%%%%%%%%%%%%%%%%%%%%%%%%%%%%%%
&&a_{2L}={G_{_F}\alpha_{_e}m_{_l}^2m_{_{\rm w}}^2\tan\beta\over8\sqrt{2}\pi^3s_{_{\rm w}}^2m_{_F}^2}
\Big[A+B\ln{m_{_F}^2\over m_{_{\rm W}}^2}\Big]\cos\theta_\mu\;,
\nonumber\\
&&d_{2L}={G_{_F}\alpha_{_e}em_{_l}m_{_{\rm w}}^2\tan\beta\over16\sqrt{2}\pi^3s_{_{\rm w}}^2m_{_F}^2}
\Big[C+D\ln{m_{_F}^2\over m_{_{\rm W}}^2}\Big]\sin\theta_\mu\;.
%%%%%%%%%%%%%%%%%%%%%%%%%%%%%%%%%%%%%%%%%%%%%%%%%%%%%%%%%%%%%%%%%%%%%%
\label{tan-beta3}
\end{eqnarray}
Here, the form factors $A,\;B,\;C,\;D$ depend on the masses of
higgs and the mixing matrix of neutral CP-even higgs.

%%%%%%%%%%%%%%%%%%%%%%%%%%%%%%%%%%%%%%%%%%%%%%%%%%%%%%%%%%%%%%%%%%%
\begin{figure}[t]
\setlength{\unitlength}{1mm}
\begin{center}
\begin{picture}(0,80)(0,0)
\put(-80,-105){\includegraphics{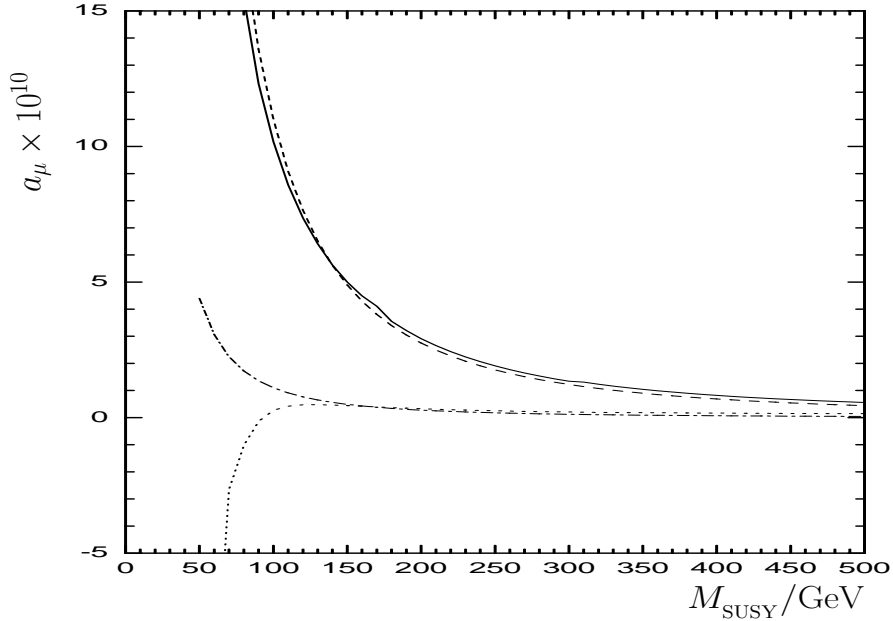}}
\end{picture}
\caption[]{The supersymmetric corrections to the muon MDM $a_\mu$ and
varies with the supersymmetric scale $M_{_{\rm SUSY}}$
when $\mu_{_{\rm H}}=m_2=(3c_{_{\rm w}}^2/5s_{_{\rm w}}^2)m_1
=m_{_A}=M_{_{\rm SUSY}}$ and $\tan\beta=5,\;50$.
Here the solid line stands for the two-loop contributions
from neutralino/chargino sector with $\tan\beta=50$,
the dash line stands for the results of Eq.(\ref{2L-MSSM1}) with $\tan=50$;
the dot line stands for the two-loop contributions
from neutralino/chargino sector with $\tan\beta=5$,
the dash-dot line stands for the results of Eq.(\ref{2L-MSSM1}) with $\tan=5$.}
\label{fig10}
\end{center}
\end{figure}
%%%%%%%%%%%%%%%%%%%%%%%%%%%%%%%%%%%%%%%%%%%%%%%%%%%%%%%%%%%%%%%%%%%

The two loop corrections to lepton MDMs and EDMs are proportional to
$1/m_{_F}^2$ in large $\tan\beta$ limit, which are consistent with the result
presented in Ref.\cite{heinemeyer} qualitatively.
Using HME approximation and projection operator method, Ref.\cite{heinemeyer}
approaches the corrections to
the muon MDM from two loop diagrams in which a closed chargino/neutralino loop is
inserted into those two Higgs doublet one loop diagrams as
\begin{eqnarray}
%%%%%%%%%%%%%%%%%%%%%%%%%%%%%%%%%%%%%%%%%%%%%%%%%%%%%%%%%%%%%%%%%%%%%%
&&a_{2L}^{MSSM}([5])=11\times10^{-10}\Big({\tan\beta\over50}\Big)
\Big({100{\rm GeV}\over M_{_{\rm SUSY}}}\Big)^2{\it sign}(\mu_{_{\rm H}})\;,
%%%%%%%%%%%%%%%%%%%%%%%%%%%%%%%%%%%%%%%%%%%%%%%%%%%%%%%%%%%%%%%%%%%%%%
\label{2L-MSSM1}
\end{eqnarray}
under the assumption $\mu_{_{\rm H}}=m_2=(3c_{_{\rm w}}^2/5s_{_{\rm w}}^2)m_1
=m_{_A}=M_{_{\rm SUSY}}$ and large $\tan\beta$ limit, where $m_{_A}$ denotes
the mass of neutral CP-odd Higgs.
%Since we adopt a
%different method to obtain the corrections to the anomalous dipole
%moments of leptons and make a different assumption on the parameter
%space of supersymmetry\footnote{We cannot simplify the dominant supersymmetric
%corrections with large $\tan\beta$ in a concise form presented in Eq.(\ref{tan-beta3})
%if we assume $\mu_{_{\rm H}}=m_2=(3c_{_{\rm w}}^2/5s_{_{\rm w}}^2)m_1
%=m_{_A}=m_{_F}$ because the mass spectrum of four neutralinos is not degenerate.},
%it is challenging to recover Eq.(\ref{2L-MSSM1}) exactly.

To compare our result with that presented in Ref.\cite{heinemeyer} numerically,
we take the same assumption $\mu_{_{\rm H}}=m_2=(3c_{_{\rm w}}^2/5s_{_{\rm w}}^2)m_1
=m_{_A}=M_{_{\rm SUSY}}$ on the parameter space of supersymmetry.
In addition, the existence of a CP-even SM like Higgs with mass above
$115\;{\rm GeV}$ sets a strong constraint on the parameter space
of the employed model. To address this problem, we include all loop corrected effects
in the Higgs potential\cite{Higgs-potential},
and choose the Yukawa couplings of the 3rd generation sfermions
as $A_{_t}=A_{_b}=A_{_\tau}=M_{_{\rm SUSY}}$.
%In order to evade the
%experimental constraint from LEP data on lightest chargino mass,
%we take $M_{_{\rm SUSY}}\ge100$ GeV in our numerical discussion.
Considering those points above, we present
the numerical results in Fig.\ref{fig10}. For large $\tan\beta$ case,
our numerical results agree with the approximation presented in
Eq.(\ref{2L-MSSM1}) very well. It implies that the equation
Eq.(\ref{2L-MSSM1}) fits the exact result perfectly
for large $\tan\beta$ and $\mu_{_{\rm H}}=m_2=(3c_{_{\rm w}}^2/5s_{_{\rm w}}^2)m_1
=m_{_A}=M_{_{\rm SUSY}}$.

%When the mass scale of new physics $M_{_{\rm SUSY}}\simeq400$ GeV,
%the difference between our numerical result and that from Eq.(\ref{2L-MSSM1})
%is about $1\%$ since we take the effective Lagrangian approach differing
%from the HME adopted in Ref.\cite{heinemeyer}. Along with the increasing
%of $M_{_{\rm SUSY}}$, the difference turns more and more minor.
%In other words, the results of effective Lagrangian approach
%fit that of HME approximation very well in the limit $M_{_{\rm SUSY}}\gg\Lambda_{_{\rm EW}}$.
%Furthermore, there is a similar result
%in the next to minimal supersymmetric extension of SM\cite{Domingo}.

\subsection{the littlest Higgs with T-parity}
\indent\indent
In the framework of the littlest Higgs with T-parity, all the SM
particles are even, as well as the corresponding mirror fields
are odd under the discrete T-transformation \cite{Hubisz}. In order to avoid
dangerous contributions to the Higgs mass from one-loop quadratic
divergences, we introduce additionally one T-even top quark $T_+$ together with
its mirror partner, the T-odd top quark $T_-$ besides the SM fermions $f^i$
and their mirror partners $f_H^i$. The couplings among the SM gauge
bosons and heavy fermions in Eq.(\ref{ZFF}) and Eq.(\ref{WFF}) are
respectively given as \cite{Blanke}
\begin{eqnarray}
%%%%%%%%%%%%%%%%%%%%%%%%%%%%%%%%%%%%%%%%%%%%%%%%%%%%%%%%%%%%%%%%%%%%%%
&&\xi^L_{_{T_+T_+}}=-\eta_{_L}^2{\upsilon^2\over f^2}+{4\over3}s_{_{\rm w}}^2\;,\;\;\;
\xi^R_{_{T_+T_+}}={4\over3}s_{_{\rm w}}^2
\;,\nonumber\\
&&\xi^L_{_{T_+t}}=-\eta_{_L}{\upsilon\over f}\;,\;\;\;
\xi^R_{_{T_+t}}=0
\;,\nonumber\\
&&\xi^L_{_{T_-T_-}}=\xi^R_{_{T_-T_-}}={4\over3}s_{_{\rm w}}^2
\;,\nonumber\\
&&\xi^L_{_{u_{H}^iu_{H}^i}}=\xi^R_{_{u_{H}^iu_{H}^i}}=-1+{4\over3}s_{_{\rm w}}^2
\;,\nonumber\\
&&\xi^L_{_{d_{H}^id_{H}^i}}=\xi^R_{_{d_{H}^id_{H}^i}}=1-{2\over3}s_{_{\rm w}}^2
\;,\nonumber\\
&&\xi^L_{_{\nu_{H}^i\nu_{H}^i}}=\xi^R_{_{\nu_{H}^i\nu_{H}^i}}=-1
\;,\nonumber\\
&&\xi^L_{_{e_{H}^ie_{H}^i}}=\xi^R_{_{e_{H}^ie_{H}^i}}=1-2s_{_{\rm w}}^2
\;,\nonumber\\
&&\zeta^L_{_{T_+b}}={V_{_{tb}}\over\sqrt{2}}\eta_{_L}{\upsilon\over f}\;,\;\;\;
\zeta^R_{_{T_+b}}=0
\;,\nonumber\\
&&\zeta^L_{_{u_{H}^id_{H}^j}}=\zeta^R_{_{u_{H}^id_{H}^j}}={\delta_{_{ij}}\over\sqrt{2}}
\;,\nonumber\\
&&\zeta^L_{_{\nu_{H}^ie_{H}^j}}=\zeta^R_{_{\nu_{H}^ie_{H}^j}}={\delta_{_{ij}}\over\sqrt{2}}\;.
%%%%%%%%%%%%%%%%%%%%%%%%%%%%%%%%%%%%%%%%%%%%%%%%%%%%%%%%%%%%%%%%%%%%%%
\label{LHT-couplings}
\end{eqnarray}
Here, $f$ is the breaking scale of a large $SU(5)/SO(5)$ symmetry,
and
\begin{eqnarray}
%%%%%%%%%%%%%%%%%%%%%%%%%%%%%%%%%%%%%%%%%%%%%%%%%%%%%%%%%%%%%%%%%%%%%%
&&\eta_{_L}={\lambda_1^2\over\lambda_1^2+\lambda_2^2}\;,
%%%%%%%%%%%%%%%%%%%%%%%%%%%%%%%%%%%%%%%%%%%%%%%%%%%%%%%%%%%%%%%%%%%%%%
\label{LHT-parameters}
\end{eqnarray}
with $\lambda_{1,2}$ represent the Yukawa couplings of top quark sector.
Additionally, the relations among the masses of heavy fermions are
presented as
\begin{eqnarray}
%%%%%%%%%%%%%%%%%%%%%%%%%%%%%%%%%%%%%%%%%%%%%%%%%%%%%%%%%%%%%%%%%%%%%%
&&m_{_{T_+}}={f\over\upsilon}{m_{_t}\over\sqrt{\eta_{_L}(1-\eta_{_L})}}
\;,\nonumber\\
&&m_{_{T_-}}={f\over\upsilon}{m_{_t}\over\sqrt{\eta_{_L}}}
\;,\nonumber\\
&&m_{_{u_{H}^i}}=m_{_{d_{H}^i}}(1-{\upsilon^2\over8f^2})\;,
\nonumber\\
&&m_{_{\nu_{H}^i}}=m_{_{e_{H}^i}}(1-{\upsilon^2\over8f^2})\;,\;\;(i=1,2,3)\;.
%%%%%%%%%%%%%%%%%%%%%%%%%%%%%%%%%%%%%%%%%%%%%%%%%%%%%%%%%%%%%%%%%%%%%%
\label{LHT-masses}
\end{eqnarray}
Applying the equations above, we give the leading corrections from
heavy fermions to lepton MDMs in the limit $f=m_{_{e_{H}^i}}
=m_{_{d_{H}^i}}=m_{_F}\gg m_{_{\rm w}}$ as
\begin{eqnarray}
%%%%%%%%%%%%%%%%%%%%%%%%%%%%%%%%%%%%%%%%%%%%%%%%%%%%%%%%%%%%%%%%%%%%%%
&&a_{2L}^{LHT}=a_{2L}^{SM}
+{5G_{_F}\alpha_{_e}m_{_l}^2\over6\sqrt{2}\pi^3s_{_{\rm w}}^2}
-{G_{_F}m_{_l}^2\over12\sqrt{2}\pi^2s_{_{\rm w}}^2}\Big(
1-2s_{_{\rm w}}^2+2s_{_{\rm w}}^4\Big)\Delta \rho_{_{LHT}}\;,
%%%%%%%%%%%%%%%%%%%%%%%%%%%%%%%%%%%%%%%%%%%%%%%%%%%%%%%%%%%%%%%%%%%%%%
\label{2L-LHT}
\end{eqnarray}
the 1-loop corrections to $\rho$-parameter from heavy fermions
can be written as \cite{rho-LHT}
\begin{eqnarray}
%%%%%%%%%%%%%%%%%%%%%%%%%%%%%%%%%%%%%%%%%%%%%%%%%%%%%%%%%%%%%%%%%%%%%%
&&\Delta \rho_{_{LHT}}=\eta_{_L}{\upsilon^2\over f^2}
\Big[-{1\over2}+\ln{m_{_{T_+}}^2\over m_{_{\rm w}}^2}\Big]\Delta \rho_{_{SM}}\;.
%%%%%%%%%%%%%%%%%%%%%%%%%%%%%%%%%%%%%%%%%%%%%%%%%%%%%%%%%%%%%%%%%%%%%%
\label{rho-LHT}
\end{eqnarray}

\subsection{the universal extra dimension}
\indent\indent
If all particles of the SM are zero modes of corresponding
5-dimension bulk fields \cite{Georgi}, the KK excitations
of fermion acquire the masses
\begin{eqnarray}
%%%%%%%%%%%%%%%%%%%%%%%%%%%%%%%%%%%%%%%%%%%%%%%%%%%%%%%%%%%%%%%%%%%%%%
&&m_{_{f_{i(n)}}}=\sqrt{m_{_{f_i}}^2+{n^2\over R^2}}\;,(n=1,\;2,\;\cdots)\;,
%%%%%%%%%%%%%%%%%%%%%%%%%%%%%%%%%%%%%%%%%%%%%%%%%%%%%%%%%%%%%%%%%%%%%%
\label{KK-excitation-mass}
\end{eqnarray}
where $m_{_{f_i}}$ denotes the mass of corresponding SM field, and $R$ is
the compactification radius. To fit the present experimental data,
we choose $1/R\geq200\;{\rm GeV}$. Furthermore, we formulate the couplings
among the zero modes of weak gauge bosons and the KK excitations of
fermions as \cite{Buras}
\begin{eqnarray}
%%%%%%%%%%%%%%%%%%%%%%%%%%%%%%%%%%%%%%%%%%%%%%%%%%%%%%%%%%%%%%%%%%%%%%
&&\xi^L_{_{{\cal Q}_{_i}^{(n)}{\cal Q}_{_i}^{(n)}}}=
\xi^R_{_{{\cal Q}_{_i}^{(n)}{\cal Q}_{_i}^{(n)}}}={4\over3}s_{_{\rm w}}^2-c_{_{i(n)}}^2
\;,\nonumber\\
&&\xi^L_{_{{\cal U}_{_i}^{(n)}{\cal U}_{_i}^{(n)}}}=
\xi^R_{_{{\cal U}_{_i}^{(n)}{\cal U}_{_i}^{(n)}}}={4\over3}s_{_{\rm w}}^2-s_{_{i(n)}}^2
\;,\nonumber\\
&&\xi^L_{_{{\cal D}_{_i}^{(n)}{\cal D}_{_i}^{(n)}}}=
\xi^R_{_{{\cal D}_{_i}^{(n)}{\cal D}_{_i}^{(n)}}}=-{2\over3}s_{_{\rm w}}^2+s_{_{i(n)}}^2
\;,\nonumber\\
&&\xi^L_{_{{\cal Q}_{_i}^{(n)}{\cal U}_{_i}^{(n)}}}=
-\xi^R_{_{{\cal Q}_{_i}^{(n)}{\cal U}_{_i}^{(n)}}}=c_{_{i(n)}}s_{_{i(n)}}
\;,\nonumber\\
&&\zeta^L_{_{{\cal Q}_{_i}^{(n)}d_{_j}}}=c_{_{i(n)}}{V_{_{ij}}\over\sqrt{2}}\;,\;\;\;
\zeta^R_{_{{\cal Q}_{_i}^{(n)}d_{_j}}}=0
\;,\nonumber\\
&&\zeta^L_{_{{\cal U}_{_i}^{(n)}d_{_j}}}=-s_{_{i(n)}}{V_{_{ij}}\over\sqrt{2}}\;,\;\;\;
\zeta^R_{_{{\cal U}_{_i}^{(n)}d_{_j}}}=0
\;,\nonumber\\
&&\zeta^L_{_{{\cal L}_{_i}^{(n)}\nu_{_j}}}={\delta_{_{ij}}\over\sqrt{2}}\;,\;\;\;
\zeta^R_{_{{\cal L}_{_i}^{(n)}\nu_{_j}}}=0
\;,\nonumber\\
%%%%%%%%%%%%%%%%%%%%%%%%%%%%%%%%%%%%%%%%%%%%%%%%%%%%%%%%%%%%%%%%%%%%%%
\label{UED-couplings}
\end{eqnarray}
with
\begin{eqnarray}
%%%%%%%%%%%%%%%%%%%%%%%%%%%%%%%%%%%%%%%%%%%%%%%%%%%%%%%%%%%%%%%%%%%%%%
&&\tan2\alpha_{_{i(n)}}={m_{_{f_i}}\over n/R}\;,\;\;\;(f_i=u_{_i},\;d_{_i},\;
\nu_{_i},\;e_{_i},\;\;i=1,2,3)\;,
\nonumber\\
&&c_{_{i(n)}}=\cos\alpha_{_{i(n)}},\;\;s_{_{i(n)}}=\sin\alpha_{_{i(n)}}\;.
%%%%%%%%%%%%%%%%%%%%%%%%%%%%%%%%%%%%%%%%%%%%%%%%%%%%%%%%%%%%%%%%%%%%%%
\label{UED-parameters}
\end{eqnarray}
Using Eq.(\ref{KK-excitation-mass}), Eq.(\ref{UED-couplings}) and
Eq.(\ref{UED-parameters}), we formulate the leading contributions from the KK
excitations of fermions as
\begin{eqnarray}
%%%%%%%%%%%%%%%%%%%%%%%%%%%%%%%%%%%%%%%%%%%%%%%%%%%%%%%%%%%%%%%%%%%%%%
&&a_{2L}^{UED}=a_{2L}^{SM}
-{G_{_F}\alpha_{_e}m_{_l}^2\over2\sqrt{2}\pi^3s_{_{\rm w}}^2}\;.
%%%%%%%%%%%%%%%%%%%%%%%%%%%%%%%%%%%%%%%%%%%%%%%%%%%%%%%%%%%%%%%%%%%%%%
\label{2L-UED}
\end{eqnarray}

\section{Conclusion\label{sec4}}
\indent\indent
In this work, we have investigated the electroweak corrections to
the lepton MDMs and EDMs from some two loop diagrams in which a closed
heavy fermion loop is inserted into those two Higgs doublet diagrams.
Adopting on-shell scheme, we subtract the ultraviolet divergence
caused by the subdiagrams and  get the theoretical
predictions on lepton MDMs. As the masses of virtual fermions in
inner loop are much heavier than the electroweak scale, we verify
the final results satisfying the decoupling theorem explicitly
if the interactions among Higgs and heavy fermions do not contain
the nondecoupling couplings.
Our results are universal for all extensions of the SM
where the interactions among the electroweak gauge bosons and
heavy fermions are renormalizable. As application of our analysis, we present
the leading corrections to lepton MDMs in some popular extensions
of the SM, such as the fourth generation, supersymmetry, universal
extra dimension, and the littlest higgs with T-parity.

\begin{acknowledgments}
\indent\indent
The work has been supported by the National Natural Science Foundation of China (NNSFC)
with Grant No. 10675027.
\end{acknowledgments}
\vspace{1.6cm}
%%%%%%%%%%%%%%%%%%%%%%%%%%%% APPENDICES %%%%%%%%%%%%%%%%%%%%%%%%%%%%%%%%
\appendix

\section{The form factors}
\indent\indent
\begin{eqnarray}
%%%%%%%%%%%%%%%%%%%%%%%%%%%%%%%%%%%%%%%%%%%%%%%%%%%%%%%%%%%%%%%%%%%%%%
&&{\cal N}_{_{\rm ww}}^{(1)}=-{24\over(q_1^2-m_{_{\rm w}}^2)^3}\Big[-{(q_1^2)^2q_1\cdot
q_2-(q_1^2)^2q_2^2\over D+2} +4\cdot{q_1^2(q_1\cdot q_2)^2-(q_1^2)^2q_2^2\over D(D+2)}\Big]
\nonumber\\
&&\hspace{1.4cm}
+{6\over D+2}{q_1^2(q_2^2)^2-q_1\cdot q_2(q_2^2)^2\over(q_2^2-m_{_{F_\beta}}^2)^3}
+{18\over D+2}{(q_1^2)^2q_2^2-q_1^2q_1\cdot q_2q_2^2\over(q_1^2-m_{_{\rm w}}^2)^2
(q_2^2-m_{_{F_\beta}}^2)}
\nonumber\\
&&\hspace{1.4cm}
+{12\over D+2}{q_1^2q_1\cdot q_2q_2^2-q_1^2(q_2^2)^2\over(q_1^2-m_{_{\rm w}}^2)
(q_2^2-m_{_{F_\beta}}^2)^2}
-{2\over(q_2^2-m_{_{F_\beta}}^2)^2}\Big[\Big(1-{Q_\beta\over D}\Big)q_1^2q_2^2
\nonumber\\
&&\hspace{1.4cm}
-\Big(1+{1\over D}-{2\over D}Q_\beta\Big)q_1\cdot q_2q_2^2\Big]
-{3\over(q_1^2-m_{_{\rm w}}^2)^2}\Big\{{3D\over D+2}(q_1^2)^2
-{2(D+5)\over D+2}q_1^2q_1\cdot q_2
\nonumber\\
&&\hspace{1.4cm}
-{D-10\over D+2}q_1^2q_2^2
+4(D^2-D+6)\cdot{(q_1\cdot q_2)^2-q_1^2q_2^2\over D(D-1)(D+2)}\Big\}
\nonumber\\
&&\hspace{1.4cm}
-{1\over(q_1^2-m_{_{\rm w}}^2)(q_2^2-m_{_{F_\beta}}^2)}\Big[
\Big(3-{2\over D}Q_\beta\Big)q_1^2q_1\cdot q_2-\Big(4+{4\over D}(1-Q_\beta)\Big)q_1^2q_2^2
\nonumber\\
&&\hspace{1.4cm}
+\Big(1+{2\over D}\Big)q_1\cdot q_2q_2^2\Big]
\nonumber\\
&&\hspace{1.4cm}
+{1\over q_2^2-m_{_{F_\beta}}^2}\Big[({D\over4}-{Q_\beta\over2})q_1^2-\Big({D\over4}+{1\over2}
+{1\over D}-(1+{1\over D})Q_\beta\Big)q_1\cdot q_2\Big]
\nonumber\\
&&\hspace{1.4cm}
-{1\over q_1^2-m_{_{\rm w}}^2}\Big[\Big({D\over2}+2-(1+{2\over D})Q_\beta\Big)q_1^2
-\Big({D\over2}+1+{2\over D}-{4\over D}Q_\beta\Big)q_1\cdot q_2\Big]
\;,\nonumber\\
%%%%%%%%%%%%%%%%%%%%%%%%%%%%%%%%%%%%%%%%%%%%%%%%%%%%%%%%%%%%%%%%%%%%%%
&&{\cal N}_{_{\rm ww}}^{(2)}=
{2\over D}{Q_\beta q_1^2q_2^2-q_1\cdot q_2q_2^2\over(q_2^2-m_{_{F_\beta}}^2)^2}
-{1\over q_2^2-m_{_{F_\beta}}^2}\Big[{Q_\beta\over2}q_1^2
-\Big({1\over2}+{1\over D}-{1\over D}Q_\beta\Big)q_1\cdot q_2\Big]
\nonumber\\
&&\hspace{1.4cm}
+{2\over D}{Q_\beta q_1^2q_1\cdot q_2-q_1^2q_2^2\over(q_1^2-m_{_{\rm w}}^2)
(q_2^2-m_{_{F_\beta}}^2)}+{q_1^2-q_1\cdot q_2\over q_1^2-m_{_{\rm w}}^2}
\;,\nonumber\\
%%%%%%%%%%%%%%%%%%%%%%%%%%%%%%%%%%%%%%%%%%%%%%%%%%%%%%%%%%%%%%%%%%%%%%
&&{\cal N}_{_{\rm ww}}^{(3)}=
{6(D-2)\over D(D+2)}{q_1\cdot q_2q_2^2\over(q_2^2-m_{_{F_\beta}}^2)^3}
-\Big(1-{2\over D}-{2\over D}Q_\beta\Big){q_1\cdot q_2\over(q_2^2-m_{_{F_\beta}}^2)^2}
\nonumber\\
&&\hspace{1.4cm}
+{24(D-2)\over D(D+2)}{(q_1^2)^2\over(q_1^2-m_{_{\rm w}}^2)^3}
-{1\over D}(Q_\beta-1){q_1\cdot q_2\over(q_2^2-m_{_{F_\beta}}^2)((q_2-q_1)^2-m_{_{F_\alpha}}^2)}
\nonumber\\
&&\hspace{1.4cm}
+{18(D-2)\over D(D+2)}{q_1^2q_1\cdot q_2\over(q_1^2-m_{_{\rm w}}^2)^2
(q_2^2-m_{_{F_\beta}}^2)}
+{12(D-2)\over D(D+2)}{q_1^2q_2^2\over(q_1^2-m_{_{\rm w}}^2)
(q_2^2-m_{_{F_\beta}}^2)^2}
\nonumber\\
&&\hspace{1.4cm}
-3\Big(1-{6\over D}\Big){q_1^2\over(q_1^2-m_{_{\rm w}}^2)^2}
-{1\over(q_1^2-m_{_{\rm w}}^2)(q_2^2-m_{_{F_\beta}}^2)}\Big[\Big({3D-6\over D+2}
-{2\over D}Q_\beta\Big)q_1^2
\nonumber\\
&&\hspace{1.4cm}
-\Big({D^2+20\over D(D+2)}-{2\over D}Q_\beta\Big)q_1\cdot q_2\Big]
-{2\over D}{(Q_\beta-1)q_1\cdot q_2\over(q_1^2-m_{_{\rm w}}^2)
((q_2-q_1)^2-m_{_{F_\alpha}}^2)}
\;,\nonumber\\
%%%%%%%%%%%%%%%%%%%%%%%%%%%%%%%%%%%%%%%%%%%%%%%%%%%%%%%%%%%%%%%%%%%%%%
&&{\cal N}_{_{\rm ww}}^{(4)}
={2\over D}Q_\beta\Big[{q_1^2\over(q_1^2-m_{_{\rm w}}^2)(q_2^2-m_{_{\chi_\beta^\pm}}^2)}
+{q_1\cdot q_2\over(q_2^2-m_{_{\chi_\beta^\pm}}^2)^2}\Big]
\nonumber\\
&&\hspace{1.4cm}
+{2\over D}(Q_\beta-1)\Big[{q_1^2\over(q_1^2-m_{_{\rm z}}^2)((q_2-q_1)^2-m_{_{\chi_\alpha^\pm}}^2)}
\nonumber\\
&&\hspace{1.4cm}
+{1\over 2}{q_1\cdot q_2\over(q_2^2-m_{_{\chi_\beta^\pm}}^2)
((q_2-q_1)^2-m_{_{\chi_\alpha^\pm}}^2)}\Big]\;.
%%%%%%%%%%%%%%%%%%%%%%%%%%%%%%%%%%%%%%%%%%%%%%%%%%%%%%%%%%%%%%%%%%%%%%
\label{aeq1}
\end{eqnarray}

\section{The functions\label{ap1}}
\indent\indent

The definition of $\Psi(x,y,z)$ is written as:
\begin{itemize}
\item $\lambda^2>0,\;\sqrt{y}+\sqrt{z}<\sqrt{x}$:
\begin{eqnarray}
&&\Psi(x,y,z)=2\ln\Big({x+y-z-\lambda\over2x}\Big)
\ln\Big({x-y+z-\lambda\over2x}\Big)-\ln{y\over x}\ln{z\over x}
\nonumber\\
&&\hspace{2.2cm}
-2L_{i_2}\Big({x+y-z-\lambda\over2x}\Big)
-2L_{i_2}\Big({x-y+z-\lambda\over2x}\Big)+{\pi^2\over3}\;,
\label{aeq2}
\end{eqnarray}
where $L_{i_2}(x)$ is the spence function;

\item $\lambda^2>0,\;\sqrt{x}+\sqrt{z}<\sqrt{y}$:
\begin{eqnarray}
&&\Psi(x,y,z)={\rm Eq.}(\ref{aeq2})(x\leftrightarrow y)\;;
\label{aeq3}
\end{eqnarray}

\item $\lambda^2>0,\;\sqrt{x}+\sqrt{y}<\sqrt{z}$:
\begin{eqnarray}
&&\Psi(x,y,z)={\rm Eq.}(\ref{aeq2})(x\leftrightarrow z)\;;
\label{aeq4}
\end{eqnarray}

\item $\lambda^2<0$:
\begin{eqnarray}
&&\Psi(x,y,z)=2\Big\{Cl_2\Big(2\arccos(
{-x+y+z\over2\sqrt{yz}})\Big)
+Cl_2\Big(2\arccos({x-y+z\over2\sqrt{xz}})\Big)
\nonumber\\
&&\hspace{2.2cm}
+Cl_2\Big(2\arccos({x+y-z\over2\sqrt{xy}})\Big)\Big\}\;,
\label{aeq12}
\end{eqnarray}
where $Cl_2(x)$ denotes the Clausen function.
\end{itemize}

The expressions of $\varphi_0(x,y),\;\varphi_1(x,y),\;\varphi_2(x,y)$
and $\varphi_3(x,y)$ are given as
\begin{eqnarray}
%%%%%%%%%%%%%%%%%%%%%%%%%%%%%%%%%%%%%%%%%%%%%%%%%%%%%%%%%%%%%%%%%%%%%%
&&\varphi_0(x,y)=\left\{\begin{array}{ll}(x+y)\ln x\ln y+(x-y)\Theta(x,y)
\;,&x>y\;;\\
2x\ln^2x\;,&x=y\;;\\
(x+y)\ln x\ln y+(y-x)\Theta(y,x)
\;,&x<y\;.\end{array}\right.
%%%%%%%%%%%%%%%%%%%%%%%%%%%%%%%%%%%%%%%%%%%%%%%%%%%%%%%%%%%%%%%%%%%%%%
\label{varphi0}
\end{eqnarray}

\begin{eqnarray}
%%%%%%%%%%%%%%%%%%%%%%%%%%%%%%%%%%%%%%%%%%%%%%%%%%%%%%%%%%%%%%%%%%%%%%
&&\varphi_1(x,y)=\left\{\begin{array}{ll}-\ln x\ln y-{x+y\over x-y}\Theta(x,y)
\;,&x>y\;;\\
4-2\ln x-\ln^2x\;,&x=y\;;\\
-\ln x\ln y-{x+y\over y-x}\Theta(y,x)
\;,&x<y\;,\end{array}\right.
%%%%%%%%%%%%%%%%%%%%%%%%%%%%%%%%%%%%%%%%%%%%%%%%%%%%%%%%%%%%%%%%%%%%%%
\label{varphi1}
\end{eqnarray}

\begin{eqnarray}
%%%%%%%%%%%%%%%%%%%%%%%%%%%%%%%%%%%%%%%%%%%%%%%%%%%%%%%%%%%%%%%%%%%%%%
&&\varphi_2(x,y)=\left\{\begin{array}{ll}
{(2x^2+6xy)\ln x-(6xy+2y^2)\ln y\over(x-y)^3}-{4xy\over(x-y)^3}\Theta(x,y)
\;,&x>y\;;\\
-{5\over9x}+{2\over3x}\ln x\;,&x=y\;;\\
{(2x^2+6xy)\ln x-(6xy+2y^2)\ln y\over(x-y)^3}-{4xy\over(y-x)^3}\Theta(y,x)
\;,&x<y\;,\end{array}\right.
%%%%%%%%%%%%%%%%%%%%%%%%%%%%%%%%%%%%%%%%%%%%%%%%%%%%%%%%%%%%%%%%%%%%%%
\label{varphi2}
\end{eqnarray}

\begin{eqnarray}
%%%%%%%%%%%%%%%%%%%%%%%%%%%%%%%%%%%%%%%%%%%%%%%%%%%%%%%%%%%%%%%%%%%%%%
&&\varphi_3(x,y)=\left\{\begin{array}{ll}-{12xy(x+y)\over(x-y)^5}\Theta(x,y)
-{2(x^2+6xy+y^2)\over(x-y)^4} & \\
+{2(x^3+20x^2y+11xy^2)\ln x-2(y^3+20xy^2+11x^2y)\ln y\over(x-y)^5}
\;,&x>y\;;\\
-{53\over150x^2}+{1\over5x^2}\ln x\;,&x=y\;;\\
-{12xy(x+y)\over(y-x)^5}\Theta(y,x)-{2(x^2+6xy+y^2)\over(x-y)^4} & \\
+{2(x^3+20x^2y+11xy^2)\ln x-2(y^3+20xy^2+11x^2y)\ln y\over(x-y)^5}
\;,&x<y\;,
\end{array}\right.
%%%%%%%%%%%%%%%%%%%%%%%%%%%%%%%%%%%%%%%%%%%%%%%%%%%%%%%%%%%%%%%%%%%%%%
\label{varphi3}
\end{eqnarray}
with
\begin{eqnarray}
%%%%%%%%%%%%%%%%%%%%%%%%%%%%%%%%%%%%%%%%%%%%%%%%%%%%%%%%%%%%%%%%%%%%%%
&&\Theta(x,y)=\ln x\ln{y\over x}-2\ln(x-y)\ln{y\over x}-2Li_2({y\over x})+{\pi^2\over3}\;.
%%%%%%%%%%%%%%%%%%%%%%%%%%%%%%%%%%%%%%%%%%%%%%%%%%%%%%%%%%%%%%%%%%%%%%
\label{theta}
\end{eqnarray}

The functions adopted in the text are written as
\begin{eqnarray}
%%%%%%%%%%%%%%%%%%%%%%%%%%%%%%%%%%%%%%%%%%%%%%%%%%%%%%%%%%%%%%%%%%%%%%
&&\varrho_{_{i,j}}(x,y)={x^i\ln^jx-y^i\ln^jy\over x-y}\;,\;\;
%\nonumber\\
%%%%%%%%%%%%%%%%%%%%%%%%%%%%%%%%%%%%%%%%%%%%%%%%%%%%%%%%%%%%%%%%%%%%%%
\Omega_i(x,y;u,v)={x^i\Phi(x,u,v)-y^i\Phi(y,u,v)\over x-y}\;,
\nonumber\\
%%%%%%%%%%%%%%%%%%%%%%%%%%%%%%%%%%%%%%%%%%%%%%%%%%%%%%%%%%%%%%%%%%%%%%
&&T_1(x,y,z)={(y-z)^2\over3x^3}(1+\varrho_{_{1,1}}(y,z))
-{1\over48x^2}\Big[10\varrho_{_{2,1}}(y,z)+(31-12Q_\beta)y
\nonumber\\
&&\hspace{2.8cm}
+(19+12Q_\beta)z-(22+12Q_\beta)y\ln y-(16-12Q_\beta)z\ln z\Big]
\nonumber\\
&&\hspace{2.8cm}
-{1\over8x}\Big[3(1-Q_\beta)-(3-2Q_\beta)\ln z\Big]
\nonumber\\
%---------------------------------------------------------------------
&&\hspace{2.8cm}
-{1\over48}\Big\{{\partial^4\over\partial x^4}\Big[(y-z)^2\Phi
-x^2\Phi\Big](x,y,z)
\nonumber\\
%---------------------------------------------------------------------
&&\hspace{2.8cm}
-6{\partial^4\over\partial x^3\partial z}\Big[z(y-z)
\Phi+xz\Phi\Big](x,y,z)
\nonumber\\
%---------------------------------------------------------------------
&&\hspace{2.8cm}
+6{\partial^4\over\partial x^2\partial z^2}\Big[z(y+z)
\Phi-xz\Phi\Big](x,y,z)
\nonumber\\
%---------------------------------------------------------------------
&&\hspace{2.8cm}
-2{\partial^4\over\partial x\partial z^3}\Big[z^2(y-z)
{\Phi(x,y,z)-\varphi_0(y,z)\over x}+z^2\Phi(x,y,z)\Big]
\nonumber\\
%---------------------------------------------------------------------
&&\hspace{2.8cm}
+3{\partial^3\over\partial x^3}\Big[(y-z)^2{\Phi(x,y,z)-\varphi_0(y,z)\over x}
+\Big(4y-4z+3x\Big)\Phi(x,y,z)\Big]
\nonumber\\
%---------------------------------------------------------------------
&&\hspace{2.8cm}
+6{\partial^3\over\partial x\partial z^2}\Big[({5\over2}-Q_{_\beta})z(y-z)
{\Phi(x,y,z)-\varphi_0(y,z)\over x}+{3\over2}z\Phi(x,y,z)\Big]
\nonumber\\
%---------------------------------------------------------------------
&&\hspace{2.8cm}
-3{\partial^3\over\partial x^2\partial z}\Big[3z(y-z)
{\Phi(x,y,z)-\varphi_0(y,z)\over x}
+\Big((6-Q_{_\beta})y
\nonumber\\
&&\hspace{2.8cm}
+(11-3Q_{_\beta})z\Big)\Phi(x,y,z)-(6-Q_{_\beta})x\Phi(x,y,z)\Big]
\nonumber\\
&&\hspace{2.8cm}
-3{\partial^2\over\partial x\partial z}\Big[(7-5Q_{_\beta})(y-z)
{\Phi(x,y,z)-\varphi_0(y,z)\over x}+(1+Q_{_\beta})\Phi(x,y,z)\Big]
\nonumber\\
%---------------------------------------------------------------------
&&\hspace{2.8cm}
+6{\partial^2\over\partial x^2}\Big[({7\over2}-Q_{_\beta})(y-z)
{\Phi(x,y,z)-\varphi_0(y,z)\over x}+({9\over2}-2Q_{_\beta})\Phi(x,y,z)\Big]\Big\}\;,
\nonumber\\
%%%%%%%%%%%%%%%%%%%%%%%%%%%%%%%%%%%%%%%%%%%%%%%%%%%%%%%%%%%%%%%%%%%%%%
&&T_2(x,y,z)=-{1\over16}\Big\{{2\ln z\over x}
-{4\over x^2}\Big(y-z+y\ln y-z\ln z\Big)
\nonumber\\
%---------------------------------------------------------------------
&&\hspace{2.8cm}
+{\partial^3\over\partial x\partial z^2}
\Big[(1-2Q_{_\beta})z\Phi(x,y,z)-z(y-z){\Phi(x,y,z)-\varphi_0(y,z)\over x}\Big]
\nonumber\\
%---------------------------------------------------------------------
&&\hspace{2.8cm}
-{\partial^2\over\partial x\partial z}\Big[(3-5Q_{_\beta})\Phi(x,y,z)
-(3-Q_{_\beta})(y-z){\Phi(x,y,z)-\varphi_0(y,z)\over x}\Big]
\nonumber\\
%---------------------------------------------------------------------
&&\hspace{2.8cm}
-{\partial^3\over\partial x^2\partial z}\Big[Q_{_\beta}x\Phi
-(Q_{_\beta}y+(2-Q_{_\beta})z)\Phi\Big](x,y,z)
\nonumber\\
%---------------------------------------------------------------------
&&\hspace{2.8cm}
-2{\partial^2\over\partial x^2}\Big[\Phi(x,y,z)+(y-z)
{\Phi(x,y,z)-\varphi_0(y,z)\over x}\Big]\Big\}\;,
\nonumber\\
%%%%%%%%%%%%%%%%%%%%%%%%%%%%%%%%%%%%%%%%%%%%%%%%%%%%%%%%%%%%%%%%%%%%%%
&&T_3(x,y,z)=
{5\over12x^2}\varrho_{_{1,1}}(y,z)+{7\over6x^2}+{1-3Q_{_\beta}\over24xz}
-{1-Q_{_\beta}\over8x^2}\ln y+{4-Q_\beta\over8x^2}\ln z
\nonumber\\
&&\hspace{2.8cm}
-{1\over48}\Big\{{\partial^4\over\partial x\partial z^3}
\Big[z(y-z)\Omega_{_0}-z\Omega_{_1}\Big](x,y;y,z)
\nonumber\\
%---------------------------------------------------------------------
&&\hspace{2.8cm}
-3(1-Q_\beta){\partial^3\over\partial x\partial z^2}
\Big[(y-z){\Phi(x,y,z)-\varphi_0(y,z)\over x}-\Phi(x,y,z)\Big]
\nonumber\\
%---------------------------------------------------------------------
&&\hspace{2.8cm}
+3(1-Q_\beta){\partial^3\over\partial x\partial y\partial z}\Big[
(y-z){\Phi(x,y,z)-\varphi_0(y,z)\over x}-\Phi(x,y,z)\Big]
\nonumber\\
%---------------------------------------------------------------------
&&\hspace{2.8cm}
-2{\partial^4\over\partial x^4}(x\Phi)(x,y,z)
+3{\partial^4\over\partial x^3\partial z}\Big[(y-z)\Phi
-x\Phi\Big](x,y,z)
\nonumber\\
%---------------------------------------------------------------------
&&\hspace{2.8cm}
-6{\partial^4\over\partial x^2\partial z^2}\Big(z\Phi
(x,y,z)\Big)-6{\partial^3\Phi\over\partial x^3}(x,y,z)
\nonumber\\
%---------------------------------------------------------------------
&&\hspace{2.8cm}
+3{\partial^3\over\partial x^2\partial z}\Big[(3-Q_\beta)(y-z)
{\Phi(x,y,z)-\varphi_0(y,z)\over x}+(1-Q_\beta)\Phi(x,y,z)\Big]
\nonumber\\
%---------------------------------------------------------------------
&&\hspace{2.8cm}
+3(1-Q_\beta){\partial^3\over\partial x^2\partial y}\Big[(y-z)
{\Phi(x,y,z)-\varphi_0(y,z)\over x}-\Phi(x,y,z)\Big]\Big\}\;,
\nonumber\\
%%%%%%%%%%%%%%%%%%%%%%%%%%%%%%%%%%%%%%%%%%%%%%%%%%%%%%%%%%%%%%%%%%%%%%
&&T_4(x,y,z)=-{1\over16}\Big\{Q_\beta\Big[{2\over zx}
-2{\partial^3\Phi\over\partial x^2\partial z}(x,y,z)
\nonumber\\
%---------------------------------------------------------------------
&&\hspace{2.8cm}
+{\partial^3\over\partial x\partial z^2}\Big((y-z){\Phi(x,y,z)-\varphi_0(y,z)\over x}
-\Phi(x,y,z)\Big)\Big]
-Q_\alpha\Big[2{\partial^3\Phi\over\partial x^2\partial y}(x,y,z)
\nonumber\\
%---------------------------------------------------------------------
&&\hspace{2.8cm}
-{\partial^3\over\partial x\partial y\partial z}
\Big((y-z){\Phi(x,y,z)-\varphi_0(y,z)\over x}-\Phi(x,y,z)\Big)\Big]\Big\}\;,
\nonumber\\
%%%%%%%%%%%%%%%%%%%%%%%%%%%%%%%%%%%%%%%%%%%%%%%%%%%%%%%%%%%%%%%%%%%%%%
&&T_5(x,y,z)
={2\over3x}+\Big(-{4\over3x^2}+{4\ln x\over3x^2}\Big)(y+z)
\nonumber\\
%---------------------------------------------------------------------
&&\hspace{2.3cm}
+\Big({7\over6x^2}+{2\over3x^2}\ln x\Big)(y\ln y+z\ln z)
\nonumber\\
%---------------------------------------------------------------------
&&\hspace{2.3cm}
+\Big({2\over3x^3}-{4\over3x^3}\ln x\Big)(y-z)^2
(1+\varrho_{_{1,1}}(y,z))
\nonumber\\
%---------------------------------------------------------------------
&&\hspace{2.3cm}
+{23\over6x^2}(y+z)\varrho_{_{1,1}}(y,z)
-{5\varrho_{_{2,1}}(y,z)\over x^2}
\nonumber\\
%---------------------------------------------------------------------
&&\hspace{2.3cm}
-{1\over3x^2}\Big(1-{2(y+z)\over x}\Big)\Big(\Phi(x,y,z)
-\varphi_0(y,z)\Big)
\nonumber\\
%---------------------------------------------------------------------
&&\hspace{2.3cm}
+{1\over3x}\Big({y+z\over x}-{2(y-z)^2\over x^2}\Big)\varphi_1(y,z)
\nonumber\\
%---------------------------------------------------------------------
&&\hspace{2.3cm}
+{1\over3x}\Big(1-{3(y+z)\over x}+{2(y-z)^2\over x^2}\Big)
{\partial\Phi\over\partial x}(x,y,z)
\nonumber\\
%---------------------------------------------------------------------
&&\hspace{2.3cm}
-{1\over3}\Big(1-{2(y+z)\over x}+{(y-z)^2\over x^2}\Big)
{\partial^2\Phi\over\partial x^2}(x,y,z)
\nonumber\\
%---------------------------------------------------------------------
&&\hspace{2.3cm}
-{(y-z)^2\over3x^2}\varphi_2(y,z)\;,
\nonumber\\
%%%%%%%%%%%%%%%%%%%%%%%%%%%%%%%%%%%%%%%%%%%%%%%%%%%%%%%%%%%%%%%%%%%%%%
&&T_6(x,y,z)={4\over x}\ln z-{4\over x^2}\Big(y-y\ln y
-z+z\ln z\Big)
+{\partial^3\over\partial x^2\partial z}
\Big[(y-3z-x)\Phi(x,y,z)\Big]
\nonumber\\
%---------------------------------------------------------------------
&&\hspace{2.3cm}
-2{\partial^3\over\partial x\partial z^2}\Big[{yz-z^2\over x}
\Big(\Phi(x,y,z)-\varphi_0(y,z)\Big)\Big]
\nonumber\\
%---------------------------------------------------------------------
&&\hspace{2.3cm}
-{\partial^2\over\partial x\partial z}\Big[\Phi(x,y,z)
-{5\over x}(y-z)\Big(\Phi(x,y,z)
-\varphi_0(y,z)\Big)\Big]
\nonumber\\
%---------------------------------------------------------------------
&&\hspace{2.3cm}
-2{\partial^2\over\partial x^2}
\Big[{y-z\over x}\Big(\Phi(x,y,z)-\varphi_0(y,z)\Big)
+2\Phi(x,y,z)\Big]\;,
\nonumber\\
%%%%%%%%%%%%%%%%%%%%%%%%%%%%%%%%%%%%%%%%%%%%%%%%%%%%%%%%%%%%%%%%%%%%%%
&&T_{7}(x,y,z)
=-{1\over x^2}\Big(\varphi_0-(y-z)
{\partial\varphi_0\over\partial z}\Big)(y,z)
+\Big[2z{\partial^3\Phi\over\partial x\partial z^2}
+{\partial^2\Phi\over\partial x^2}
-{y-z\over x^2}{\partial\Phi\over\partial z}
\nonumber\\
%---------------------------------------------------------------------
&&\hspace{2.3cm}
+(x-y+z){\partial^3\Phi\over\partial x^2\partial z}+{\Phi\over x^2}
-{1\over x}{\partial\Phi\over\partial x}
+(1+{y-z\over x}){\partial^2\Phi\over\partial x\partial z}\Big]
(x,y,z)\;,
\nonumber\\
%%%%%%%%%%%%%%%%%%%%%%%%%%%%%%%%%%%%%%%%%%%%%%%%%%%%%%%%%%%%%%%%%%%%%%
&&T_{8}(x,y,z)
=-4{\partial^3\Phi\over\partial x^2\partial z}(x,y,z)
+{4\over xz}-{4\over x^2}\Big(\ln y-\ln z\Big)
+2\Big({\partial^3\over\partial x^2\partial z}
-{\partial^3\over\partial x\partial z^2}
\nonumber\\
%---------------------------------------------------------------------
&&\hspace{2.3cm}
+{\partial^3\over\partial x^2\partial y}
+{\partial^3\over\partial x\partial y\partial z}\Big)\Big[\Phi(x,y,z)
-{y-z\over x}\Big(\Phi(x,y,z)-\varphi_0(y,z)\Big)\Big]\;,
\nonumber\\
%%%%%%%%%%%%%%%%%%%%%%%%%%%%%%%%%%%%%%%%%%%%%%%%%%%%%%%%%%%%%%%%%%%%%%
&&T_{9}(x,y,z)
=-4\Big({\partial^3\Phi\over\partial x^2\partial z}
+{\partial^3\Phi\over\partial x^2\partial y}\Big)(x,y,z)
+{4\over xz}+{2\over x^2}(2+\ln y)
\nonumber\\
%---------------------------------------------------------------------
&&\hspace{2.3cm}
+\Big(2{\partial^3\over\partial x\partial z^2}
+{\partial^3\over\partial x^2\partial y}\Big)
\Big[{y-z\over x}\Big(\Phi(x,y,z)-\varphi_0(y,z)\Big)
-\Phi(x,y,z)\Big]\;,
\nonumber\\
%%%%%%%%%%%%%%%%%%%%%%%%%%%%%%%%%%%%%%%%%%%%%%%%%%%%%%%%%%%%%%%%%%%%%%
&&T_{10}(x,y,z)={4\over x}\ln z
-{8z\over x^2}\Big({\partial\Phi\over\partial z}(x,y,z)
-{\partial\varphi_0\over\partial z}(y,z)\Big)
+{8\over x}\Big({\partial\Phi\over\partial z}
-{\partial\Phi\over\partial x}\Big)(x,y,z)
\nonumber\\
%---------------------------------------------------------------------
&&\hspace{2.3cm}
+2{\partial^2\over\partial x\partial z}\Big({y-z\over x}\cdot
(\Phi(x,y,z)-\varphi_0(y,z))-\Phi(x,y,z)\Big)
+{8z\over x}{\partial^2\Phi\over\partial x\partial z}(x,y,z)
\;,\nonumber\\
%%%%%%%%%%%%%%%%%%%%%%%%%%%%%%%%%%%%%%%%%%%%%%%%%%%%%%%%%%%%%%%%%%%%%%
&&T_{11}(x,y,z)={1\over x}\Bigg\{-4(2+\ln y)(\ln x-1)
-{\partial\over\partial z}\Big[\Big(1+2{y-z\over x}\Big)\Phi\Big]
(x,y,z)
\nonumber\\
%---------------------------------------------------------------------
&&\hspace{2.8cm}
+{\partial\over\partial z}\Big[\Big(1+2{y-z\over x}\Big)
\varphi_0+2(y-z)\varphi_1\Big](y,z)\Bigg\}\;,
\nonumber\\
%%%%%%%%%%%%%%%%%%%%%%%%%%%%%%%%%%%%%%%%%%%%%%%%%%%%%%%%%%%%%%%%%%%%%%
&&T_{12}(x,y,z)
={1\over x}\Bigg[{\partial \Phi\over\partial z}
(x,y,z)-{\partial\varphi_0\over\partial z}(y,z)\Bigg]
\;,\nonumber\\
%%%%%%%%%%%%%%%%%%%%%%%%%%%%%%%%%%%%%%%%%%%%%%%%%%%%%%%%%%%%%%%%%%%%%%
&&F_1(x,y,z,u)=2\Big((2-Q_\beta)\ln u+1-2Q_\beta\Big)\varrho_{_{0,1}}(x,y)
-{6(z-u)\over xy}
\nonumber\\
&&\hspace{2.7cm}
-{6(z\ln z-u\ln u)\over xy}+{Q_\beta xy+2(x+y)(z-u)\over x^2y^2}\varphi_0(z,u)
\nonumber\\
&&\hspace{2.7cm}
-{Q_\beta z-(2+Q_\beta)u\over xy}{\partial\varphi_0\over\partial u}(z,u)
-{u(z-u)\over xy}{\partial^2\varphi_0\over\partial u^2}(z,u)
\nonumber\\
&&\hspace{2.7cm}
+\Big(Q_\beta-(Q_\beta z-(2+Q_\beta)u){\partial\over\partial u}-u
(z-u){\partial^2\over\partial u^2}\Big)\Omega_{-1}(x,y;
z,u)
\nonumber\\
&&\hspace{2.7cm}
-\Big({\partial\over\partial u}+u{\partial^2\over\partial u^2}\Big)\Omega_0(x,y;z,u)
-\Big({\partial\over\partial x}+{\partial\over\partial y}\Big)^2\Big[\Omega_1(x,y;z,u)
\nonumber\\
&&\hspace{2.7cm}
+(z-u)\Omega_0(x,y;z,u)\Big]-2\Big({\partial\over\partial x}+{\partial\over\partial y}\Big)
\Big[{\partial\Omega_1\over\partial u}(x,y;z,u)
\nonumber\\
&&\hspace{2.7cm}
-(z+u){\partial\Omega_0\over\partial u}(x,y;z,u)\Big]-2(z-u)
\Big({\partial\over\partial x}+{\partial\over\partial y}\Big)\Omega_{-1}(x,y;z,u)\;,
\nonumber\\
%%%%%%%%%%%%%%%%%%%%%%%%%%%%%%%%%%%%%%%%%%%%%%%%%%%%%%%%%%%%%%%%%%%%%%
&&F_2(x,y,z,u)=2\Big(\ln u-1-(1-Q_\beta)(2+\ln z)\Big)\varrho_{_{0,1}}(x,y)
-{6(z-u)\over xy}
\nonumber\\
&&\hspace{2.7cm}
-{6(z\ln z-u\ln u)\over xy}-{Q_\beta xy-2(x+y)(z-u)\over x^2y^2}\varphi_0(z,u)
\nonumber\\
&&\hspace{2.7cm}
+{z+u\over xy}{\partial\varphi_0\over\partial u}(z,u)+(1-Q_\beta){z-u\over xy}
{\partial\varphi_0\over\partial z}(z,u)-{u(z-u)\over xy}
{\partial^2\varphi_0\over\partial u^2}(z,u)
\nonumber\\
&&\hspace{2.7cm}
+\Big(-Q_\beta+(z+u){\partial\over\partial u}
+(1-Q_\beta)(z-u){\partial\over\partial z}
-u(z-u){\partial^2\over\partial u^2}\Big)\Omega_{-1}(x,y;z,u)
\nonumber\\
&&\hspace{2.7cm}
+\Big(-{\partial\over\partial u}-(1-Q_\beta){\partial\over\partial z}
+u{\partial^2\over\partial u^2}\Big)\Omega_0(x,y;z,u)
\nonumber\\
&&\hspace{2.7cm}
+\Big({\partial\over\partial x}+{\partial\over\partial y}\Big)^2
\Big[\Omega_1(x,y;z,u)-(z-u)\Omega_0(x,y;z,u)\Big]
\nonumber\\
&&\hspace{2.7cm}
+\Big({\partial\over\partial x}+{\partial\over\partial y}\Big)
\Big[-2\Omega_0(x,y;z,u)+4u{\partial\Omega_0\over\partial u}(x,y;z,u)\Big]
\nonumber\\
&&\hspace{2.7cm}
-2(z-u)\Big({\partial\over\partial x}+{\partial\over\partial y}\Big)
\Omega_{-1}(x,y;z,u)\;,
\nonumber\\
%%%%%%%%%%%%%%%%%%%%%%%%%%%%%%%%%%%%%%%%%%%%%%%%%%%%%%%%%%%%%%%%%%%%%%
&&F_3(x,y,z,u)=-2(2+\ln u)\varrho_{_{0,1}}(x,y)+{1\over xy}\varphi_0(z,u)
-{z-u\over xy}{\partial\varphi_0\over\partial u}(z,u)
\nonumber\\
&&\hspace{2.7cm}
+(1-2Q_\beta){\partial\over\partial u}\Omega_0(x,y;z,u)
+\Big(1-(z-u){\partial\over\partial u}\Big)\Omega_{-1}(x,y;z,u)\;,
\nonumber\\
%%%%%%%%%%%%%%%%%%%%%%%%%%%%%%%%%%%%%%%%%%%%%%%%%%%%%%%%%%%%%%%%%%%%%%
&&F_4(x,y,z,u)=2\Big(2Q_\beta+\ln u-(1-Q_\beta)\ln z\Big)\varrho_{_{0,1}}(x,y)
\nonumber\\
&&\hspace{2.7cm}
-{Q_\beta \over xy}\varphi_0(z,u)+{z-u\over xy}\Big(
{\partial\over\partial u}+(1-Q_\beta){\partial\over\partial z}\Big)\varphi_0(z,u)
\nonumber\\
&&\hspace{2.7cm}
+\Big(-Q_\beta+(z-u){\partial\over\partial u}+(1-Q_\beta)(z-u){\partial\over\partial z}
\Big)\Omega_{-1}(x,y;z,u)
\nonumber\\
&&\hspace{2.7cm}
-\Big({\partial\over\partial u}+(1-Q_\beta){\partial\over\partial z}\Big)
\Omega_0(x,y;z,u)
\;,\nonumber\\
%%%%%%%%%%%%%%%%%%%%%%%%%%%%%%%%%%%%%%%%%%%%%%%%%%%%%%%%%%%%%%%%%%%%%%
&&F_5(x,y,z,u)={1\over xy}{\partial\over\partial u}\Big((z-u)\varphi_0\Big)(z,u)
+{1\over x-y}\Big\{{\partial\over\partial
u}\Big[\Big(1+{z-u\over x}\Big)\Phi\Big](x,z,u)
\nonumber\\
&&\hspace{2.7cm}
-{\partial\over\partial u}\Big[\Big(1+{z-u\over y}\Big)\Phi\Big](y,z,u)\Big\}\;,
\nonumber\\
%%%%%%%%%%%%%%%%%%%%%%%%%%%%%%%%%%%%%%%%%%%%%%%%%%%%%%%%%%%%%%%%%%%%%%
&&F_6(x,y,z,u)=-{1\over xy}{\partial\over\partial u}\Big((z-u)\varphi_0\Big)(z,u)
+{1\over x-y}\Big\{{\partial\over\partial
u}\Big[\Big(1-{z-u\over x}\Big)\Phi\Big](x,z,u)
\nonumber\\
&&\hspace{2.7cm}
-{\partial\over\partial u}\Big[\Big(1-{z-u\over y}\Big)\Phi\Big](y,z,u)\Big\}\;.
%%%%%%%%%%%%%%%%%%%%%%%%%%%%%%%%%%%%%%%%%%%%%%%%%%%%%%%%%%%%%%%%%%%%%%
\label{form}
\end{eqnarray}

\end{document}